\documentclass[12pt,a4paper]{article}
\pdfoutput=1
 
\usepackage{jheppub-arxiv}
\usepackage[usenames,dvipsnames,table]{xcolor}
 
\usepackage{amssymb} 
\usepackage{amsmath}
\usepackage{mathtools}
\usepackage{amsfonts}    
\usepackage{dsfont}
\usepackage{pdfpages}
\usepackage{verbatim}
\usepackage{stmaryrd}
\SetSymbolFont{stmry}{bold}{U}{stmry}{m}{n}
\usepackage{graphicx}
\usepackage{import}
\usepackage{etoc}
\usepackage{enumitem}
\usepackage{tensor}
\usepackage{mathrsfs}
\usepackage{textgreek} 
\usepackage[mathscr]{euscript}
\usepackage[smalltableaux]{ytableau}
\ytableausetup{centertableaux}
\usepackage{tikz}
\usetikzlibrary{decorations.pathreplacing, decorations.markings,calc,shapes.misc,decorations.pathmorphing,patterns.meta, math,positioning}
\usepackage{tabularx,ragged2e}
\usepackage{physics}

\usepackage{slashed}
\usepackage{datetime}
\usepackage{hyperref}
\hypersetup{
    pdfencoding=unicode,
	colorlinks=true,
	urlcolor=Maroon,
	linkcolor=RoyalBlue,
	citecolor=Maroon,
	pdftitle={Records from the S-Matrix Marathon: Asymptotic Observables},
	pdfauthor={},
	pdfdisplaydoctitle=true,
	pdfstartview=FitH,
	linktocpage=true
}

\usetikzlibrary{shapes}
\usepackage{enumitem}
\usepackage[export]{adjustbox}
\usepackage{nccmath}
\usepackage{bbm}
\usepackage{braket}
\usepackage{empheq}
\usepackage{cancel}
\usepackage{cleveref}
\usepackage[most]{tcolorbox}
\newcommand{\defQ}[1]{\textcolor{black}{\emph{#1}}}
\definecolor{charcoal}{HTML}{343837}
\newcommand{\sumint}{
  \mathop{%
    \mathchoice%
    {\ooalign{$\displaystyle\sum$\cr\hidewidth$\displaystyle\int$\hidewidth\cr}}%
    {\ooalign{$\textstyle\sum$\cr\hidewidth$\textstyle\int$\hidewidth\cr}}%
    {\ooalign{$\scriptstyle\sum$\cr\hidewidth$\scriptstyle\int$\hidewidth\cr}}%
    {\ooalign{$\scriptscriptstyle\sum$\cr\hidewidth$\scriptscriptstyle\int$\hidewidth\cr}}%
  }\displaylimits
}

\usepackage{bm}
\definecolor{yellowish}{rgb}{0.880722,0.611041,0.142051}

\newcommand{\ba}{\begin{align}}

\newcommand{\be}{\begin{equation}}
\newcommand{\ee}{\end{equation}}
\def\bd{\begin{tikzpicture}}
\def\ed{\end{tikzpicture}}

\renewcommand{\vec}[1]{\mathbf{#1}}
\allowdisplaybreaks

\def\XXint#1#2#3{{\setbox0=\hbox{$#1{#2#3}{\int}$}
     \vcenter{\hbox{$#2#3$}}\kern-.5\wd0}}

\definecolor{light-gray}{gray}{0.75}

\renewcommand\d{\text{d}}

\newcommand{\e}{\mathrm{e}}

\newcommand{\R}{\mathrm{R}}

\newcommand{\D}{\mathbb{D}}

\newcommand{\eps}{\varepsilon}

\renewcommand{\R}{\mathbb{R}}

\renewcommand{\le}{\leqslant}
\renewcommand{\ge}{\geqslant}
\renewcommand{\leq}{\leqslant}
\renewcommand{\geq}{\geqslant}

\def\Dden{\mathcal{D}}
\def\bary{\bar{y}}
\def\barv{\bar{v}}
\def\barp{\bar{p}}
\def\barm{\bar{m}}
\def\barw{\bar{w}}

\definecolor{dpurple}  {RGB} {189,  147,  249}

\usepackage{ntheorem}
\usepackage[framemethod=tikz,footnoteinside=false,hidealllines=true,topline=true,bottomline=true,linecolor=black!10!white,linewidth=1pt,innerleftmargin=0em,innerrightmargin=0em,frametitlerule=true,ntheorem]{mdframed}
\theorembodyfont{\upshape}

\newmdtheoremenv{mtheorem}{Theorem}[section]
\newmdtheoremenv[]{mdexample}{Example}[section]
\newmdtheoremenv{mdremark}{Remark}[section]
\newmdtheoremenv{mddefinition}{Definition}[section]
\newmdtheoremenv{mdcorollary}{Corollary}[section]
\newmdtheoremenv{mdproposition}{Proposition}[section]
\newmdtheoremenv{QA}{Audience question}[section]

\title{{\Large\normalfont Records from the S-Matrix Marathon:}\\ Asymptotic Observables: The Analytic S-Matrix Revisited}

\author{{\normalfont Lecturers:}}
\author[1]{Simon~Caron-Huot,}
\author[1]{Mathieu~Giroux}

\author{\\ {\normalfont Notes written by:}}
\author[1]{Mathieu~Giroux,}
\author[2]{Holmfridur~S.~Hannesdottir,}
\author[2,3,4]{Sebastian~Mizera,}
\author[1]{Celina~Pasiecznik}

\affiliation[1]{Department of Physics, McGill University, 3600 Rue University,\\ Montr\'eal, H3A 2T8, QC Canada}
\affiliation[2]{Institute for Advanced Study, Princeton, NJ 08540, USA}
\affiliation[3]{Department of Physics, Princeton University, Princeton, NJ 08544, USA}
\affiliation[4]{Princeton Center for Theoretical Science,\\ Princeton University, Princeton, NJ 08544, USA}

\abstract{
These lectures introduce the notion of \emph{asymptotic observables}, which are classes of measurable quantities predicted by quantum field theory.
In gapped theories with trivial infrared dynamics, 
these include scattering amplitudes, expectation values of operators approaching infinity, inclusive cross sections, etc. We argue for a unified picture in which various asymptotic observables are related by analytic continuation embodying new versions of crossing symmetry. As an application, we discuss the exponentiation of infrared divergences for inclusive observables in Quantum Electrodynamics using time-folded contours. Finally, we outline prospects for a systematic study of analyticity properties of asymptotic observables using the Fourier--Bros--Iagolnitzer transform.

These notes are based on a series of lectures held during the S-Matrix Marathon workshop at the Institute for Advanced Study on 11--22 March 2024.
}

\begin{document}

\maketitle
\setcounter{page}{1}

\setcounter{tocdepth}{4}
\setcounter{secnumdepth}{4}

\makeatletter
\g@addto@macro\bfseries{\boldmath}
\makeatother

\newpage
\section*{Preface}

This article is a chapter from the \emph{Records from the S-Matrix Marathon}, a series of lecture notes covering selected topics on scattering amplitudes~\cite{RecordsBook}. They are based on lectures delivered during a workshop on 11--22 March 2024 at the Institute for Advanced Study in Princeton, NJ. We hope that they can serve as a pedagogical introduction to the topics surrounding the S-matrix theory.

These lecture notes were prepared by the above-mentioned note-writers in collaboration with the lecturers. 

\vfill
\section*{Acknowledgments}

S.C.H.'s, M.G.’s, and C.P.'s work is supported in parts by the National Science and Engineering Council of Canada (NSERC) and the Canada Research
Chair program, reference number CRC-2022-00421.
S.C.H.'s work is additionally supported by a Simons Fellowships in Theoretical Physics
and by the Simons Collaboration on the Nonperturbative Bootstrap.
C.P. is also supported by the Walter C. Sumner Memorial Fellowship.
H.S.H. gratefully acknowledges funding provided by the J. Robert Oppenheimer Endowed Fund of the Institute for Advanced Study.
S.M. gratefully acknowledges funding provided by the Sivian Fund and the Roger Dashen Member Fund at the Institute for Advanced Study. 
This material is based upon work supported by the U.S. Department of Energy, Office of Science, Office of High Energy Physics under Award Number DE-SC0009988.

The S-Matrix Marathon workshop was sponsored by the Institute for Advanced Study and the Carl P. Feinberg Program in Cross-Disciplinary Innovation.

\def\D{\mathrm{D}}
\def\e{\mathrm{e}}
\def\T{\mathrm{T}}
\def\sgn{\mathrm{sgn}}
\def\as{\mathrm{as}}
\def\inn{\mathrm{in}}
\def\out{\mathrm{out}}
\definecolor{myblue}{rgb}{.7, .7, 1}
\newcommand\mybluebox[1]{\colorbox{myblue}{\vspace{2em}\hspace{2em}#1\hspace{2em}\vspace{2em}}}
\def\vacR{|0\>}
\def\vacL{\<0|}
\def\vac#1{\<0|#1|0\>}
\def\adag{a^\dagger}
\def\bdag{b^\dagger}
\def\>{\rangle}
\def\<{\langle}
\def\be{\begin{equation}}
\def\ee{\end{equation}}
\def\ot{\leftarrow}
\def\cM{\mathcal{M}}
\def\bary{\bar{y}}
\def\barv{\bar{v}}
\def\barp{\bar{p}}
\def\barm{\bar{m}}
\def\barw{\bar{w}}

\makeatletter
\newenvironment{sqcases}{
  \matrix@check\sqcases\env@sqcases
}{
  \endarray\right.
}
\def\env@sqcases{
  \let\@ifnextchar\new@ifnextchar
  \left\lbrack
  \def\arraystretch{1.2}
  \array{@{}l@{\quad}l@{}}
}
\makeatother

\newcommand{\la}{\langle}
\newcommand{\ra}{\rangle}

\usetikzlibrary{decorations.markings}
\usetikzlibrary{decorations.pathmorphing}
\def\bdelta{\bm{\delta}}
\def\Dden{\mathcal{D}}
\def\cC{\mathcal{C}}
\def\cO{\mathcal{O}}
\def\cR{R}
\def\cT{T}
\def\cTbar{\overline{T}}
\def\rone{\mathrm{I}}
\def\rtwo{\mathrm{II}}
\def\rthree{\mathrm{III}}
\def\rdiff{{\text{diff}}}
\def\ravg{\mathrm{avg}}
\def\GN{G}
\def\Tb#1{\overline{T}(#1)}
\def\T#1{T(#1)}
\def\R#1#2{R_{#1}(#2)}
\newcommand{\Lim}[1]{\raisebox{0.5ex}{\scalebox{0.8}{$\displaystyle \lim_{#1}\;$}}}

\tikzset{
    vector/.style={
        decoration={snake, aspect=0.75, mirror, segment length=2mm},
        decorate
    },
	photon/.style={decorate, decoration={snake, amplitude=1pt, segment length=6pt}
	}
}

\tikzset{gradGtoB/.style={
    postaction={
        decorate,
        decoration={
            markings,
            mark=at position \pgfdecoratedpathlength-0.5pt with {\arrow[ForestGreen,line width=#1] {}; },
            mark=between positions 0 and \pgfdecoratedpathlength-0pt step 0.5pt with {
                \pgfmathsetmacro\myval{multiply(divide(
                    \pgfkeysvalueof{/pgf/decoration/mark info/distance from start}, \pgfdecoratedpathlength),100)};
                \pgfsetfillcolor{RoyalBlue!\myval!ForestGreen!};
                \pgfpathcircle{\pgfpointorigin}{#1};
                \pgfusepath{fill};}
}}}}
\tikzset{gradRtoG/.style={
    postaction={
        decorate,
        decoration={
            markings,
            mark=at position \pgfdecoratedpathlength-0.5pt with {\arrow[Maroon,line width=#1] {}; },
            mark=between positions 0 and \pgfdecoratedpathlength-0pt step 0.5pt with {
                \pgfmathsetmacro\myval{multiply(divide(
                    \pgfkeysvalueof{/pgf/decoration/mark info/distance from start}, \pgfdecoratedpathlength),100)};
                \pgfsetfillcolor{ForestGreen!\myval!Maroon!};
                \pgfpathcircle{\pgfpointorigin}{#1};
                \pgfusepath{fill};}
}}}}
\tikzset{gradBlackToR/.style={
    postaction={
        decorate,
        decoration={
            markings,
            mark=at position \pgfdecoratedpathlength-0.5pt with {\arrow[black,line width=#1] {}; },
            mark=between positions 0 and \pgfdecoratedpathlength-0pt step 0.5pt with {
                \pgfmathsetmacro\myval{multiply(divide(
                    \pgfkeysvalueof{/pgf/decoration/mark info/distance from start}, \pgfdecoratedpathlength),100)};
                \pgfsetfillcolor{Maroon!\myval!black!};
                \pgfpathcircle{\pgfpointorigin}{#1};
                \pgfusepath{fill};}
}}}}

\tikzset{gradBtoBlack/.style={
    postaction={
        decorate,
        decoration={
            markings,
            mark=at position \pgfdecoratedpathlength-0.5pt with {\arrow[blue,line width=#1] {}; },
            mark=between positions 0 and \pgfdecoratedpathlength-0pt step 0.5pt with {
                \pgfmathsetmacro\myval{multiply(divide(
                    \pgfkeysvalueof{/pgf/decoration/mark info/distance from start}, \pgfdecoratedpathlength),100)};
                \pgfsetfillcolor{black!\myval!RoyalBlue!};
                \pgfpathcircle{\pgfpointorigin}{#1};
                \pgfusepath{fill};}
}}}}

\tikzset{gradRtoB/.style={
    postaction={
        decorate,
        decoration={
            markings,
            mark=at position \pgfdecoratedpathlength-0.5pt with {\arrow[RoyalBlue,line width=#1] {}; },
            mark=between positions 0 and \pgfdecoratedpathlength-0pt step 0.5pt with {
                \pgfmathsetmacro\myval{multiply(divide(
                    \pgfkeysvalueof{/pgf/decoration/mark info/distance from start}, \pgfdecoratedpathlength),100)};
                \pgfsetfillcolor{RoyalBlue!\myval!Maroon!};
                \pgfpathcircle{\pgfpointorigin}{#1};
                \pgfusepath{fill};}
}}}}

\newcommand{\tikzxmark}{%
\tikz[scale=5.23] {
    \draw[line width=0.7,line cap=round] (0,0) to [bend left=6] (1,1);
    \draw[line width=0.7,line cap=round] (0.2,0.95) to [bend right=3] (0.8,0.05);
}}

\newcommand{\cloud}{\tikzset{every picture/.style={line width=0.75pt}}  
\begin{tikzpicture}[x=0.75pt,y=0.75pt,yscale=-0.5,xscale=0.5]
 Cloud
\draw[color=charcoal,fill=white  ,fill opacity=1 ]   (107.28,126.75) .. controls (106.64,121.92) and (108.75,117.13) .. (112.73,114.42) .. controls (116.7,111.71) and (121.85,111.56) .. (125.97,114.03) .. controls (127.43,111.21) and (130.1,109.27) .. (133.18,108.79) .. controls (136.26,108.3) and (139.38,109.33) .. (141.6,111.57) .. controls (142.84,109.02) and (145.29,107.31) .. (148.06,107.04) .. controls (150.84,106.77) and (153.55,107.98) .. (155.24,110.25) .. controls (157.49,107.55) and (161.07,106.41) .. (164.42,107.33) .. controls (167.78,108.25) and (170.32,111.06) .. (170.93,114.55) .. controls (173.69,115.31) and (175.98,117.27) .. (177.22,119.9) .. controls (178.47,122.53) and (178.53,125.58) .. (177.41,128.27) .. controls (180.12,131.88) and (180.75,136.68) .. (179.07,140.89) .. controls (177.39,145.11) and (173.65,148.09) .. (169.24,148.74) .. controls (169.21,152.69) and (167.09,156.32) .. (163.7,158.22) .. controls (160.3,160.12) and (156.16,160.01) .. (152.88,157.91) .. controls (151.48,162.65) and (147.54,166.13) .. (142.76,166.86) .. controls (137.99,167.59) and (133.23,165.43) .. (130.55,161.32) .. controls (127.26,163.34) and (123.31,163.93) .. (119.59,162.94) .. controls (115.88,161.95) and (112.71,159.46) .. (110.8,156.05) .. controls (107.44,156.45) and (104.19,154.67) .. (102.67,151.59) .. controls (101.14,148.51) and (101.66,144.79) .. (103.98,142.27) .. controls (100.98,140.47) and (99.45,136.89) .. (100.18,133.4) .. controls (100.92,129.92) and (103.76,127.31) .. (107.21,126.94) ; \draw[color=charcoal]   (103.98,142.27) .. controls (105.39,143.13) and (107.03,143.51) .. (108.66,143.38)(110.8,156.05) .. controls (111.51,155.96) and (112.2,155.79) .. (112.85,155.52)(130.55,161.32) .. controls (130.05,160.56) and (129.64,159.75) .. (129.31,158.9)(152.88,157.91) .. controls (153.13,157.05) and (153.3,156.16) .. (153.37,155.26)(169.24,148.74) .. controls (169.28,144.53) and (166.94,140.67) .. (163.23,138.83)(177.41,128.27) .. controls (176.81,129.7) and (175.89,130.97) .. (174.73,131.98)(170.93,114.55) .. controls (171.04,115.12) and (171.08,115.71) .. (171.07,116.3)(155.24,110.25) .. controls (154.68,110.92) and (154.22,111.68) .. (153.87,112.49)(141.6,111.57) .. controls (141.3,112.18) and (141.08,112.83) .. (140.93,113.5)(125.97,114.03) .. controls (126.84,114.55) and (127.65,115.18) .. (128.37,115.9)(107.28,126.75) .. controls (107.37,127.42) and (107.51,128.08) .. (107.7,128.72) ;
\end{tikzpicture}}

\def\centerarc[#1](#2)(#3:#4:#5)%
    { \draw[#1] ($(#2)+({#5*cos(#3)},{#5*sin(#3)})$) arc (#3:#4:#5); }

\newpage

\section*{\label{ch:caronHuotGiroux}Asymptotic Observables: The Analytic S-Matrix Revisited\\
\normalfont{\textit{Simon Caron-Huot, Mathieu Giroux}}}

\setcounter{section}{0}

\noindent\rule{\textwidth}{0.25pt}
\vspace{-0.8em}
\etocsettocstyle{\noindent\textbf{Contents}\vskip0pt}{}
\localtableofcontents
\vspace{0.5em}
\noindent\rule{\textwidth}{0.25pt}
\vspace{1em}

\section[Asymptotic observables and crossing ]{Asymptotic observables and crossing \\ \normalfont{\textit{Simon Caron-Huot}}}

In this first lecture, we introduce a minimal set of axioms that allow us to enumerate asymptotic observables of a given multiplicity. We will then explain why these are, remarkably, interconnected through crossing symmetry. The main references on which these lectures are based are \cite{Caron-Huot:2023vxl,Caron-Huot:2023ikn}.

\subsection{Summary of conventions}

First, we spell out here the conventions assumed in this chapter, which follow \cite{Caron-Huot:2023ikn}. Unless specified otherwise, we work in $\D$ spacetime dimensions and take the momentum components of $p^\mu$ to be
\begin{equation}
    p^\mu = (p^0, \underbracket[0.4pt]{p^1, \ldots, p^{\D-1}}_{=\vec{p}})\,.
\end{equation}
We work in mostly-plus signature such that $p^2=-(p^0)^2+(p^1)^2 + \ldots + (p^{\D-1})^2$ and all-outgoing conventions, in which $p_i^0 < 0$ for the incoming particles and $p_i^0 > 0$ for the outgoing ones. When using lightcone coordinates, we define
\begin{equation}
    p^{\pm} = p^0\pm p^{\D-1}\,, \quad \vec{p}^\perp = (p^1, \ldots, p^{\D-2}) \,,
\end{equation}
such that $p^2 = - p^+ p^- + (\vec{p}^{\perp})^2$.
Furthermore, we use the notation $p_{ij \cdots k}^\mu = p_i^\mu + p_j^\mu + \ldots + p_k^\mu$, and $s_{ij \cdots k} = - p_{ij \cdots k}^2$ for Mandelstam invariants.

In what follows, the \defQ{connected interacting part} $T$ of the S-matrix is obtained in the usual way:
\be
    S=\mathbbm{1}+ i T+\ldots\,,
\ee 
where ``$\ldots$'' include disconnected products of $T$'s and are only present for multiplicities greater than $2\ot 2$.
The interacting part of the \defQ{scattering amplitude}, $\mathcal{M}$, is defined as the matrix element of $T$ with the overall momentum-conserving delta function stripped out: 
\begin{equation}
    \mathcal{M}_{f \ot i} = \langle f | T | i \rangle (2\pi)^\D \delta^\D \Big(\sum_i p_i^\mu\Big)\,.
\end{equation} 
When drawing diagrams, we will follow the operator ordering in bra-ket notation (see, e.g., \cref{eq:5obs2,eq:6ptsEx1,eq:6ptsEx2} below).
This means (thanks Dirac!) that time flows toward the \emph{left} in scattering amplitudes $S$ or $\cM$,
and toward the right in conjugated amplitudes $S^\dag$ or $\cM^\dag$ factors.

Finally, ``in'' scattering states are denoted without explicit labels: $\bra{...}={}_{\text{in}}\bra{...}$ and $\ket{...}=\ket{...}_{\text{in}}$. In contrast, the ``out'' states will remain labeled: ${}_{\text{out}}\bra{...}$ and $\ket{...}_{\text{out}}$.

\subsection{Asymptotic observables}

The discussion that follows is motivated by the observation that, in local quantum field theory, causality admits a precise statement in terms of \defQ{microcausality}. It asserts that operators corresponding to observables at spacelike separated points must commute:
\begin{equation}\label{eq:causality0}
    [O(x),O(y)]=0 \qquad \text{(spacelike)}\,,
\end{equation}
ensuring that it is impossible to send signals faster than light.

Vacuum matrix elements of such relations,
$\bra{0}[\Phi(y),\Phi^\dagger(x)]\ket{0} = 0$ 
are well known to imply that any particle must have an antiparticle of the same mass and spin. That is, the two terms in the commutator can only cancel out
if the amplitude for a $\Phi$ particle to move from $x$ to $y$ (which is nonvanishing spacelike) coincides with that of a corresponding antiparticle to go from $y$ to $x$:
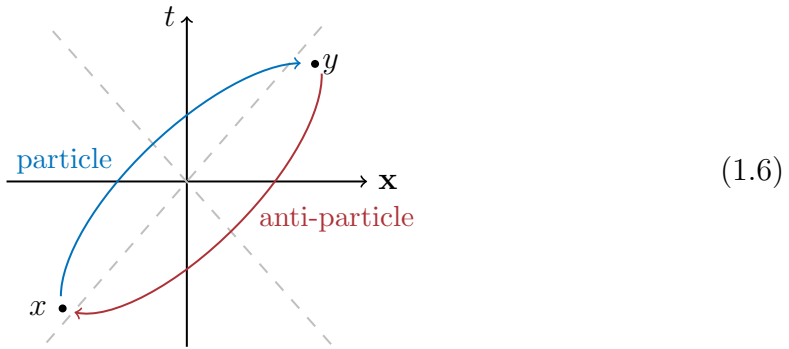
\begin{equation}\label{eq:path02}
         \adjustbox{valign=c}{
\tikzset{every picture/.style={line width=0.75pt}} 
\begin{tikzpicture}[x=0.75pt,y=0.75pt,yscale=-1,xscale=1]
 
\draw[<-]    (240,49) -- (240,215.29) node[pos=0,left]{$t$};
 
\draw[->]    (150,132.14) -- (330,132.14) node[right]{$\vec{x}$};
 
\draw  [draw opacity=0][fill=black  ,fill opacity=1 ] (176,196) .. controls (176,194.9) and (176.9,194) .. (178,194) .. controls (179.1,194) and (180,194.9) .. (180,196) .. controls (180,197.1) and (179.1,198) .. (178,198) .. controls (176.9,198) and (176,197.1) .. (176,196) node[left]{$x$} -- cycle ;
 
\draw  [dash pattern={on 4.5pt off 4.5pt},color=gray!50]  (170,213.14) -- (310,51.14) ;
 
\draw  [dash pattern={on 4.5pt off 4.5pt},color=gray!50]  (312,214) -- (170,53) ;
 
\draw  [draw opacity=0][fill=black  ,fill opacity=1 ] (302,73) .. controls (302,71.9) and (302.9,71) .. (304,71) .. controls (305.1,71) and (306,71.9) .. (306,73) .. controls (306,74.1) and (305.1,75) .. (304,75) .. controls (302.9,75) and (302,74.1) .. (302,73) node[right]{$y$} -- cycle ;
 Arc 
\draw  [draw opacity=0] (177.13,189.85) .. controls (177.03,171.04) and (194.03,141.25) .. (221.8,114.5) .. controls (248.66,88.62) and (278,72.92) .. (296.89,72.65) -- (246.89,140.55) -- cycle ; \draw[->,color=RoyalBlue]   (177.13,189.85) .. controls (177.03,171.04) and (194.03,141.25) .. (221.8,114.5) .. controls (248.66,88.62) and (278,72.92) .. (296.89,72.65) node[midway,xshift=-2.2cm,yshift=-1cm]{\small particle};  
 Arc 
\draw  [draw opacity=0] (307.32,77.87) .. controls (307.96,96.66) and (291.82,126.93) .. (264.83,154.46) .. controls (235.19,184.7) and (202,201.91) .. (183.9,197.9) -- (239,129.14) -- cycle ; \draw[->,color=Maroon]   (307.32,77.87) .. controls (307.96,96.66) and (291.82,126.93) .. (264.83,154.46) .. controls (235.19,184.7) and (202,201.91) .. (183.9,197.9) node[midway,xshift=2.5cm,yshift=1cm]{\small anti-particle};  
\end{tikzpicture}
}
\end{equation}

The question we wish to explore below is the following: What are the implications of \eqref{eq:causality0} inside a general scattering state?
As we shall see, this will relate the amplitudes of particles and those of antiparticles: \defQ{crossing symmetry}, but the answer will also reveal new and non-trivial relations among \defQ{out-of-time-order} (OTO) observables.  The OTO nature of crossing can already be anticipated, since commutators alter the ordering of the operators.

In time-ordered perturbation theory, the relation \eqref{eq:causality0} (and the analogous anti-commutation relation for fermions) ensures Lorentz invariance of T-products \cite{Weinberg:1995mt}.  Here we will directly use \eqref{eq:causality0}, which seems to buy us more mileage at the end of the day.

Next, let us outline the assumptions used to make this discussion more precise and to derive the results below.

\subsubsection{S-matrix axioms}
The idealized setup that epitomizes the idea of the S-matrix is a quantum field theory (QFT) with trivial infrared dynamics (e.g., pion scattering in $\D=4$, QED and QCD in dimensional regularization, etc.).  In this setup, it makes sense to \emph{assume} that any finite energy excitation decays at late times into a finite set of stable particles. These particles then separate from each other and effectively become free as they cease to overlap. This assumption and its consequences are encoded in the following set of axioms \cite{Caron-Huot:2023ikn} (similar sets of axioms have a long history \cite{Omnes:1960oya}):
\begin{enumerate}[label=(\roman*)]
\item The algebra of asymptotic measurements in the \emph{far past} is generated by \defQ{creation} and \defQ{annihilation operators} of stable particles,
satisfying the canonical relation:
\begin{equation} \label{canonical}
 [a_i,a^\dagger_{j}] =
\delta_{i,j}\ 2p_i^{0}\ (2\pi)^{\D-1} \delta^{\D-1}(\vec{p}_i-\vec{p}_j) ~~ \text{and} ~~  [a_i,a_{j}]= [a_i^\dagger,a^\dagger_{j}]=0\,.
\end{equation}
Here, $p_i^{0}$ is the (positive) energy of the $i^{\text{th}}$ particle and
$\delta_{i,j}$ denotes a Kronecker delta in the flavor and spin indices.
\item[(i')] There is an equivalent algebra of ``out'' measurements in the
\emph{far future}, which we denote by $b$ and $\bdag$'s.
Translated to textbook conventions, the above creation and annihilation operators correspond to
\be
a_i \equiv a_i^{\mathrm{in}} \quad \text{and} \quad b_i \equiv a_i^{\mathrm{out}}\, .
\ee
\item These operators act on equivalent Hilbert spaces and are related by a unitary \defQ{evolution operator} $S$:
\begin{equation} \label{b from a}
    b_i = S^\dagger a_i S, \qquad
    b_i^\dagger = S^\dagger a_i^\dagger S\,.
\end{equation}
It follows from \eqref{b from a} that $[a,b]$ can be a complicated object. This is expected because measurements in the past affect the future in a complicated way.
\item There exists a \emph{time-invariant vacuum}, $\vacR$, that does not contain particles:
\begin{equation} \label{Svac}
    a_i\vacR=b_i\vacR=0,\qquad S\vacR=\vacR\,.
\end{equation}
\item One-particle states evolve trivially, which means that they are \emph{stable}
\begin{equation}
    \bdag_i \vacR = S^\dagger\adag_i \vacR = \adag_i \vacR\,. \label{stability0}
\end{equation}
This also implies $S a_i^\dag \vacR = a_i^\dag \vacR$ from the unitarity condition $S^\dagger S=\mathbbm{1}$.
\end{enumerate}
Given these axioms and the algebra of measurements, we can now ask:
what can be measured asymptotically?
We organize the discussion by multiplicity, starting at four-point.  

\subsubsection{Enumerating observables: \emph{four-point}} The possible four-point measurements are of the form 
\begin{equation}
    \vac{...} \quad \text{where ``...'' is any length-four string of}~ a, \adag, b, \bdag\,.
\end{equation}
Naturally, numerous of these combinations result in trivial observables. This is because $a\ket{0} = b\ket{0} = 0$, and both $a_ia_j^\dagger\ket{0}$ and $b_ib_j^\dagger\ket{0}$ are proportional to a delta function, which vanishes for non-forward momenta (and in any rate is a $c$-number times
$\ket{0}$ according to \eqref{canonical}: an uninteresting state!).
Therefore, only two non-trivial states can be generated by acting with two asymptotic operators on the vacuum:
\begin{equation}
    a_i^\dagger a_j^\dagger\vacR\equiv \ket{ij} \quad \text{and} \quad b_i^\dagger a_j^\dagger\vacR=b_i^\dagger b_j^\dagger\vacR\equiv \ket{ij}_\text{out}\,.
\end{equation}
These are the \emph{only} two options! Multiplying by the analogous classification on the left, we conclude that the possible observables are therefore the conventional (causal) scattering amplitude
\begin{equation}\label{eq:4obs1}
\begin{split}
    \vac{b_4 b_3 \adag_2\adag_1}&=\vac{a_4 a_3 S \adag_2\adag_1}=\bra{34}S\ket{12}={}_{\text{out}}\braket{34|12}\,,
\end{split}
\end{equation}
the Hermitian conjugated (anti-causal) scattering amplitude
\begin{equation}\label{eq:4obs2}
\begin{split}
    \vac{a_4 a_3 \bdag_2\bdag_1}&=\vac{a_4 a_3 S^\dagger \adag_2\adag_1}=\bra{34}S^\dagger\ket{12}\,,
\end{split}
\end{equation}
and the forwards terms
\begin{equation}
\begin{split}
    \vac{a_4 a_3 \adag_2\adag_1}&=\bra{34}\mathbbm{1}\ket{12}=\delta_{12,34}=\vac{b_4 b_3 \bdag_2\bdag_1}\,.
\end{split}
\end{equation}
Restricting two non-forward kinematics, we find only \emph{two} four-point observables: \eqref{eq:4obs1} and \eqref{eq:4obs2}.  Given the age of this subject, it is reassuring that we did not discover anything new at four points.
The fact that there are two observables and not just a single one is significant. 
As we will elaborate on later, crossing symmetry relates these two objects---amplitudes to their Hermitian conjugate, and vice versa!

\begin{equation}\label{eq:crossingFig2to2}
    \adjustbox{valign=c}{
   \begin{tikzpicture}[line width=1,draw=charcoal, scale=0.5]
  \draw[->] (-6.5, 0) -- (6.5, 0);
  \draw[->] (0, -0.3) -- (0, 4.5);
  \node[] at (6,4.0) {$s$};
  \draw[shift={(6,4)},scale=0.7] (-0.5,0.5) -- (-0.5,-0.5) -- (0.5,-0.5);
  \centerarc[<->,Maroon,thick](0,-1.5)(35:145:5);
\node[] at (-4.5,0.7) {$\cM_{1\bar{3} \to \bar{2}4}^\dag$};
\node[] at (4.5,0.7) {$\cM_{12 \to 34}$};
\end{tikzpicture}
    }
\end{equation}

Before moving on, let us stress that crossing is not simply a matter of ``flipping the sign of energies in formulas''.  Consider for example the integrated expression for an $s$-channel massless bubble integral in minimal subtraction:
\begin{equation}
 {\rm Bub}(s) \propto \begin{sqcases}
     \log \frac{-s}{\mu^2} \quad & s<0 \\
 \log \frac{s}{\mu^2}-i\pi & s>0 
 \end{sqcases}\,.
\end{equation}
To go from one expression to the other, we have to follow a precise \emph{continuation path}.  The arc in the upper-half $s$-plane tells us exactly the correct branch choice.

\subsubsection{Enumerating observables: \emph{five-point}}  As should now be clear, the task of cataloging all possible observables is a combinatorial exercise. Hence, as the number of points increases, we should encounter increasingly non-trivial possibilities, leading to more exotic observables beyond the conventional scattering amplitudes.
This becomes manifest already for five points. For example, while we still have the $3\ot 2$ scattering amplitudes
\begin{equation}
    \vac{b_5 b_4 b_3 \adag_2\adag_1}=\bra{345}S\ket{12}\,,
\end{equation}
we also have observables of the form
\begin{equation}\label{eq:5obs1}
    \vac{a_5 a_4 b_3 \adag_2\adag_1}=\bra{45}b_3\ket{12}=\bra{45}S^\dagger a_3 S\ket{12}\,.
\end{equation}
As the writing makes manifest,  \eqref{eq:5obs1} computes an expectation value for observing particle 3 in a particular \emph{in} state that contains two particles, and is totally agnostic as to what happens to these particles \emph{after} particle 3 is emitted. (Both the bra and ket are \emph{in} states: no future boundary conditions are imposed.)
We can make \eqref{eq:5obs1} look more like a conventional quantum-mechanical expectation value by integrating the on-shell momenta $p_1$, $p_2$ against a wavepacket:
\begin{equation} \label{eq:5obs1a}
\begin{aligned}
 \ket{\psi}_\text{in} &\equiv \int_{p_1,p_2} \psi(p_1,p_2) \ket{12} \\
\Rightarrow\,\, {}_\text{in}\!\bra{\psi} b_3 \ket{\psi}_\text{in} &= \int_{p_1,p_2,p_4,p_5}
 \psi^*(p_4,p_5)\psi(p_1,p_2) \bra{45}b_3\ket{12}\,.
\end{aligned}
\end{equation}
Note that \eqref{eq:5obs1a} is \emph{not} measuring the type 3 particle number; such an observable would be quadratic in field 3. Instead, \eqref{eq:5obs1a} measures the linear (leading order) response of the scattering background in field 3.
In which realistic situations could this arise?  For example, \eqref{eq:5obs1a} could represent (i) the expectation value of the electromagnetic field associated with radiation after the collision of two charged particles (further discussed in Sec.~\ref{sec:eikonal}), or (ii) the expectation value of the graviton field (gravitational wave signal) generated by the scattering of two massive objects (e.g., black holes), which is further discussed in Sec.~\ref{sec:crossing}. Both are examples of radiative \defQ{waveforms}.
In this context, the wavefunction $\psi$ sets the initial energy and impact parameter of the colliding bodies (see \cite{Kosower:2018adc}).

Let us reiterate a basic but important point: the signal observed at a detector such as LIGO/VIRGO/KAGRA is \emph{linear} in metric perturbations.
Linear measurements of electric fields, while not necessarily important nor natural at colliders, are familiar in low-energy physics: this is what a voltmeter measures.
Although the definition \eqref{eq:5obs1a} makes no reference to coherence effects,
such linear measurements are typically only practical when the final state contains a coherent state of many photons or gravitons that makes the expectation value \eqref{eq:5obs1a} large.

That said, it may not yet be clear from \eqref{eq:5obs1a} \emph{how} the observable should be computed in practice (e.g., using perturbation theory). To get started, it is useful to insert in \eqref{eq:5obs1} a complete basis of on-shell states 
\begin{equation}\label{eq:completeBasis}
    \mathbbm{1}=\sumint_X\ket{X}_{\mathrlap{\text{in}}}{\hspace{-0.45cm}\phantom{\bra{X}}}_{\mathrlap{\text{in}}\hspace{0.15cm}}\bra{X}=\sumint_X\ket{X}_{\mathrlap{\text{out}}}{\hspace{-0.2cm}\phantom{\bra{X}}}_{\mathrlap{\text{out}}\hspace{0.35cm}}\bra{X}
\,,
\end{equation}
as follows:
\begin{equation}\label{eq:5obs2}
    \begin{split}
          \vac{a_5 a_4 b_3 \adag_2\adag_1}&=\sumint_X\bra{45}S^\dagger\ket{X}\bra{X3} S\ket{12}
          \\&=
           \adjustbox{valign=c,scale={1}{1}}{
\begin{tikzpicture}[line width=1]
\draw[->, color=gray!55,xshift=30pt] (1.5,-1.2) -- (0.5,-1.2) node[above,midway]{\small{time}};
\draw[<-, color=gray!55,xshift=-30pt] (1.5,-1.2) -- (0.5,-1.2) node[above,midway]{\small{time}};
\draw[] (0,0.3) -- (-1.2,0.3) node[left] {\small$4$};
\draw[->] (0,0.3) -- (-1,0.3);
\draw[] (0,-0.3) -- (-1.2,-0.3) node[left] {\small$5$};
\draw[->] (0,-0.3) -- (-1,-0.3);
\draw[] (2,0.3) -- (3.2,0.3) node[right] {\small$1$};
\draw[-<] (2,0.3) -- (3,0.3);
\draw[] (2,-0.3) -- (3.2,-0.3) node[right] {\small$2$};
\draw[-<] (2,-0.3) -- (3,-0.3);
\draw[] (2,0) -- (1,1) node[left] {\small$3$};
\draw[->] (2,0) -- (1.2,0.8);
\filldraw[fill=gray!30](0,-0.3) rectangle (2,0.3);
\draw (2,0.3) -- (0.95,0.3);
\draw (2,-0.3) -- (0.95,-0.3);
\draw[] (1,0) node {$\medmath{X}$};
\filldraw[fill=gray!5, line width=1.3](0,0) circle (0.6) node[yshift=1] {$S^\dag$};
\filldraw[fill=gray!5, line width=1.3](2,0) circle (0.6) node {$S$};
\draw[dashed,orange] (1,1.2) -- (1,-0.8);
\end{tikzpicture}
}\,.
\end{split}
\end{equation}
To be clear, \eqref{eq:completeBasis} is a sum over the complete basis of states $\ket{X}\bra{X}$ integrated over the \defQ{on-shell phase space} 
\begin{equation}\label{eq:sumintDef}
\hspace{-0.4cm}
    \sumint_X\ket{X}\bra{X}{=}\int \frac{\d^{\D-1}p_1}{(2\pi)^{\D-1}2E_{p_1}}\ket{p_1}\bra{p_1}{+}\int \prod_{j=1}^2\frac{\d^{\D-1}p_j}{(2\pi)^{\D-1}2E_{p_j}}\frac{\ket{p_1p_2}\bra{p_1p_2}}{2!}{+}\ldots
\end{equation}
Now, from \eqref{eq:5obs2}, one way to practically compute the waveform observable becomes apparent: it is the product of S-matrix elements (each computable separately using e.g. Feynman diagrams) connected by an on-shell phase-space integral. 
Note that this calculation is completely inclusive in the final state $X$:
as far as we care, the colliding bodies could annihilate each other and we would still get a non-trivial waveform!

All other observables at five-point are either forward or obtained by Hermitian conjugation or time reversal of those listed above. Thus, we move on to six-point. 

\subsubsection{Enumerating observables: \emph{six-point}} Similar to the previous instances, conventional six-point scattering amplitudes (i.e., $\bra{3456}S\ket{12}$, $\bra{456}S\ket{123}$, relevant for scattering of one parton against two partons and their Hermitian conjugates) remain. More intriguingly, we have observables of the form 
\begin{subequations}
    \begin{align}
       &\label{eq:6ptsEx1} \vac{a_6 a_5\bdag_4 b_3\adag_2\adag_1}
=\adjustbox{valign=c}{
\begin{tikzpicture}[line width=1,scale=0.9]
\draw[] (0,0.3) -- (-1.2,0.3) node[left] {\small$5$};
\draw[->] (0,0.3) -- (-1,0.3);
\draw[] (0,-0.3) -- (-1.2,-0.3) node[left] {\small$6$};
\draw[->] (0,-0.3) -- (-1,-0.3);
\draw[] (2,0.3) -- (3.2,0.3) node[right] {\small$1$};
\draw[-<] (2,0.3) -- (3,0.3);
\draw[] (2,-0.3) -- (3.2,-0.3) node[right] {\small$2$};
\draw[-<] (2,-0.3) -- (3,-0.3);
\draw[] (2,0) -- (1,-1) node[left] {\small$3$};
\draw[->] (2,0) -- (1.2,-0.8);
\draw[] (1,1) node[right] {\small$4$} -- (0,0);
\draw[>-] (0.8,0.8) -- (0,0);
\filldraw[fill=gray!30](0,-0.3) rectangle (2,0.3);
\draw (2,0.3) -- (0.95,0.3);
\draw (2,-0.3) -- (0.95,-0.3);
\draw[] (1,0) node {$\medmath{X}$};
\filldraw[fill=gray!5, line width=1.3](0,0) circle (0.6) node[yshift=1] {$S^\dag$};
\filldraw[fill=gray!5, line width=1.3](2,0) circle (0.6) node {$S$};
\draw[dashed,orange] (1,1.2) -- (1,-1.2);
\end{tikzpicture}
}\,,
\\&
\label{eq:6ptsEx2}\vac{a_6 b_5a_4 \bdag_3\adag_2\bdag_1}
=\adjustbox{valign=c}{
\begin{tikzpicture}[line width=1,scale=0.9]
\draw[] (0,0.3) -- (-1.2,0.3) node[left] {\small$5$};
\draw[->] (0,0.3) -- (-1,0.3);
\draw[] (0,-0.3) -- (-1.2,-0.3) node[left] {\small$6$};
\draw[->] (0,-0.3) -- (-1,-0.3);
\draw[] (3,0.3) -- (4.2,0.3) node[right] {\small$1$};
\draw[-<] (3,0.3) -- (4,0.3);
\draw[] (3,-0.3) -- (4.2,-0.3) node[right] {\small$2$};
\draw[-<] (3,-0.3) -- (4,-0.3);
\draw[] (1.5,0) -- (2.5,-1) node[right] {\small$3$};
\draw[-<] (1.5,0) -- (2.3,-0.8);
\draw[] (0.5,1) node[left] {\small$4$} -- (1.5,0);
\draw[<-] (0.7,0.8) -- (1.5,0);
\filldraw[fill=gray!30](0,-0.3) rectangle (1.5,0.3);
\filldraw[fill=gray!30](1.5,-0.3) rectangle (3,0.3);
\filldraw[fill=gray!5, line width=1.3](0,0) circle (0.6) node {$S$};
\filldraw[fill=gray!5, line width=1.3](1.5,0) circle (0.6) node[yshift=1] {$S^\dag$};
\filldraw[fill=gray!5, line width=1.3](3,0) circle (0.6) node {$S$};
\draw (2.25,0.3) -- (2.26,0.3);
\draw (2.25,-0.3) -- (2.26,-0.3);
\draw (0.75,0.3) -- (0.76,0.3);
\draw (0.75,-0.3) -- (0.76,-0.3);
\draw[dashed,orange] (0.75,1.2) -- (0.75,-1.2);
\draw[dashed,orange] (2.25,1.2) -- (2.25,-1.2);
\draw[] (0.75,0) node {\small $\medmath{X}$};
\draw[] (2.25,0) node {\small $\medmath{Y}$};
\end{tikzpicture}
}\,.
\end{align}
\end{subequations}
In the forward limit ($6,5,4\to 1,2,3$), the observable \eqref{eq:6ptsEx1} is well known: it is the inclusive cross-section measuring the number of particles of type 3 produced in the scattering of particles 1 and 2: $\vac{a_6 a_5\bdag_4 b_3\adag_2\adag_1}\to \bra{12}N_3\ket{12}$, with $N_3=\bdag_3b_3$ quadratic in the field.

The same observable \eqref{eq:6ptsEx1} also appears in more exotic situations.  For center-of-mass energies well above the Planck mass $\sqrt{s_{12}}\gg M_{\text{pl}}$, one could imagine \eqref{eq:6ptsEx1} as measuring the ``square'' of the following process:
\begin{equation}\label{eq:sqrtHawking}
\adjustbox{valign=c}{\tikzset{every picture/.style={line width=0.75pt}}       
\begin{tikzpicture}[x=0.75pt,y=0.75pt,yscale=-1,xscale=1]
\draw [line width=2.25,line cap=round]    (201.35,158.96) -- (140.68,159.7) node[midway, above] {$\text{BH}$};
\draw    (240.04,178.58) node[right] {2} .. controls (237.8,179.31) and (236.31,178.56) .. (235.58,176.32) .. controls (234.85,174.08) and (233.36,173.33) .. (231.12,174.06) .. controls (228.88,174.79) and (227.39,174.03) .. (226.66,171.79) .. controls (225.93,169.55) and (224.44,168.8) .. (222.2,169.53) .. controls (219.96,170.26) and (218.47,169.51) .. (217.74,167.27) .. controls (217.01,165.03) and (215.52,164.28) .. (213.28,165.01) .. controls (211.04,165.74) and (209.55,164.99) .. (208.82,162.75) .. controls (208.09,160.51) and (206.6,159.76) .. (204.36,160.49) -- (201.35,158.96) -- (201.35,158.96);
\draw    (201.35,158.96) .. controls (202.18,156.75) and (203.69,156.07) .. (205.9,156.9) .. controls (208.11,157.73) and (209.63,157.05) .. (210.46,154.84) .. controls (211.29,152.63) and (212.8,151.95) .. (215.01,152.78) .. controls (217.22,153.61) and (218.74,152.93) .. (219.57,150.72) .. controls (220.4,148.51) and (221.91,147.83) .. (224.12,148.66) .. controls (226.33,149.49) and (227.85,148.81) .. (228.68,146.6) .. controls (229.51,144.39) and (231.03,143.71) .. (233.24,144.54) .. controls (235.45,145.37) and (236.96,144.69) .. (237.79,142.48) -- (241.99,140.58) -- (241.99,140.58) node[right] {1};
\draw [color=gray]    (100.04,178.08) .. controls (100.87,175.87) and (102.38,175.19) .. (104.59,176.02) .. controls (106.8,176.85) and (108.32,176.17) .. (109.15,173.96) .. controls (109.98,171.75) and (111.49,171.07) .. (113.7,171.9) .. controls (115.91,172.73) and (117.43,172.05) .. (118.26,169.84) .. controls (119.09,167.63) and (120.6,166.95) .. (122.81,167.78) .. controls (125.02,168.61) and (126.54,167.93) .. (127.37,165.72) .. controls (128.2,163.51) and (129.72,162.83) .. (131.93,163.66) .. controls (134.14,164.49) and (135.65,163.81) .. (136.48,161.6) -- (140.68,159.7) -- (140.68,159.7) ;
\draw   [color=gray] (111.39,191.07) .. controls (111.3,188.71) and (112.44,187.49) .. (114.8,187.41) .. controls (117.15,187.33) and (118.29,186.11) .. (118.21,183.76) .. controls (118.13,181.41) and (119.27,180.19) .. (121.62,180.11) .. controls (123.98,180.03) and (125.12,178.81) .. (125.04,176.45) .. controls (124.96,174.1) and (126.1,172.88) .. (128.45,172.8) .. controls (130.8,172.71) and (131.94,171.49) .. (131.86,169.14) .. controls (131.78,166.79) and (132.92,165.57) .. (135.27,165.49) .. controls (137.63,165.41) and (138.77,164.19) .. (138.69,161.83) -- (140.68,159.7) -- (140.68,159.7) ;
\draw [color=gray]  (99.36,172.08) .. controls (100.48,170.01) and (102.08,169.53) .. (104.15,170.65) .. controls (106.22,171.76) and (107.82,171.28) .. (108.94,169.21) .. controls (110.06,167.14) and (111.66,166.66) .. (113.73,167.78) .. controls (115.8,168.89) and (117.4,168.41) .. (118.52,166.34) .. controls (119.64,164.27) and (121.24,163.79) .. (123.31,164.91) .. controls (125.38,166.02) and (126.98,165.54) .. (128.1,163.47) .. controls (129.22,161.4) and (130.82,160.92) .. (132.89,162.04) .. controls (134.96,163.15) and (136.56,162.67) .. (137.68,160.6) -- (140.68,159.7) -- (140.68,159.7) ;
\draw  [color=gray]  (102.33,145.08) .. controls (104.48,144.11) and (106.03,144.71) .. (107,146.86) .. controls (107.96,149.01) and (109.52,149.61) .. (111.67,148.64) .. controls (113.82,147.67) and (115.38,148.27) .. (116.35,150.42) .. controls (117.31,152.57) and (118.87,153.17) .. (121.02,152.2) .. controls (123.17,151.24) and (124.73,151.84) .. (125.69,153.99) .. controls (126.65,156.14) and (128.21,156.74) .. (130.36,155.77) .. controls (132.51,154.8) and (134.07,155.4) .. (135.03,157.55) .. controls (136,159.7) and (137.56,160.3) .. (139.71,159.33) -- (140.68,159.7) -- (140.68,159.7) ;
\draw   [color=gray] (109.32,134.07) .. controls (111.67,133.84) and (112.96,134.89) .. (113.19,137.24) .. controls (113.42,139.59) and (114.71,140.64) .. (117.06,140.4) .. controls (119.41,140.16) and (120.7,141.21) .. (120.93,143.56) .. controls (121.16,145.91) and (122.45,146.96) .. (124.8,146.73) .. controls (127.15,146.49) and (128.44,147.54) .. (128.68,149.89) .. controls (128.91,152.24) and (130.2,153.29) .. (132.55,153.05) .. controls (134.9,152.82) and (136.19,153.87) .. (136.42,156.22) .. controls (136.65,158.57) and (137.94,159.62) .. (140.29,159.38) -- (140.68,159.7) -- (140.68,159.7) ;
\draw  [color=gray]  (104.32,138.08) .. controls (106.61,137.49) and (108.04,138.34) .. (108.62,140.63) .. controls (109.2,142.92) and (110.63,143.77) .. (112.92,143.19) .. controls (115.2,142.61) and (116.63,143.46) .. (117.21,145.74) .. controls (117.79,148.03) and (119.22,148.88) .. (121.51,148.3) .. controls (123.8,147.72) and (125.23,148.57) .. (125.81,150.86) .. controls (126.39,153.15) and (127.82,154) .. (130.11,153.41) .. controls (132.4,152.83) and (133.83,153.68) .. (134.4,155.97) .. controls (134.98,158.26) and (136.41,159.11) .. (138.7,158.52) -- (140.68,159.7) -- (140.68,159.7) ;
\draw  [color=gray]  (99.34,152.08) .. controls (101.28,150.75) and (102.92,151.05) .. (104.26,152.99) .. controls (105.59,154.93) and (107.23,155.23) .. (109.17,153.9) .. controls (111.11,152.56) and (112.75,152.86) .. (114.09,154.8) .. controls (115.43,156.74) and (117.07,157.04) .. (119.01,155.71) .. controls (120.95,154.37) and (122.59,154.67) .. (123.92,156.61) .. controls (125.26,158.55) and (126.9,158.85) .. (128.84,157.52) .. controls (130.78,156.18) and (132.42,156.48) .. (133.76,158.42) .. controls (135.1,160.36) and (136.74,160.66) .. (138.68,159.33) -- (140.68,159.7) -- (140.68,159.7) ;
\draw  [color=gray]  (104.38,185.08) .. controls (104.79,182.75) and (106.16,181.8) .. (108.48,182.21) .. controls (110.8,182.62) and (112.16,181.67) .. (112.57,179.35) .. controls (112.98,177.03) and (114.35,176.07) .. (116.67,176.48) .. controls (118.99,176.89) and (120.36,175.94) .. (120.77,173.62) .. controls (121.18,171.3) and (122.55,170.34) .. (124.87,170.75) .. controls (127.19,171.16) and (128.56,170.21) .. (128.97,167.89) .. controls (129.38,165.57) and (130.74,164.61) .. (133.06,165.02) .. controls (135.38,165.43) and (136.75,164.48) .. (137.16,162.16) -- (140.68,159.7) -- (140.68,159.7) ;
\draw    (171.01,159.33) .. controls (171.24,161.68) and (170.18,162.96) .. (167.83,163.18) .. controls (165.48,163.41) and (164.42,164.69) .. (164.64,167.04) .. controls (164.87,169.39) and (163.81,170.67) .. (161.46,170.89) .. controls (159.11,171.11) and (158.05,172.39) .. (158.27,174.74) .. controls (158.49,177.09) and (157.43,178.37) .. (155.08,178.6) .. controls (152.73,178.82) and (151.67,180.1) .. (151.9,182.45) .. controls (152.12,184.8) and (151.06,186.08) .. (148.71,186.3) .. controls (146.36,186.53) and (145.3,187.81) .. (145.52,190.16) -- (144,192) -- (144,192)  node[left] {3} ;
\draw (95.36,169) node [anchor=north west][inner sep=0.75pt]  [font=\small,rotate=-270,color=gray] [align=left] {...} node [anchor=north west, xshift=-60,yshift=12.5][inner sep=0.75pt]  [font=\small,rotate=0,color=gray] [align=left] {
$\mathcal{O}(S)~\text{states}$};
\end{tikzpicture}}
\end{equation}
where particles 1 and 2 collide and form a black hole, which eventually fully evaporates by emitting Hawking radiation.  In a strict $S$-matrix approach, the exact final state is a complicated combination of $\mathcal{O}(S)$ quanta where $S$ is the entropy of the intermediate black hole, with matrix elements that are typically exponentially small $\sim \e^{-S/2}$ by general statistical considerations.  The calculation of this quantum state remains an unsolved task, even in principle (although there was notable recent progress in understanding its statistical properties using replica wormholes).
However, \eqref{eq:6ptsEx1} is a much simpler object because it is agnostic about the complicated cloud of on-shell final states and only cares about the particular Hawking quantum labeled $3$ in the figure.
(By making the momenta $p_3$ and $p_4$ slightly non-forward and performing a Wigner transform, the observable is also sensitive to the arrival time of the measured radiation.)
Viewing \eqref{eq:6ptsEx1} as a two-point function in a black hole geometry also makes it clear that it can be reliably calculated without knowing the exact final state---in fact this is exactly how Hawking \cite{Hawking:1975vcx} calculated it! 

Let us finally comment on \eqref{eq:6ptsEx2}, which is at face value a ``maximal scrambling'' of $a$'s and $b$'s. It is useful to consider particles 1 and 6 as defining a background state and to interpret this observable as a \emph{four-point} OTOC (commutator squared) in this background. By transforming to the time domain and applying a large-time translation to $\bdag_3$ and $b_5$, this process potentially probes the Lyapunov exponent characterizing the chaotic growth of small perturbation $\bdag_3$ (see \cite{Maldacena:2015waa} for more references).
For a more complete list of six-point observables, see \cite{Caron-Huot:2023vxl}.

\subsection{(Some) crossing moves}\label{sec:crossing}
We now sketch how crossing symmetry bridges some of the above observables through analytic continuations in the kinematic space. 
The basic object (further discussed in Sec.~\ref{sec:reduction}) is the retarded product of currents $j\propto (-\partial^2+m^2)\phi$,
\begin{equation}\label{eq:retFct}
    \begin{split}
        \mathcal{R}(p_2,p_3)=\int \d^\D x_2\d^\D x_3 \e^{-i(p_2\cdot x_2+p_3\cdot x_3)}{}_\text{out\hspace{-0.1cm}}\bra{B} \R{3}{j_3j_2}\ket{A}\,,
    \end{split}
\end{equation}
where $\R{3}{j_3j_2}=[j_3,j_2]\theta(x_3^0{>}x_2^0)$ denotes the retarded product.
The current $j$ amputates the external propagators and following the logic of the LSZ reduction formula one can show that for real on-shell momenta :
\begin{equation}\label{eq:bdryCOnd}
    \mathcal{R}(p_2,p_3)=\begin{sqcases}
        \bra{B}[b_3,\adag_2]\ket{A} & p_2^0<0\,,~ p_3^0>0\,, \\
        -\bra{B}[a_2,\bdag_3]\ket{A} & p_2^0>0\,, ~ p_3^0<0\,. \\
    \end{sqcases} ~~ \text{(on-shell)}
\end{equation}

The challenge of crossing is to parameterize a \emph{continuous} path of analytic continuation that (i) is wholly supported on the mass shell and (ii) connects the
two preceding real configurations \eqref{eq:bdryCOnd}:
\begin{equation}\label{eq:path0}
         \adjustbox{valign=c}{
\tikzset{every picture/.style={line width=0.75pt}}  
\begin{tikzpicture}[x=0.75pt,y=0.75pt,yscale=-1,xscale=1]
\draw[<-]    (178,73.5) -- (178,178.5) ;
\draw[->]    (38.5,163.86) -- (317,163.86) ;
\draw[yshift=-13]   (312,109.5) -- (301,109.5) -- (301,98.5);
\draw  [draw opacity=0] (98.43,139.55) .. controls (108.82,105.6) and (140.4,80.92) .. (177.75,80.92) .. controls (214.83,80.92) and (246.22,105.25) .. (256.84,138.81) -- (177.75,163.86) -- cycle ; \draw[<-]  [color=Maroon  ,draw opacity=1 ] (98.43,139.55) .. controls (108.82,105.6) and (140.4,80.92) .. (177.75,80.92) .. controls (214.83,80.92) and (246.22,105.25) .. (256.84,138.81) ;  
\draw[yshift=-13] (302,98.9) node [anchor=north west][inner sep=0.75pt]  [font=\normalsize]  {$z$};
\draw (210,141.4) node [anchor=north west, yshift=3pt][inner sep=0.75pt]  [font=\normalsize]  {$\bra{B}[b_3,\adag_2]\ket{A}$};
\draw (39,141.4) node [anchor=north west, yshift=3pt][inner sep=0.75pt]  [font=\normalsize]  {$ -\bra{B}[a_2,\bdag_3]\ket{A} $};
\draw (60,80) node [anchor=north west][inner sep=0.75pt]  [font=\normalsize,color=Maroon  ,opacity=1]  {$\mathcal{R}( p_{2} ,p_{3})$};
 \draw[decorate, decoration={zigzag, segment length=6, amplitude=2}, RoyalBlue!80] (178,100) -- (178,148.5);
\draw (190,130) node [font=\normalsize] {$?$};
\draw (170,110) node [font=\normalsize] {$?$};
 \draw[RoyalBlue!80,fill=RoyalBlue!80,thick] (178,100) circle (2);
 \draw[RoyalBlue!80,fill=RoyalBlue!80,thick] (178,148.5) circle (2);
  \draw[decorate, decoration={zigzag, segment length=6, amplitude=2}, RoyalBlue!80] (138,100) -- (168,128.5);
  \draw[RoyalBlue!80,fill=RoyalBlue!80,thick] (138,100) circle (2);
  \draw[RoyalBlue!80,fill=RoyalBlue!80,thick] (168,128.5) circle (2);
  \draw[decorate, decoration={zigzag, segment length=6, amplitude=2}, RoyalBlue!80] (188.5,108) -- (200,138);
  \draw[RoyalBlue!80,fill=RoyalBlue!80,thick] (188.5,108) circle (2);
  \draw[RoyalBlue!80,fill=RoyalBlue!80,thick] (200,138) circle (2);
  \draw[decorate, decoration={zigzag, segment length=6, amplitude=2}, RoyalBlue!80]    (38.5,163.86) -- (317,163.86);
\end{tikzpicture}
    }
\end{equation}
In the case of gapped (and infrared-trivial) quantum field theories, such a path is guaranteed to exist for $2\ot 2$ and $3\ot 2$ scattering thanks to the work of Bros, Epstein and Glaser initiated in the 1960's \cite{Bros:1964iho,Bros:1965kbd,Bros:1972jh,Bros:1985gy}. In gapless theories and at higher multiplicities, the existence of such a path remains conjectural to this day, although recent progress in this direction has been made \cite{Mizera:2021fap,Caron-Huot:2023ikn}. The $z$-path we will consider is simply a complex boost applied to two of the external momenta; see \cite{Caron-Huot:2023ikn} for details.

To develop some intuition about why the mass gap plays an important role
without going into the technical details of \cite{Bros:1964iho,Bros:1965kbd,Bros:1972jh,Bros:1985gy}, we can consider the following example.

\begin{mdexample} \textbf{(Intuition for analyticity)} 
Under the change of variables $x_3\mapsto x_2+\Delta x$, \eqref{eq:retFct} becomes
\begin{equation}\label{eq:retFct2}
    \begin{split}
        \mathcal{R}(p_2,p_3)=\int_{\Delta x \in \bar{V}^+} \d^\D \Delta x\, \e^{-ip_3\cdot \Delta x} \int \d^\D x_3 \e^{-i(p_2+p_3)\cdot x_2}{}_\text{out\hspace{-0.1cm}}\bra{B} \R{3}{j_3j_2}\ket{A}\,.
    \end{split}
\end{equation}
The Fourier transform converges provided that (i) $\text{Re}(-ip_3\cdot \Delta x)\leq 0$ $\forall\Delta x$ and (ii) $(p_2+p_3)^\mu\in\mathbbm{R}^{1,3}$.
Since the retarded product is supported in the (closure of) the
foward lightcone $\Delta x\in \bar{V}^+$, we conclude that the integral is analytic
for ${\rm Im}\,p_3^\mu\in V^+$ since then the Fourier transform converges exponentially.
This is the basic reason why we expect analyticity in some upper-half plane.

The difficulty is that a timelike imaginary part for $p_3^\mu$
is incompatible with the mass-shell condition $p_3^2+m^2=0$.  Consider for example the on-shell momentum
\begin{equation}
    p_3^\mu=E(1,v,0,0)+i\varepsilon(v,1,0,0) \qquad (0<v<1\,{\rm real}, ~E\gg \varepsilon>0)\,.
\end{equation}
The imaginary part is spacelike and point (i) is only satisfied if $\Delta x^1< v \Delta t$, where $\Delta t=(x_3-x_2)^0>0$ and $\Delta x^1=(x_3-x_2)^1$. Thus, we have a decaying exponential only for displacements $\Delta x$ \emph{inside} the shaded region:
\begin{equation}\label{eq:lc3}
\adjustbox{valign=c}{\tikzset{every picture/.style={line width=0.75pt}}      
\begin{tikzpicture}[x=0.75pt,y=0.75pt,yscale=-1,xscale=1]
\draw[->]    (51,107) -- (149,107) node[right]{$\Delta x$};
\draw[->]    (100,114) -- (100,58) node[above,left]{$\Delta t$};
\draw  [color=Gray  ,draw opacity=0.25 ] (65,67) -- (135,67) -- (100,107) -- cycle ;
\draw  [draw opacity=0][fill={rgb, 255:red, 155; green, 155; blue, 155 }  ,fill opacity=0.41 ] (101.17,107) -- (65,67) -- (114,67) -- cycle ;
\draw[->][color=Maroon]    (100,106) -- (114,67) node[right, yshift=7pt]{$v\Delta t$};
\end{tikzpicture}}
\end{equation}
We see that while we got analyticity in $V^+$ just from the causal support $|\Delta x^1|< \Delta t$ of the retarded product, analyticity on-shell will only hold if the retarded product is supported within the smaller shaded region $|\Delta x^1|< v\Delta t$. 
Physically, this is a constraint on the velocity of the intermediate particles produced.

This naive analysis suggests that analyticity is safe for sufficiently boosted external momenta in gapped theories, but can fail in gapless theories whenever \emph{external} particles move more slowly than internal ones.

We hope that this intuition can be formalized in the near future using the concept of \defQ{essential support}, described in Sec.~\ref{sec:FBI} below.
For now, it remains unproven whether the domain of analyticity of \eqref{eq:retFct} intersects with the mass shell, in general. We will proceed under the assumption that it does.

\end{mdexample}
\begin{mdexample}
\textbf{(Five-point tree-level with massive resonances)}

We now borrow a five-point example from \cite{Caron-Huot:2023ikn} to illustrate how the inclusive observable \eqref{eq:5obs2} (equivalent to a product of conventional (time-ordered) amplitudes) is related to a five-point scattering amplitude through analytical continuation. Here, we work at tree level, which is (perhaps surprisingly) sufficient to highlight the non-trivial conceptual aspects of this exercise without the unnecessary complications arising from branch cuts.

Consider a cubic theory where the following tree-level process involving a heavy intermediate particle of mass $M$ is possible
\begin{equation}\label{eq:M_12_345}
i\cM_{345 \ot 12}^{\text{start}}=
\begin{gathered}
\adjustbox{valign=c}{
\begin{tikzpicture}[baseline= {($(current bounding box.base)+(10pt,10pt)$)},line width=1, scale=0.7]
\coordinate (a) at (0,0) ;
\coordinate (b) at (1,0) ;
\coordinate (c) at ($(b)+(-40:1)$);
\draw[] (a) -- (b);
\draw[] (b) -- (c);
\draw[Maroon] (b) -- ++ (30:1) node[right] {\footnotesize$2$};
\draw[Maroon] (c) -- ++ (-150:1) node[left]{\footnotesize$3$};
\draw[RoyalBlue] (c) -- ++ (-30:1) node[right] {\footnotesize$1$};
\draw[RoyalBlue] (a) -- (-150:1) node[left] {\footnotesize$4$};
\draw[RoyalBlue] (a) -- (150:1) node[left] {\footnotesize$5$};
\fill[black,thick] (a) circle (0.07);
\fill[black,thick] (b) circle (0.07);
\fill[black,thick] (c) circle (0.07);
\end{tikzpicture}
}
\end{gathered}
=
\frac{-ig^3}{(-s_{45}+M^2-i\eps)(-s_{13}+M^2)}\,.
\end{equation}
Note that the $i\varepsilon$ only matters if it is kinematically possible to produce $M$, in which case it is necessarily unstable (``if it can be produced, it can decay'').  The tree-level diagrams here should be understood as a narrow-width approximation to this resonance. 

In \eqref{eq:M_12_345} we dropped the $i\varepsilon$ prescription for $s_{13}$ since this invariant is spacelike ($s_{13}<0$) in the kinematics under consideration and therefore its propagator cannot go on shell. Nevertheless, the $i\varepsilon$ must be retained for the timelike Mandelstam invariant $s_{45}$. We proceed now to   analytically continue to the kinematic region where the energies of particles 2 and 3 flip signs. The invariant $s_{13}$, initially negative, rotates in the counterclockwise direction, ending its journey below the positive real axis, while the invariant $s_{45}$ remains unchanged:

\begin{equation}
\adjustbox{valign=c}{
\begin{tikzpicture}
  \draw[->,thick] (-1.5, 0) -- (1.5, 0);
  \draw[->,thick,white] (0, -1.6) -- (0, 1.6);
  \draw[->,thick] (0, -1.25) -- (0, 1.25);
  \node[] at (1.1,1.05) {$s_{ij}$};
  \draw[RoyalBlue,fill=RoyalBlue,thick] (0.5,0.1) circle (0.05);
  \node[] at (0.35,0.4) {$\textcolor{RoyalBlue}{s_{45}}$};
  \draw[] (1.35,0.85) -- (0.8,0.85) -- (0.8,1.25);
  \draw[Maroon,fill=Maroon,thick] (-1.1,0.0) circle (0.05);
  \draw[->,Maroon,thick] (-1.1,0.0) arc (0:178:-1.1);
  \node[] at (-0.35,-0.6) {$\textcolor{Maroon}{s_{13}}$};
\end{tikzpicture}
}
\end{equation}
The result of this analytic continuation is written as
\begin{equation}
 \left[i\cM_{345 \ot 12}\right]_{ 
 \raisebox{\depth}{\scalebox{1}[-1]{$\curvearrowright$}}s_{13}} =
 \frac{-ig^3}{(-s_{45}+M^2-i\eps)(-s_{13}+M^2+i\eps)}\,, \label{5pt tree crossing}
\end{equation}
where now $s_{13} > 0$ and its $-i\eps$ prescription matters.
Observe that the second propagator acquires the ``wrong'' $i\varepsilon$ prescription. This detail is important because after the crossing we are now in $245 \ot 13$ kinematics where this pole is accessible.
The fact that $s_{13}$ is on the ``wrong side'' of its cut mirrors what happens for $2\ot 2$ scattering, where the large arc (see \eqref{eq:path0}) lands us on the
wrong side of the $u$-channel cut.  However, here the $s_{45}$ pole remained on the correct side, preventing identifying the right-hand side with a complex conjugated amplitude. \emph{It is a different object}!

Our claim now is that the blob pattern in \eqref{eq:5obs2} is a shortcut to produce the same result as the analytic continuation we just went through. To see this, substitute $S^\dagger=\mathbbm{1}-iT^\dagger$ and $S=\mathbbm{1}+iT\to iT$ into \eqref{eq:5obs2}; for $S$ we could drop the $\mathbbm{1}$ part since 1 and 2 must interact in order that 3 be emitted, but for $S^\dagger$ we have both a disconnected and connected term.
They give respectively the complex-conjugated five-point amplitude, plus a non-linear term where the heavy particle $M$ is produced:
\begin{subequations}
\begin{align}
\hspace{-0.3cm}
 \left[i\cM_{345 \ot 12}\right]_{\raisebox{\depth}{\scalebox{1}[-1]{$\curvearrowright$}}s_{13}}
& \stackrel{?}{=} i\cM^{\dag}_{245\ot 13} +
 [i\cM_{45\ot M}]\, 2\pi\delta(-s_{45}+M^2)\,[i\cM^{\dag}_{M2\ot 13}]
\label{5pt crossing tree test}
\\ &=
\frac{-ig^3}{(-s_{45}+M^2+i\eps)(-s_{13}+M^2+i\eps)}
+\frac{2\pi \delta(-s_{45}+M^2) g^3}{-s_{13}+M^2+i\eps}
\\ &= \left( \frac{1}{-s_{45}+M^2+i\eps}+ 2\pi i \delta(-s_{45}+M^2)\right) \frac{-ig^3}{-s_{13}+M^2+i\eps}\,.
\label{eq:5pttree_c}
\end{align}
\end{subequations}
Using the identity $\frac{1}{x\pm i\eps}=\text{PV}\frac{1}{x}\mp i\pi\delta(x)$, the result manifestly agrees with \eqref{5pt tree crossing}! Diagrammatically, we can summarize the discussion as follows:
\begin{equation}
\left[
\begin{gathered}
\adjustbox{valign=c}{
\begin{tikzpicture}[baseline= {($(current bounding box.base)+(10pt,10pt)$)},line width=1, scale=0.7]
\coordinate (a) at (0,0) ;
\coordinate (b) at (1,0) ;
\coordinate (c) at ($(b)+(-40:1)$);
\draw[] (a) -- (b);
\draw[] (b) -- (c);
\draw[Maroon] (b) -- ++ (30:1) node[right] {\footnotesize$2$};
\draw[Maroon] (c) -- ++ (-150:1) node[left]{\footnotesize$3$};
\draw[RoyalBlue] (c) -- ++ (-30:1) node[right] {\footnotesize$1$};
\draw[RoyalBlue] (a) -- (-150:1) node[left] {\footnotesize$4$};
\draw[RoyalBlue] (a) -- (150:1) node[left] {\footnotesize$5$};
\fill[black,thick] (a) circle (0.07);
\fill[black,thick] (b) circle (0.07);
\fill[black,thick] (c) circle (0.07);
\end{tikzpicture}
}
\end{gathered}
\right]_{\raisebox{\depth}{\scalebox{1}[-1]{$\curvearrowright$}}s_{13}}
=
\adjustbox{valign=c}{
\begin{tikzpicture}[baseline= {($(current bounding box.base)+(10pt,10pt)$)},line width=1, scale=0.7]
\coordinate (a) at (0,0) ;
\coordinate (b) at (1,0) ;
\coordinate (c) at ($(b)+(-40:1)$);
\draw[] (a) -- (b);
\draw[] (b) -- (c);
\draw[Maroon] (b) -- ++ (150:1) node[yshift=3,left] {\footnotesize$\bar{2}$};
\draw[Maroon] (c) -- ++ (30:1) node[right]{\footnotesize$\bar{3}$};
\draw[RoyalBlue] (c) -- ++ (-30:1) node[right] {\footnotesize$1$};
\draw[RoyalBlue] (a) -- (-150:1) node[left] {\footnotesize$4$};
\draw[RoyalBlue] (a) -- (150:1) node[left] {\footnotesize$5$};
\fill[black,thick] (a) circle (0.07);
\fill[black,thick] (b) circle (0.07);
\fill[black,thick] (c) circle (0.07);
\draw[dashed,orange] (-0.5,-1.2) -- (-0.5,1);
\draw[dashed,orange] (2.3,-1.2) -- (2.3,1);
\end{tikzpicture}
}
+
\adjustbox{valign=c}{
\begin{tikzpicture}[baseline= {($(current bounding box.base)+(10pt,10pt)$)},line width=1, scale=0.7]
\coordinate (a) at (0,0) ;
\coordinate (b) at (1,0) ;
\coordinate (c) at ($(b)+(-40:1)$);
\draw[] (a) -- (b);
\draw[] (b) -- (c);
\draw[Maroon] (b) -- ++ (150:1) node[yshift=3,left] {\footnotesize$\bar{2}$};
\draw[Maroon] (c) -- ++ (30:1) node[right]{\footnotesize$\bar{3}$};
\draw[RoyalBlue] (c) -- ++ (-30:1) node[right] {\footnotesize$1$};
\draw[RoyalBlue] (a) -- (-150:1) node[left] {\footnotesize$4$};
\draw[RoyalBlue] (a) -- (150:1) node[left] {\footnotesize$5$};
\fill[black,thick] (a) circle (0.07);
\fill[black,thick] (b) circle (0.07);
\fill[black,thick] (c) circle (0.07);
\draw[dashed,orange] (0.5,-1.2) -- (0.5,1);
\draw[dashed,orange] (2.3,-1.2) -- (2.3,1);
\end{tikzpicture}
}
\end{equation}

Similar computations at loop level are discussed in \cite{Caron-Huot:2023ikn}, where a general diagrammatic statement of \eqref{eq:path0} is also given.
In particular, we found that the type of computation described above worked for purely massless scattering (without a mass gap). However, we also point out counterexamples in cases where external particles move slower than internal particles, verifying the intuition developed below \eqref{eq:lc3}.
\end{mdexample}
In the next lecture, we will show how the observables introduced above follow from the reduction of (O)TO products of on-shell currents.

\section[Analyticity and IR properties of timefolded observables]{Analyticity and IR properties of timefolded observables \\ \normalfont{\textit{Mathieu Giroux}}}
The purpose of this lecture is to revisit the application of reduction formulae for observables with unconventional time ordering through explicit examples. As motivation, we quickly revisit the so-called Lehmann--Symanzik--Zimmermann (LSZ) reduction formula \cite{LSZ}---which relates conventional time-ordered scattering amplitudes to time-ordered correlators---from a modern perspective. Later on, we generalize this concept to out-of-time order correlators, which reduce to the aforementioned out-of-time order (OTO) observables.

\subsection{LSZ reduction revisited}\label{sec:reduction}
The idea behind the LSZ reduction formula is simple: it makes precise the idea that \emph{\textcolor{Maroon}{time-ordered products of fields}}
encode more information than is needed to describe in-out particle collisions characterized by \emph{\textcolor{RoyalBlue}{scattering amplitudes}}, namely

\begin{align}\label{eq:LSZamp}
   \hspace{-0.4cm} \textcolor{RoyalBlue}{_\inn\!\bra{(j{+}1) ...n  } S \ket{1...j}_\inn}
   {=} \textcolor{black}{\Big[\prod_{k=1}^n i\int \d^{\D}x_k~\e^{-ip_k\cdot x_k}\big[{-}\partial^2_{x_k}{+}m_k^2\big]\Big]}\textcolor{Maroon}{\bra{0} T(\phi_1\phi_2...\phi_n) \ket{0}}\,.
\end{align}
Indeed, because $\big[{-}\partial^2_{x}{+}m^2\big]\phi(x)=0$ for asymptotic (free) fields, the physical content of \eqref{eq:LSZamp} is that the S-matrix \emph{projects out} (or ``amputates'') such fields from the time-ordered product. This means that only terms with a pole of the form $p^2+m^2=0$ in momentum space ultimately contribute to the scattering amplitude. For what follows, it is useful to revisit \emph{why} \eqref{eq:LSZamp} holds from a modern perspective.

The first thing we do is to express the right-hand side of \eqref{eq:LSZamp} in terms of the currents defined by 
\begin{equation}
    \cT(j(x)\cdots) \equiv i(-\partial_x^2+m^2) \cT(\phi(x)\cdots)\,.
\end{equation}
Given this definition, the right-hand side of \eqref{eq:LSZamp} reduces to the Fourier transformed (momentum space) time-ordered correlator of currents 
\begin{equation}\label{eq:intStep0}
\begin{split}
      \text{RHS of~\eqref{eq:LSZamp}}&=\textcolor{black}{\Big[\prod_{k=1}^n \int \d^{\D}x_k~\e^{-ip_k\cdot x_k}}\Big]\textcolor{black}{\bra{0} T(j_1j_2...j_n) \ket{0}}\\&=\textcolor{black}{\bra{0} T(j(p_1)j(p_2)...j(p_n)) \ket{0}}\,.
\end{split}
\end{equation}
From there, the key observation is that the on-shell limit of the currents is a total derivative. In fact, an elementary computation shows that
\be \label{total derivative}
j(p)\equiv\int \text{d}^\D x\, \e^{-ip{\cdot} x} j(x)
\xrightarrow{\stackrel{\text{on-shell:}}{p^2\to -m^2}}  \int \text{d}^\D x \frac{\partial}{\partial x^\mu}
\left[ \e^{-ip{\cdot} x}\ (-i\partial^\mu_x+p^\mu) \phi(x)\right]\,.
\ee

Now, we observe that the Fourier phase $\e^{-ip \cdot x}$ oscillates rapidly as $x \to \infty$, except possibly along the direction of the external particles: $x^\mu \propto \pm p^\mu$. Along such directions, particles are eventually detected by a detector infinitely far in the future (or past), which is only possible if \emph{constructive interference} occurs between the rapidly oscillatory Fourier phase and the phase of the field
\begin{equation}\label{eq:expectAsym}
    \lim_{t\to \pm \infty}\e^{-ip{\cdot} x}\phi = \text{\emph{non}-oscillatory quantity along $x^\mu\propto \pm p^\mu$}\,.
\end{equation}
In other words, \eqref{eq:expectAsym} is the minimal condition that ensures non-zero asymptotic measurements ($j(p)\neq0$). Thus, the on-shell current in \eqref{total derivative} reduces to non-trivial surface terms (which \emph{defines} for us the $a,b,\adag$ and $\bdag$ introduced earlier) \emph{only} along the trajectory of the particle. This is schematically summarized as follows:
\be\label{eq:expKID}
 \adjustbox{valign=c,scale={0.9}{0.9}}{\tikzset{every picture/.style={line width=0.75pt}} 
\begin{tikzpicture}[x=0.75pt,y=0.75pt,yscale=-1.3,xscale=1.3]
  \draw (287.31,95.26) -- (232.12,95.26) -- (232.08,190.72) -- (253.5,190.74);
    \fill[RoyalBlue] (287.31,95.26) circle (0.7pt);
    \fill[Maroon] (253.5,190.74) circle (0.7pt);
\draw [color=Maroon, draw opacity=1]   (253.5,190.74) .. controls (255.19,189.1) and (256.86,189.13) .. (258.5,190.82) .. controls (260.14,192.51) and (261.81,192.54) .. (263.5,190.91) .. controls (265.19,189.27) and (266.86,189.3) .. (268.49,190.99);
\draw (302.31,95.63) -- (327.5,95.65) -- (327.46,191.11) -- (268.49,191.05);
\fill[RoyalBlue] (302.31,95.63) circle (0.7pt);
    \fill[Maroon] (268.49,191.05) circle (0.7pt);
\draw [color=RoyalBlue, draw opacity=1]   (287.31,95.32) .. controls (289,93.68) and (290.67,93.71) .. (292.31,95.4) .. controls (293.95,97.09) and (295.62,97.12) .. (297.31,95.49) .. controls (299,93.85) and (300.67,93.88) .. (302.31,95.57);
\draw[<->,color=gray!, line width=0.5]    (296.61,100.3) -- (281.95,137.24) ;
\draw[<->,color=gray!, line width=0.5]    (277.06,149.78) -- (262.4,186.71) ;
\draw[->, line width=0.5]    (184.5,142.5) -- (204.5,142.5) ;
\draw[->, line width=0.5]    (184.5,142.75) -- (184.5,121.75) ;
\draw  [dashed,color=gray!15]  (312.4,108.33) -- (247.18,178.04) ;
\draw  [dashed,color=gray!15]  (246.22,109.26) -- (313.36,177.11) ;
\draw (174.5,127.15) node [anchor=north west][inner sep=0.75pt]    {$t$};
\draw (187.5,144.15) node [anchor=north west][inner sep=0.75pt]    {$x$};
\draw[color=RoyalBlue] (200,52.4) node [anchor=north west][inner sep=0.75pt]    {$\Lim{t\to+\infty}\e^{-ip\cdot x}\phi\propto \begin{sqcases}
b & \text{(outgoing)}\\
\bdag & \text{(incoming)}
\end{sqcases}
$};
\draw[color=Maroon] (200,193.4) node [anchor=north west][inner sep=0.75pt]    {$\Lim{t\to-\infty}\e^{-ip\cdot x}\phi\propto  \begin{sqcases}
a & \text{(outgoing)}\\
\adag & \text{(incoming)}\\
\end{sqcases}$};
\end{tikzpicture}}\hspace{-1cm}\leadsto\lim_{p^2\to -m^2} j(p) \equiv \begin{sqcases}
 b - a\,, \quad & p^0>0\,,\\
   \adag - \bdag\,, \quad & p^0<0\,.
 \end{sqcases}
\ee
The cartoon on the left illustrates the asymptotic measurements made in the far past and future ($a,\adag$ and $b,\bdag$, respectively) and emphasizes that they follow from constructive interference (see \eqref{eq:expectAsym}) at late times along the direction of particles; along other directions, destructive interference occurs and no measurement is made. (The dotted lines show light-cone axes centered around the collision region at finite time.)
The sign of $p^0$ determines which case in \eqref{eq:expKID} is applicable.

The LSZ reduction formula of a time-ordered correlator easily follows: apply \eqref{eq:expKID} to each current then follow the $T$-ordering instruction to bring all 
$b$ and $\bdag$ to the left of the $a$ and $\adag$ (assuming here that $p_i^0<0$ for $i\leq j$)
\begin{align}
   \hspace{-0.4cm} \bra{0} T(j_1(p_1)j_2(p_2)...j_n(p_n)) \ket{0}
   &=\bra{0} T(a_1^\dagger \cdots a_j^\dagger b_{j+1} \cdots b_n)\ket{0}
   \\& = \bra{0} b_{j+1} \cdots b_n a_1^\dagger \cdots a_j^\dagger \ket{0}\notag
   \\&=\textcolor{RoyalBlue}{_\inn\!\bra{(j{+}1) ...n  } S \ket{1...j}_\inn}\notag\,.
\end{align}
Here we have also used the fact that $a\vacR=0=\bra{0}\bdag$ to eliminate all terms with an $a$ or $\bdag$ in \eqref{eq:expKID}. Note that this argument (and the LSZ reduction formula itself, as far as we understand) requires non-forward kinematics: $(p_i+p_j)^\mu\neq 0$ for each pair.

For out-of-time-ordered observables (which we study next), we must generally keep all the terms in \eqref{eq:expKID}. This leads to generalized LSZ reduction formulae!

\subsection{Generalized reduction formulae}\label{sec:genRed}
As reviewed above, reduction formulae connect scattering amplitudes with the on-shell limit of amputated Green functions. Below, we illustrate how generalized reduction formulae can be used to reveal non-trivial abstract features of either time-ordered or out-of-time-ordered (OTO) observables (such as analyticity). In this chapter, we examine how applying LSZ-like reductions to generic OTO correlators (OTOC) reveals a wide range of asymptotic observables. We do so from two perspectives: from an algebraic point of view (involving formal manipulations of various time ordering operators) and visually using the Schwinger--Keldysh formalism.

\subsubsection{OTOC: definitions and generalities}

The type of correlators we wish to reduce will always be expressible in terms of products of \defQ{time-ordered products} of the following form (introduced in the 1960s by Ruelle and Araki--Burgoyne \cite{ruelle1961,araki1960properties})
\begin{equation}
\la0|T(\phi_1...\phi_{i_1})T(\phi_{i_1+1}...\phi_{i_2})...T(\phi_{i_{n-1}+1}...\phi_{i_n})|0\ra\,,\label{eq:TOP}
\end{equation}
where $\phi_{n}=\phi(x_n)$ is a local operator (field)\footnote{Note that the notation is abused here; the fields are allowed to be of different nature, which is not apparent as we suppressed this label.} and $T(\phi_1...\phi_n)$ is the conventional time-ordered product of $n$ local operators
\begin{equation}\label{eq:defTprod}
\begin{split}
T(\phi_1...\phi_n)&=\sum_{\sigma\in\text{S}_n}\theta(x_{\sigma(1)}^0>x_{\sigma(2)}^0>...>x_{\sigma(n)}^0)\epsilon(\sigma)\phi_{\sigma(1)}\phi_{\sigma(2)}...\phi_{\sigma(n)}\\&= \sum_{\sigma\in\text{S}_n}\Big[\prod_{m=1}^{n-1}\theta(x_{\sigma(m)}^0>x_{\sigma(m+1)}^0)\Big]\epsilon(\sigma)\phi_{\sigma(1)}\phi_{\sigma(2)}...\phi_{\sigma(n)}\,,
\end{split}
\end{equation}
where $\text{S}_n$ denotes the symmetric group of degree $n$ and 
\begin{equation}
    \epsilon(\sigma)=\begin{sqcases}
        1 & \text{bosonic operators}\\
        \text{sign}(\sigma) & \text{fermionic operators}
    \end{sqcases}\,.
\end{equation}
For example, for the bosonic two-point function, we get the familiar expression
\begin{equation}
    T(\phi_1\phi_2)=\phi_1\phi_2\theta(x_1^0>x_2^0)+\phi_2\phi_1\theta(x_2^0>x_1^0)\,.
\end{equation}
Interestingly, linear combinations of products of $T$'s span other useful and physical objects such as \defQ{anti-time-ordered products} $\overline{T}(...)$, \defQ{retarded products} $R_i(...)$ and \defQ{advanced products} $A_i(...)$. At $n$-point, the former is simply defined by reversing the inequalities in \eqref{eq:defTprod}, while the second and third are defined as follows 
\begin{subequations}
    \begin{align}
        \label{eq:defRprod}
        \hspace{-0.3cm} R_{\textcolor{Maroon}{0}}(\textcolor{Maroon}{\phi_0} \phi_1 ...  \phi_n)&{=} \sum_{\sigma\in\text{S}_n} \theta(\textcolor{Maroon}{x_{0}^0}{>}x_{\sigma(1)}^0{>}{...}{>}x_{\sigma(n)}^0)
    \underbracket[0.4pt]{\big[ \big[ {...} \big[ \big[}_{(n{-}1)\text{-fold}} \hspace{-0.2cm}\textcolor{Maroon}{\phi_0}, \phi_{\sigma(1)}  \big], \phi_{\sigma(2)} \big], {...} \big],\phi_{\sigma(n)}\big]\,,\\
     \hspace{-0.3cm} A_{\textcolor{Maroon}{0}}(\textcolor{Maroon}{\phi_0} \phi_1 ...  \phi_n)&{=} \sum_{\sigma\in\text{S}_n} \theta(\textcolor{Maroon}{x_{0}^0}{<}x_{\sigma(1)}^0{<}...{<}x_{\sigma(n)}^0)
\big[\phi_{\sigma(n)},\big[...,\big[\phi_{\sigma(2)},\big[\phi_{\sigma(1)},\textcolor{Maroon}{\phi_0}\hspace{-0.2cm}\underbracket[0.4pt]{\big] \big] {...} \big] \big]}_{(n{-}1)\text{-fold}} \textcolor{Maroon}\,,
    \end{align}
\end{subequations}
with the special case $R_{\textcolor{Maroon}{0}}(\textcolor{Maroon}{\phi_0})= \textcolor{Maroon}{\phi_0}$. The $A$-product is like the $R$ product with commutators nested in the second entry rather than in the first one, and with the inequalities in the $\theta$-functions reversed.

Importantly, the $R$-product (as well as the $A$-product) is symmetric in $\{1,2,{...},n\}$, but one field $\textcolor{Maroon}{\phi_0}$ is singled out as the one with the latest time; it is \textcolor{Maroon}{\emph{pinned}} in the future of everyone. That is,
\begin{equation}
    R_{\textcolor{Maroon}{0}}(\textcolor{Maroon}{\phi_0}\phi_1...\phi_n) = 0 \qquad \text{ if } \textcolor{Maroon}{x_0^0} < \max (x_1^0, x_2^0, ..., x_n^0) \,.
\end{equation}

\begin{mdexample} To digest these definitions, let us look at some low-$n$ examples:
\begin{equation}\label{eq:rpExs}
\begin{split}
      R_{\textcolor{Maroon}{2}}(\textcolor{Maroon}{\phi_2}\phi_1)&=R_{\textcolor{Maroon}{2}}(\phi_1\textcolor{Maroon}{\phi_2}) = [\textcolor{Maroon}{\phi_2},\phi_1]\theta(\textcolor{Maroon}{x_2^0}>x_1^0)\,,\\
            R_{\textcolor{Maroon}{3}}(\textcolor{Maroon}{\phi_3}\phi_2\phi_1)&=R_{\textcolor{Maroon}{3}}(\phi_1\phi_2\textcolor{Maroon}{\phi_3}) = [[\textcolor{Maroon}{\phi_3},\phi_2],\phi_1]\theta(\textcolor{Maroon}{x_3^0}>x_2^0)\theta(\textcolor{Maroon}{x_2^0}>x_1^0)+(1\leftrightarrow 2)\,.
\end{split}
\end{equation}
Note that both \eqref{eq:defTprod} and \eqref{eq:defRprod} involve the sum over $(n-1)!$ terms. Above, we indicated that both the time reversal of $\eqref{eq:defTprod}$ and \eqref{eq:defRprod} can be expressed solely in terms of standard time-ordered products. This extends to any number of fields, as we can show by a $\theta$-functions bookkeeping. Examining the simple case of $n=2$ is sufficient to understand why. Indeed, introducing the shorthand notation 
\begin{equation}
    \theta_{ij}=\theta(x_i^0>x_j^0)\,,
\end{equation}
we have
\begin{equation}
    \begin{split}
\overline{T}(\phi_1\phi_2)&=\phi_1\phi_2\theta_{21}+\phi_2\phi_1\theta_{12}\\&=\phi_1\phi_2(\theta_{21}+\theta_{12})-\phi_1\phi_2\theta_{12}+\phi_2\phi_1(\theta_{12}+\theta_{21})-\phi_2\phi_1\theta_{21}
\\&=\phi_1\phi_2-\phi_1\phi_2\theta_{12}+\phi_2\phi_1-\phi_2\phi_1\theta_{21}
\\&=T(\phi_1)T(\phi_2)+T(\phi_2)T(\phi_1)-T(\phi_1\phi_2)\,,
    \end{split}
\end{equation}
and 
\begin{equation}
    \begin{split}
        R_1(\phi_1\phi_2)&=[\phi_1,\phi_2]\theta_{12}=\phi_1\phi_2\theta_{12}-\phi_2\phi_1\theta_{12}\\&=\phi_1\phi_2\theta_{12}-\phi_2\phi_1(\theta_{12}+\theta_{21})+\phi_2\phi_1\theta_{21}\\&=T(\phi_1\phi_2)-T(\phi_2)T(\phi_1)\,.
    \end{split}
\end{equation}
We emphasize once more that such relations generalize to any number of points and are invertible. Consequently, replacing $T$ by $\overline{T}$ or $R$ in \eqref{eq:TOP} would generate the same linear span of correlators. 
\end{mdexample}
In the following, we revisit the Schwinger--Keldysh formalism, as described in the contribution \cite{chapterHaehlRangamani}. This formalism provides a useful visual approach to understanding OTOCs.

\subsubsection{Schwinger--Keldysh formalism}
In order to better understand OTOCs and their evaluation in many situations, it is useful to turn to a path-integral representation. This is precisely what \defQ{Schwinger--Keldysh formalism} (SK) provides. The idea behind this formalism is quite simple:

This approach allows us to write down any combination
of anti- and time-ordered products by linking the corresponding Lorentzian (real) time axes
by infinitesimal Euclidean (imaginary) time shifts. Ultimately, this results in a path integral that features numerous time-folds that oscillate between the past and the future.
\begin{mdexample}
To illustrate this in the context of correlators, we can consider the following seven-field correlator represented by a path integral over a Schwinger--Keldysh contour with three timefolds (I, II, and III):
\begin{align}
   \<0|\, \underbracket[0.4pt]{\cT(FG)}_{\text{fold III}}\, \underbracket[0.4pt]{\overline{\cT}(CDE)}_{\text{fold II}}\, \underbracket[0.4pt]{\cT(AB)}_{\text{fold I}}\,|0\>
&\equiv \<0| \cC(A^{\rone}B^{\rone} C^{\rtwo}D^{\rtwo}E^{\rtwo} F^{\rthree}G^{\rthree})|0\> 
\\&= \adjustbox{valign=c}{\tikzset{every picture/.style={line width=0.85pt}}
\begin{tikzpicture}[x=0.75pt,y=0.75pt,yscale=-1,xscale=1]
\tikzset{ma/.style={decoration={markings,mark=at position 0.5 with {\arrow[scale=0.7]{>}}},postaction={decorate}}}
\tikzset{ma2/.style={decoration={markings,mark=at position 0.3 with {\arrow[scale=0.7,Maroon!70!black]{>}}},postaction={decorate}}}
\tikzset{ma3/.style={decoration={markings,mark=at position 0.7 with {\arrow[scale=0.7,RoyalBlue!80!black]{>}}},postaction={decorate}}}
\tikzset{mar/.style={decoration={markings,mark=at position 0.5 with {\arrowreversed[scale=0.7]{>}}},postaction={decorate}}}
\draw[gradBlackToR=0.5]    (331.43,66.57) -- (331.43,81.57) ;
\draw[mar] [color=Maroon!  ,draw opacity=1 ]   (90.43,81.57) -- (331.43,81.57)
node [pos=0.189, inner sep=0.75pt] [font=\small, color=Maroon!, opacity=1] {$\tikzxmark$}
node [pos=0.595, inner sep=0.75pt] [font=\small, color=Maroon!, opacity=1] {$\tikzxmark$}
node [pos=0, left][inner sep=0.75pt]  [font=\small,color=Maroon!  ,opacity=1 ]  {$\text{I}\;$};
\draw[ma] [color=ForestGreen!  ,draw opacity=1 ]   (90.43,95) -- (331.43,95) 
node [pos=0.27, inner sep=0.75pt] [font=\small, color=ForestGreen!, opacity=1] {$\tikzxmark$}
node [pos=0.40, inner sep=0.75pt] [font=\small, color=ForestGreen!, opacity=1] {$\tikzxmark$}
node [pos=0.87, inner sep=0.75pt] [font=\small, color=ForestGreen!, opacity=1] {$\tikzxmark$}
node [pos=0, left][inner sep=0.75pt]  [font=\small,color=ForestGreen!  ,opacity=1 ]  {$\text{II}\;$};
\draw[gradRtoG=0.5]    (90.43,81.57) -- (90.43,95) ;
\draw [gradGtoB=0.5]    (331.43,95) -- (331.43,108.43) ;
\draw[mar] [color=RoyalBlue!  ,draw opacity=1 ]   (90.43,108.43) -- (331.43,108.43) node [pos=0.76, inner sep=0.75pt] [font=\small, color=RoyalBlue!, opacity=1] {$\tikzxmark$} node [pos=0.02, inner sep=0.75pt] [font=\small, color=RoyalBlue!, opacity=1] {$\tikzxmark$}
node [pos=0, left][inner sep=0.75pt]  [font=\small,color=RoyalBlue!  ,opacity=1 ]  {$\text{III}\;$};
\draw[gradBtoBlack=0.5]    (90.43,108.43) -- (90.43,123.43) ;
\draw[<-,line width =0.5, color=gray!]    (355,90) -- (371,90) ;
\draw[<-,line width =0.5, color=gray!]    (370.95,75) -- (370.95,90.25) ;
\draw (356,94.4) node [anchor=north west][inner sep=0.75pt, color=gray!]  [font=\tiny]  {$\text{Re}\, t$};
\draw (372.8,80.4) node [anchor=north west][inner sep=0.75pt, color=gray!]  [font=\tiny]  {$\text{Im}\, t$};
\draw (237,69.4) node [anchor=north west][inner sep=0.75pt]  [font=\scriptsize]  {$A$};
\draw (140,69.4) node [anchor=north west][inner sep=0.75pt]  [font=\scriptsize]  {$B$};
\draw (158,83.97) node [anchor=north west][inner sep=0.75pt]  [font=\scriptsize]  {$C$};
\draw (190.93,83.97) node [anchor=north west][inner sep=0.75pt]  [font=\scriptsize]  {$D$};
\draw (303.93,83.97) node [anchor=north west][inner sep=0.75pt]  [font=\scriptsize]  {$E$};
\draw (276.93,96.97) node [anchor=north west][inner sep=0.75pt]  [font=\scriptsize]  {$F$};
\draw (98.93,96.97) node [anchor=north west][inner sep=0.75pt]  [font=\scriptsize]  {$G$};
\end{tikzpicture}}.\nonumber
\label{example 3fold}
\end{align}
The \defQ{contour-ordering symbol} $\cC$ exemplified here is a natural generalization of the time-ordering symbol. The superscript on an operator labels on which timefold it is inserted. The total action includes the contribution from past directed branches with a minus sign (in our convention, the even ones: $\rtwo$, $\mathrm{IV}$, \ldots) since $\d t<0$, such that $\e^{iS_{\mathcal{C}}}=\e^{iS^{\rone}-iS^{\rtwo}+iS^{\rthree}-iS^{\mathrm{IV}}+\ldots}$.
The path integral measure is simply a product of the usual one: \(\int\mathcal{D}\phi=\int \mathcal{D}\phi^{\rone} \int \mathcal{D}\phi^{\rtwo}\cdots\).
\end{mdexample}

\begin{mdexample} {\bf (Path integral for retarded products)} Let us consider what kind of formula we might expect for retarded functions
by considering the simplest case of $n=2$:
\begin{equation}\label{eq:retP1}
    \vac{R_{2}(\phi_1\phi_2)} = \vac{[\phi_2,\phi_1]\theta_{21}}=\vac{(\phi_2\phi_1-\phi_1\phi_2)\theta_{21}}\,.
\end{equation}
If we were to represent the two terms in parentheses, ignoring the $\theta$-function, one would obtain the following pictures
\begin{equation}
   \hspace{-0.5cm} \adjustbox{valign=c}{\tikzset{every picture/.style={line width=0.85pt}}
\begin{tikzpicture}[x=0.75pt,y=0.75pt,yscale=-1,xscale=1]
\tikzset{ma/.style={decoration={markings,mark=at position 0.5 with {\arrow[scale=0.7]{>}}},postaction={decorate}}}
\tikzset{ma2/.style={decoration={markings,mark=at position 0.3 with {\arrow[scale=0.7,Maroon!70!black]{>}}},postaction={decorate}}}
\tikzset{ma3/.style={decoration={markings,mark=at position 0.7 with {\arrow[scale=0.7,RoyalBlue!80!black]{>}}},postaction={decorate}}}
\tikzset{mar/.style={decoration={markings,mark=at position 0.5 with {\arrowreversed[scale=0.7]{>}}},postaction={decorate}}}
\draw[gradBlackToR=0.5]    (301.43,66.57) -- (301.43,81.57) ;
\draw[mar] [color=Maroon!  ,draw opacity=1 ]   (90.43,81.57) -- (301.43,81.57)
node [pos=0.189, inner sep=0.75pt] [font=\small, color=Maroon!, opacity=1] {$\tikzxmark$}
node [pos=0.595, inner sep=0.75pt] [font=\small, color=Maroon!, opacity=1] {$\tikzxmark$}
node [pos=0, left][inner sep=0.75pt]  [font=\small,color=Maroon!  ,opacity=1 ]  {$\text{I}\;$};
\draw (237,69.4) node [anchor=north west][inner sep=0.75pt]  [font=\scriptsize]  {$1$};
\draw (140,69.4) node [anchor=north west][inner sep=0.75pt]  [font=\scriptsize]  {$2$};
\end{tikzpicture}} \quad \text{and} \quad \adjustbox{valign=c}{\tikzset{every picture/.style={line width=0.85pt}}
\begin{tikzpicture}[x=0.75pt,y=0.75pt,yscale=-1,xscale=1]
\tikzset{ma/.style={decoration={markings,mark=at position 0.5 with {\arrow[scale=0.7]{>}}},postaction={decorate}}}
\tikzset{ma2/.style={decoration={markings,mark=at position 0.3 with {\arrow[scale=0.7,Maroon!70!black]{>}}},postaction={decorate}}}
\tikzset{ma3/.style={decoration={markings,mark=at position 0.7 with {\arrow[scale=0.7,RoyalBlue!80!black]{>}}},postaction={decorate}}}
\tikzset{mar/.style={decoration={markings,mark=at position 0.5 with {\arrowreversed[scale=0.7]{>}}},postaction={decorate}}}
\draw[gradBlackToR=0.5]    (301.43,66.57) -- (301.43,81.57) ;
\draw[mar] [color=Maroon!  ,draw opacity=1 ]   (90.43,81.57) -- (301.43,81.57)
node [pos=0.189, inner sep=0.75pt] [font=\small, color=Maroon!, opacity=1] {$\tikzxmark$}
node [pos=0.595, inner sep=0.75pt] [font=\small, color=Maroon!, opacity=1] {$\tikzxmark$}
node [pos=0, left][inner sep=0.75pt]  [font=\small,color=Maroon!  ,opacity=1 ]  {$\text{I}\;$};
\draw (237,69.4) node [anchor=north west][inner sep=0.75pt]  [font=\scriptsize]  {$2$};
\draw (140,69.4) node [anchor=north west][inner sep=0.75pt]  [font=\scriptsize]  {$1$};
\end{tikzpicture}}
\end{equation}
However, it is clear that the second contour cannot accurately describe \eqref{eq:retP1} precisely due to the presence of the $\theta$ function, which requires that $1$ is in the past of $2$. A straightforward and correct solution to this problem is to \emph{superpose} both contours and to \emph{fold} them as follows
\begin{equation}
\begin{split}\label{eq:retProp}
       \hspace{-0.45cm}   \vac{R_{2}(\phi_1\phi_2)}{=}\adjustbox{valign=c}{\tikzset{every picture/.style={line width=0.85pt}}
\begin{tikzpicture}[x=0.75pt,y=0.75pt,yscale=-1,xscale=1]
\tikzset{ma/.style={decoration={markings,mark=at position 0.5 with {\arrow[scale=0.7]{>}}},postaction={decorate}}}
\tikzset{ma2/.style={decoration={markings,mark=at position 0.3 with {\arrow[scale=0.7,Maroon!70!black]{>}}},postaction={decorate}}}
\tikzset{ma3/.style={decoration={markings,mark=at position 0.7 with {\arrow[scale=0.7,RoyalBlue!80!black]{>}}},postaction={decorate}}}
\tikzset{mar/.style={decoration={markings,mark=at position 0.5 with {\arrowreversed[scale=0.7]{>}}},postaction={decorate}}}
\draw[gradBlackToR=0.5]    (331.43,66.57) -- (331.43,81.57) ;
\draw[mar] [color=Maroon!  ,draw opacity=1 ]   (90.43,81.57) -- (331.43,81.57)
node [pos=0.189, inner sep=0.75pt] [font=\small, color=Maroon!, opacity=1] {$\tikzxmark$}
node [pos=0.595, inner sep=0.75pt] [font=\small, color=Maroon!, opacity=1] {$\tikzxmark$}
node [pos=0, left][inner sep=0.75pt]  [font=\small,color=Maroon!  ,opacity=1 ]  {$\text{I}\;$};
\draw[ma] [color=RoyalBlue!  ,draw opacity=1 ]   (90.43,95) -- (331.43,95) 
node [pos=0.595, inner sep=0.75pt] [font=\small, color=RoyalBlue!, opacity=1] {$\tikzxmark$}
node [pos=0, left][inner sep=0.75pt]  [font=\small,color=RoyalBlue!  ,opacity=1 ]  {$\text{II}\;$};
\draw[gradBlackToR=0.5]    (331.43,66.57) -- (331.43,81.57) ;
\draw[gradRtoB=0.5]    (90.43,81.57) -- (90.43,95) ;
\draw [gradBtoBlack=0.5]    (331.43,95) -- (331.43,108.43) ;
\draw (237,69.4) node [anchor=north west][inner sep=0.75pt]  [font=\scriptsize]  {$1$};
\draw (237,99.4) node [anchor=north west][inner sep=0.75pt]  [font=\scriptsize]  {$1$};
\draw (140,69.4) node [anchor=north west][inner sep=0.75pt]  [font=\scriptsize]  {$2$};
\draw (100,99.4) node [anchor=north west][inner sep=0.75pt]  [font=\scriptsize]  {$\mathcal{C}$};
\end{tikzpicture}}&{=}\vac{\mathcal{C}(\phi_2(\phi_1^\text{I}{-}\phi_1^\text{II}))}\\&\equiv \vac{\mathcal{C}(\phi_2\phi_1^\text{diff})}\,,
\end{split}
\end{equation}
where $\phi_k^{\rdiff}(x)\equiv \phi_k^\rone(x)-\phi_k^\rtwo(x)$. Note that the minus sign comes from the commutator and that the role of the $\mathcal{C}$-operation is simply to bring any $\phi^\text{II}$ to the left of $\phi^\text{I}$.

At $n$-point, we replace the commutator in \eqref{eq:retP1} by the nested commutator in \eqref{eq:defRprod} and repeat the same exercise to get the $n$-point retarded product in terms of the $n$-point correlators of field differences, namely
\be\label{eq:npointRP}
\begin{split}
    \cR_0(\phi_0 \phi_1\cdots \phi_{n-1} ) &\equiv
 \cC( \phi_0^{\rone} \phi_1^{\rdiff}\cdots \phi_{n-1}^{\rdiff}) \\&= \adjustbox{valign=c}{\tikzset{every picture/.style={line width=0.85pt}}
\begin{tikzpicture}[x=0.75pt,y=0.75pt,yscale=-1.1,xscale=1.1]
\tikzset{ma/.style={decoration={markings,mark=at position 0.5 with {\arrow[scale=0.7]{>}}},postaction={decorate}}}
\tikzset{ma2/.style={decoration={markings,mark=at position 0.3 with {\arrow[scale=0.7,Maroon!70!black]{>}}},postaction={decorate}}}
\tikzset{ma3/.style={decoration={markings,mark=at position 0.7 with {\arrow[scale=0.7,RoyalBlue!80!black]{>}}},postaction={decorate}}}
\tikzset{mar/.style={decoration={markings,mark=at position 0.5 with {\arrowreversed[scale=0.7]{>}}},postaction={decorate}}}
\draw[line width =1, dash pattern=on 1pt off 1pt]    (136,81.4) -- (136,94.4) ;
\draw[line width =1, dash pattern=on 1pt off 1pt]    (234,81.4) -- (234,94.4) ;
\begin{scope}[yshift=0.25]
\draw[gradBlackToR=0.5]    (331.43,66.57) -- (331.43,81.57) ;
\end{scope}
\draw[mar] [color=Maroon!  ,draw opacity=1 ]   (90.43,81.57) -- (331.43,81.57) node [pos=0.069, inner sep=0.75pt] [font=\small, color=Maroon!, opacity=1] {$\tikzxmark$}
node [pos=0.189, inner sep=0.75pt] [font=\small, color=Maroon!, opacity=1] {$\tikzxmark$}
node [pos=0.595, inner sep=0.75pt] [font=\small, color=Maroon!, opacity=1] {$\tikzxmark$}
node [pos=0, left][inner sep=0.75pt]  [font=\small,color=Maroon!  ,opacity=1 ]  {$\text{I}\;$};
\draw[ma] [color=RoyalBlue!  ,draw opacity=1 ]   (90.43,95) -- (331.43,95) 
node [pos=0.189, inner sep=0.75pt] [font=\small, color=RoyalBlue!, opacity=1] {$\tikzxmark$}
node [pos=0.595, inner sep=0.75pt] [font=\small, color=RoyalBlue!, opacity=1] {$\tikzxmark$}
node [pos=0, left][inner sep=0.75pt]  [font=\small,color=RoyalBlue!  ,opacity=1 ]  {$\text{II}\;$};
\draw[gradRtoB=0.5]    (90.43,81.57) -- (90.43,95) ;
\draw[gradBtoBlack=0.5]    (331.43,95) -- (331.43,108.43) ;
\draw[<-,line width =0.5, color=gray!]    (355,90) -- (371,90) ;
\draw[<-,line width =0.5, color=gray!]    (370.95,75) -- (370.95,90.25) ;
\draw (356,94.4) node [anchor=north west][inner sep=0.75pt, color=gray!]  [font=\tiny]  {$\text{Re}\, t$};
\draw (372.8,80.4) node [anchor=north west][inner sep=0.75pt, color=gray!]  [font=\tiny]  {$\text{Im}\, t$};
\draw (237,65) node [anchor=north west][inner sep=0.75pt]  [font=\scriptsize]  {$\phi_{1}^{\text{diff}}$};
\draw (140,65) node [anchor=north west][inner sep=0.75pt]  [font=\scriptsize]  {$\phi_{n-1}^{\text{diff}}$};
\draw (109.93,66.4) node [anchor=north west][inner sep=0.75pt]  [font=\scriptsize]  {$\phi_0^\text{I}$};
\draw (174.93,85.4) node [anchor=north west][inner sep=0.75pt]  [font=\scriptsize]  {$\cdots$};
\end{tikzpicture}}\,.
\end{split}
\ee
Note that the relative positions of the fields $1, \ldots, n-1$ with respect to each other on the timefolds are arbitrary; the key point is that $\phi_0$ has to be the farthest in the future; otherwise we get zero. For advanced products, $\phi_0$ is simply positioned in the past relative to all other fields.
\end{mdexample}

Let us now illustrate how Cutkosky-like cutting rules arise from the Schwinger--Keldysh formalism. 

\begin{mdexample}
The \defQ{largest time equation} \cite{britto2024cuttingedge}
\begin{equation}\label{eq:LTEdef}
    \sum_{k=0}^n(-1)^k\sum_{\sigma\in P(k,n{-}k)}\overline{T}(\phi_{\sigma(1)}...\phi_{\sigma(k)})T(\phi_{\sigma(k+1)}...\phi_{\sigma(n)})=0\,,
\end{equation}
(where $P(k,n{-}k)$ denotes the set of partitions of $n$ labels into two sets of size $k$ and $n-k$) which reflects the fact that unitarity cuts are not independent of each other, takes a very simple form in the SK formalism
\be \label{largest time equation}
  \<0|\, \cC(\phi_1^{\rdiff}\phi_2^{\rdiff}\cdots \phi_n^{\rdiff})\,|0\>=0\,.
\ee
This equation is manifestly true thanks to the boundary condition 
\begin{equation}\label{eq:bdryCond}
    \lim_{x^0\to\infty}\phi^\rone-\phi^\rtwo=0\,,
\end{equation}
where the timefolds meet. It is not difficult to check that \eqref{largest time equation} is equivalent to \eqref{eq:LTEdef}. This is particularly easy to check explicitly on low $n$ cases, e.g., for $n=2$
\begin{align}
        \eqref{eq:LTEdef}&\leadsto (-1)^2\textcolor{RoyalBlue}{T(\phi_1\phi_2)}{+}(-1)^2\textcolor{Maroon}{\overline{T}(\phi_1\phi_2)}{+}(-1)^1\textcolor{ForestGreen}{\overline{T}(\phi_1)T(\phi_2)}{+}(-1)^1\textcolor{Orange}{\overline{T}(\phi_2)T(\phi_1)}=0\,,\notag\\
        \eqref{largest time equation}&\leadsto   \<0|\,\textcolor{RoyalBlue}{\phi_1^{\text{I}}\phi_2^{\text{I}}}{-}\textcolor{Orange}{\phi_2^{\text{II}}\phi_1^{\text{I}}}-\textcolor{ForestGreen}{\phi_1^{\text{II}}\phi_2^{\text{I}}}+\textcolor{Maroon}{\phi_1^{\text{II}}\phi_1^{\text{II}}}\,|0\>=0\,.
\end{align}
\end{mdexample}
We conclude our discussion of the Schwinger--Keldysh formalism's generalities by presenting some illustrative examples of how various products of $T$, $\bar{T}$ and $R$-products can be realized on multifold contours.
\begin{mdexample}
    Various types of time-ordering products discussed in the literature and their relations to Schwinger--Keldysh time-folds. For illustrative purposes, each panel only shows a three-time-fold contour, with the generalization to an arbitrary number of time-folds following the same obvious patterns: 
\begin{equation}\label{example 3fold}
    \begin{minipage}{0.49\textwidth}
\begin{tcolorbox}[colback=white, colframe=white, rounded corners]
\begin{equation*}
     \hspace{-0.7cm}\adjustbox{valign=c, width=\textwidth}{\tikzset{grad0/.style={
    postaction={
        decorate,
        decoration={
            markings,
            mark=at position \pgfdecoratedpathlength-0.5pt with {\arrow[Maroon!,line width=#1] {}; },
            mark=between positions 0 and \pgfdecoratedpathlength-0pt step 0.5pt with {
                \pgfmathsetmacro\myval{multiply(divide(
                    \pgfkeysvalueof{/pgf/decoration/mark info/distance from start}, \pgfdecoratedpathlength),100)};
                \pgfsetfillcolor{RedOrange!\myval!Maroon!};
                \pgfpathcircle{\pgfpointorigin}{#1};
                \pgfusepath{fill};}
}}}}

\tikzset{grad1/.style={
    postaction={
        decorate,
        decoration={
            markings,
            mark=at position \pgfdecoratedpathlength-0.5pt with {\arrow[RedOrange!,line width=#1] {}; },
            mark=between positions 0 and \pgfdecoratedpathlength-0pt step 0.5pt with {
                \pgfmathsetmacro\myval{multiply(divide(
                    \pgfkeysvalueof{/pgf/decoration/mark info/distance from start}, \pgfdecoratedpathlength),100)};
                \pgfsetfillcolor{Dandelion!\myval!RedOrange!};
                \pgfpathcircle{\pgfpointorigin}{#1};
                \pgfusepath{fill};}
}}}}

\tikzset{grad2/.style={
    postaction={
        decorate,
        decoration={
            markings,
            mark=at position \pgfdecoratedpathlength-0.5pt with {\arrow[Dandelion!,line width=#1] {}; },
            mark=between positions 0 and \pgfdecoratedpathlength-0pt step 0.5pt with {
                \pgfmathsetmacro\myval{multiply(divide(
                    \pgfkeysvalueof{/pgf/decoration/mark info/distance from start}, \pgfdecoratedpathlength),100)};
                \pgfsetfillcolor{ForestGreen!\myval!Dandelion!};
                \pgfpathcircle{\pgfpointorigin}{#1};
                \pgfusepath{fill};}
}}}}

\tikzset{grad3/.style={
    postaction={
        decorate,
        decoration={
            markings,
            mark=at position \pgfdecoratedpathlength-0.5pt with {\arrow[ForestGreen!,line width=#1] {}; },
            mark=between positions 0 and \pgfdecoratedpathlength-0pt step 0.5pt with {
                \pgfmathsetmacro\myval{multiply(divide(
                    \pgfkeysvalueof{/pgf/decoration/mark info/distance from start}, \pgfdecoratedpathlength),100)};
                \pgfsetfillcolor{RoyalBlue!\myval!ForestGreen!};
                \pgfpathcircle{\pgfpointorigin}{#1};
                \pgfusepath{fill};}
}}}}

\tikzset{grad4/.style={
    postaction={
        decorate,
        decoration={
            markings,
            mark=at position \pgfdecoratedpathlength-0.5pt with {\arrow[RoyalBlue!,line width=#1] {}; },
            mark=between positions 0 and \pgfdecoratedpathlength-0pt step 0.5pt with {
                \pgfmathsetmacro\myval{multiply(divide(
                    \pgfkeysvalueof{/pgf/decoration/mark info/distance from start}, \pgfdecoratedpathlength),100)};
                \pgfsetfillcolor{Violet!\myval!RoyalBlue!};
                \pgfpathcircle{\pgfpointorigin}{#1};
                \pgfusepath{fill};}
}}}}

\tikzset{grad5/.style={
    postaction={
        decorate,
        decoration={
            markings,
            mark=at position \pgfdecoratedpathlength-0.5pt with {\arrow[Violet!,line width=#1] {}; },
            mark=between positions 0 and \pgfdecoratedpathlength-0pt step 0.5pt with {
                \pgfmathsetmacro\myval{multiply(divide(
                    \pgfkeysvalueof{/pgf/decoration/mark info/distance from start}, \pgfdecoratedpathlength),100)};
                \pgfsetfillcolor{Black!\myval!Violet!};
                \pgfpathcircle{\pgfpointorigin}{#1};
                \pgfusepath{fill};}
}}}}

\tikzset{every picture/.style={line width=0.85pt}}
\begin{tikzpicture}[x=0.75pt,y=0.75pt,yscale=-1,xscale=1]
\tikzset{ma/.style={decoration={markings,mark=at position 0.5 with {\arrow[scale=0.7]{>}}},postaction={decorate}}}
\tikzset{ma2/.style={decoration={markings,mark=at position 0.3 with {\arrow[scale=0.7,Maroon!70!black]{>}}},postaction={decorate}}}
\tikzset{ma3/.style={decoration={markings,mark=at position 0.7 with {\arrow[scale=0.7,RoyalBlue!80!black]{>}}},postaction={decorate}}}
\tikzset{mar/.style={decoration={markings,mark=at position 0.5 with {\arrowreversed[scale=0.7]{>}}},postaction={decorate}}}

\draw[gradBlackToR=0.5]    (331.43,66.57) -- (331.43,81.57) ;
\draw[mar] [color=Maroon!  ,draw opacity=1 ]   (90.43,81.57) -- (331.43,81.57) 
node [pos=0.67, inner sep=0.75pt] [font=\small, color=Maroon!, opacity=1] {$\tikzxmark$}
node [pos=0.37, inner sep=0.75pt] [font=\small, color=Maroon!, opacity=1] {$\tikzxmark$}
node [pos=0, left][inner sep=0.75pt]  [font=\small,color=Maroon!  ,opacity=1 ]  {$\text{I}\;$}
node [midway, above, yshift=15pt, inner sep=0.75pt] [font=\small, color=black!, opacity=1] {$T\{\textcolor{RoyalBlue}{FGH}\}T\{\textcolor{Dandelion}{CDE}\}T\{\textcolor{Maroon}{AB}\}$};


\draw[ma] [color=RedOrange!  ,draw opacity=1 ]   (90.43,95) -- (331.43,95) 
node [pos=0, left][inner sep=0.75pt]  [font=\small,color=RedOrange!  ,opacity=1 ]  {$\text{II}\;$};
\draw[grad0=0.5]    (90.43,81.57) -- (90.43,95) ;

\draw [grad1=0.5]    (331.43,95) -- (331.43,108.43) ;
\draw[mar] [color=Dandelion!  ,draw opacity=1 ]   (90.43,108.43) -- (331.43,108.43) 
node [pos=0.47, inner sep=0.75pt] [font=\small, color=Dandelion!, opacity=1] {$\tikzxmark$}
node [pos=0.97, inner sep=0.75pt] [font=\small, color=Dandelion!, opacity=1] {$\tikzxmark$}
node [pos=0.27, inner sep=0.75pt] [font=\small, color=Dandelion!, opacity=1] {$\tikzxmark$}
node [pos=0, left][inner sep=0.75pt]  [font=\small,color=Dandelion!  ,opacity=1 ]  {$\text{III}\;$};

\draw[grad2=0.5]    (90.43,108.43) -- (90.43,123.43) ;
\draw[ma] [color=ForestGreen!  ,draw opacity=1 ]   (90.43,123.43) -- (331.43,123.43)
node [pos=0, left][inner sep=0.75pt]  [font=\small,color=ForestGreen!  ,opacity=1 ]  {$\text{IV}\;$};

\draw[grad3=0.5]    (331.43,123.43) -- (331.43,138.43) ;
\draw[mar] [color=RoyalBlue!  ,draw opacity=1 ]   (90.43,138.43) -- (331.43,138.43) 
node [pos=0.61, inner sep=0.75pt] [font=\small, color=RoyalBlue!, opacity=1] {$\tikzxmark$}
node [pos=0.31, inner sep=0.75pt] [font=\small, color=RoyalBlue!, opacity=1] {$\tikzxmark$}
node [pos=0.27, inner sep=0.75pt] [font=\small, color=RoyalBlue!, opacity=1] {$\tikzxmark$}
node [pos=0, left] [inner sep=0.75pt]  [font=\small,color=RoyalBlue!  ,opacity=1 ]  {$\text{V}\;$};

\draw[grad4=0.5]    (90.43,138.43) -- (90.43,153.43) ;
\draw[ma] [color=Violet!, draw opacity=1 ] (90.43,153.43) -- (331.43,153.43) 
node [pos=0, left, inner sep=0.75pt]  [font=\small, color=Violet!, opacity=1] {$\text{VI}\;$};

\draw[grad5=0.5]    (331.43,153.43) -- (331.43,168.43) ;

\end{tikzpicture}}
\end{equation*}
\end{tcolorbox}
\end{minipage}\hfill
\begin{minipage}{0.50\textwidth}
\begin{tcolorbox}[colback=white, colframe=white, rounded corners]
\begin{equation*}
     \hspace{-0.7cm}\adjustbox{valign=c, width=\textwidth}{\tikzset{grad0/.style={
    postaction={
        decorate,
        decoration={
            markings,
            mark=at position \pgfdecoratedpathlength-0.5pt with {\arrow[Maroon!,line width=#1] {}; },
            mark=between positions 0 and \pgfdecoratedpathlength-0pt step 0.5pt with {
                \pgfmathsetmacro\myval{multiply(divide(
                    \pgfkeysvalueof{/pgf/decoration/mark info/distance from start}, \pgfdecoratedpathlength),100)};
                \pgfsetfillcolor{RedOrange!\myval!Maroon!};
                \pgfpathcircle{\pgfpointorigin}{#1};
                \pgfusepath{fill};}
}}}}
\tikzset{grad1/.style={
    postaction={
        decorate,
        decoration={
            markings,
            mark=at position \pgfdecoratedpathlength-0.5pt with {\arrow[RedOrange!,line width=#1] {}; },
            mark=between positions 0 and \pgfdecoratedpathlength-0pt step 0.5pt with {
                \pgfmathsetmacro\myval{multiply(divide(
                    \pgfkeysvalueof{/pgf/decoration/mark info/distance from start}, \pgfdecoratedpathlength),100)};
                \pgfsetfillcolor{Dandelion!\myval!RedOrange!};
                \pgfpathcircle{\pgfpointorigin}{#1};
                \pgfusepath{fill};}
}}}}

\tikzset{grad2/.style={
    postaction={
        decorate,
        decoration={
            markings,
            mark=at position \pgfdecoratedpathlength-0.5pt with {\arrow[Dandelion!,line width=#1] {}; },
            mark=between positions 0 and \pgfdecoratedpathlength-0pt step 0.5pt with {
                \pgfmathsetmacro\myval{multiply(divide(
                    \pgfkeysvalueof{/pgf/decoration/mark info/distance from start}, \pgfdecoratedpathlength),100)};
                \pgfsetfillcolor{ForestGreen!\myval!Dandelion!};
                \pgfpathcircle{\pgfpointorigin}{#1};
                \pgfusepath{fill};}
}}}}

\tikzset{grad3/.style={
    postaction={
        decorate,
        decoration={
            markings,
            mark=at position \pgfdecoratedpathlength-0.5pt with {\arrow[ForestGreen!,line width=#1] {}; },
            mark=between positions 0 and \pgfdecoratedpathlength-0pt step 0.5pt with {
                \pgfmathsetmacro\myval{multiply(divide(
                    \pgfkeysvalueof{/pgf/decoration/mark info/distance from start}, \pgfdecoratedpathlength),100)};
                \pgfsetfillcolor{RoyalBlue!\myval!ForestGreen!};
                \pgfpathcircle{\pgfpointorigin}{#1};
                \pgfusepath{fill};}
}}}}

\tikzset{grad4/.style={
    postaction={
        decorate,
        decoration={
            markings,
            mark=at position \pgfdecoratedpathlength-0.5pt with {\arrow[RoyalBlue!,line width=#1] {}; },
            mark=between positions 0 and \pgfdecoratedpathlength-0pt step 0.5pt with {
                \pgfmathsetmacro\myval{multiply(divide(
                    \pgfkeysvalueof{/pgf/decoration/mark info/distance from start}, \pgfdecoratedpathlength),100)};
                \pgfsetfillcolor{Violet!\myval!RoyalBlue!};
                \pgfpathcircle{\pgfpointorigin}{#1};
                \pgfusepath{fill};}
}}}}

\tikzset{grad5/.style={
    postaction={
        decorate,
        decoration={
            markings,
            mark=at position \pgfdecoratedpathlength-0.5pt with {\arrow[Violet!,line width=#1] {}; },
            mark=between positions 0 and \pgfdecoratedpathlength-0pt step 0.5pt with {
                \pgfmathsetmacro\myval{multiply(divide(
                    \pgfkeysvalueof{/pgf/decoration/mark info/distance from start}, \pgfdecoratedpathlength),100)};
                \pgfsetfillcolor{Black!\myval!Violet!};
                \pgfpathcircle{\pgfpointorigin}{#1};
                \pgfusepath{fill};}
}}}}

\tikzset{every picture/.style={line width=0.85pt}}
\begin{tikzpicture}[x=0.75pt,y=0.75pt,yscale=-1,xscale=1]
\tikzset{ma/.style={decoration={markings,mark=at position 0.5 with {\arrow[scale=0.7]{>}}},postaction={decorate}}}
\tikzset{ma2/.style={decoration={markings,mark=at position 0.3 with {\arrow[scale=0.7,Maroon!70!black]{>}}},postaction={decorate}}}
\tikzset{ma3/.style={decoration={markings,mark=at position 0.7 with {\arrow[scale=0.7,RoyalBlue!80!black]{>}}},postaction={decorate}}}
\tikzset{mar/.style={decoration={markings,mark=at position 0.5 with {\arrowreversed[scale=0.7]{>}}},postaction={decorate}}}

\draw[gradBlackToR=0.5]    (331.43,66.57) -- (331.43,81.57) ;
\draw[mar] [color=Maroon!  ,draw opacity=1 ]   (90.43,81.57) -- (331.43,81.57) 
node [pos=0, left][inner sep=0.75pt]  [font=\small,color=Maroon!  ,opacity=1 ]  {$\text{I}\;$}
node [midway, above, yshift=15pt, inner sep=0.75pt] [font=\small, color=black!, opacity=1] {$\bar{T}\{\textcolor{Violet}{FGHI}\}\bar{T}\{\textcolor{ForestGreen}{DE}\}\bar{T}\{\textcolor{RedOrange}{ABC}\}$};

\draw[ma] [color=RedOrange!  ,draw opacity=1 ]   (90.43,95) -- (331.43,95) 
node [pos=0.67, inner sep=0.75pt] [font=\small, color=RedOrange!, opacity=1] {$\tikzxmark$}
node [pos=0.57, inner sep=0.75pt] [font=\small, color=RedOrange!, opacity=1] {$\tikzxmark$}
node [pos=0.27, inner sep=0.75pt] [font=\small, color=RedOrange!, opacity=1] {$\tikzxmark$}
node [pos=0, left][inner sep=0.75pt]  [font=\small,color=RedOrange!  ,opacity=1 ]  {$\text{II}\;$};
\draw[grad0=0.5]    (90.43,81.57) -- (90.43,95) ;

\draw [grad1=0.5]    (331.43,95) -- (331.43,108.43) ;
\draw[mar] [color=Dandelion!  ,draw opacity=1 ]   (90.43,108.43) -- (331.43,108.43) 
node [pos=0, left][inner sep=0.75pt]  [font=\small,color=Dandelion!  ,opacity=1 ]  {$\text{III}\;$};

\draw[grad2=0.5]    (90.43,108.43) -- (90.43,123.43) ;
\draw[ma] [color=ForestGreen!  ,draw opacity=1 ]   (90.43,123.43) -- (331.43,123.43)
node [pos=0.86, inner sep=0.75pt] [font=\small, color=ForestGreen!, opacity=1] {$\tikzxmark$}
node [pos=0.16, inner sep=0.75pt] [font=\small, color=ForestGreen!, opacity=1] {$\tikzxmark$}
node [pos=0, left][inner sep=0.75pt]  [font=\small,color=ForestGreen!  ,opacity=1 ]  {$\text{IV}\;$};

\draw[grad3=0.5]    (331.43,123.43) -- (331.43,138.43) ;
\draw[mar] [color=RoyalBlue!  ,draw opacity=1 ]   (90.43,138.43) -- (331.43,138.43) 
node [pos=0, left] [inner sep=0.75pt]  [font=\small,color=RoyalBlue!  ,opacity=1 ]  {$\text{V}\;$};

\draw[grad4=0.5]    (90.43,138.43) -- (90.43,153.43) ;
\draw[ma] [color=Violet!, draw opacity=1 ] (90.43,153.43) -- (331.43,153.43) 
node [pos=0.67, inner sep=0.75pt] [font=\small, color=Violet!, opacity=1] {$\tikzxmark$}
node [pos=0.17, inner sep=0.75pt] [font=\small, color=Violet!, opacity=1] {$\tikzxmark$}
node [pos=0.23, inner sep=0.75pt] [font=\small, color=Violet!, opacity=1] {$\tikzxmark$}
node [pos=0.77, inner sep=0.75pt] [font=\small, color=Violet!, opacity=1] {$\tikzxmark$}
node [pos=0, left, inner sep=0.75pt]  [font=\small, color=Violet!, opacity=1] {$\text{VI}\;$};

\draw[grad5=0.5]    (331.43,153.43) -- (331.43,168.43) ;

\end{tikzpicture}}
\end{equation*}
\end{tcolorbox}
\end{minipage}
\end{equation}
\begin{equation*}
    \begin{minipage}{0.49\textwidth}
\begin{tcolorbox}[colback=white, colframe=white, rounded corners]
\begin{equation*}
     \hspace{-0.7cm}\adjustbox{valign=c, width=\textwidth}{\tikzset{grad0/.style={
    postaction={
        decorate,
        decoration={
            markings,
            mark=at position \pgfdecoratedpathlength-0.5pt with {\arrow[Maroon!,line width=#1] {}; },
            mark=between positions 0 and \pgfdecoratedpathlength-0pt step 0.5pt with {
                \pgfmathsetmacro\myval{multiply(divide(
                    \pgfkeysvalueof{/pgf/decoration/mark info/distance from start}, \pgfdecoratedpathlength),100)};
                \pgfsetfillcolor{RedOrange!\myval!Maroon!};
                \pgfpathcircle{\pgfpointorigin}{#1};
                \pgfusepath{fill};}
}}}}
\tikzset{grad1/.style={
    postaction={
        decorate,
        decoration={
            markings,
            mark=at position \pgfdecoratedpathlength-0.5pt with {\arrow[RedOrange!,line width=#1] {}; },
            mark=between positions 0 and \pgfdecoratedpathlength-0pt step 0.5pt with {
                \pgfmathsetmacro\myval{multiply(divide(
                    \pgfkeysvalueof{/pgf/decoration/mark info/distance from start}, \pgfdecoratedpathlength),100)};
                \pgfsetfillcolor{Dandelion!\myval!RedOrange!};
                \pgfpathcircle{\pgfpointorigin}{#1};
                \pgfusepath{fill};}
}}}}

\tikzset{grad2/.style={
    postaction={
        decorate,
        decoration={
            markings,
            mark=at position \pgfdecoratedpathlength-0.5pt with {\arrow[Dandelion!,line width=#1] {}; },
            mark=between positions 0 and \pgfdecoratedpathlength-0pt step 0.5pt with {
                \pgfmathsetmacro\myval{multiply(divide(
                    \pgfkeysvalueof{/pgf/decoration/mark info/distance from start}, \pgfdecoratedpathlength),100)};
                \pgfsetfillcolor{ForestGreen!\myval!Dandelion!};
                \pgfpathcircle{\pgfpointorigin}{#1};
                \pgfusepath{fill};}
}}}}

\tikzset{grad3/.style={
    postaction={
        decorate,
        decoration={
            markings,
            mark=at position \pgfdecoratedpathlength-0.5pt with {\arrow[ForestGreen!,line width=#1] {}; },
            mark=between positions 0 and \pgfdecoratedpathlength-0pt step 0.5pt with {
                \pgfmathsetmacro\myval{multiply(divide(
                    \pgfkeysvalueof{/pgf/decoration/mark info/distance from start}, \pgfdecoratedpathlength),100)};
                \pgfsetfillcolor{RoyalBlue!\myval!ForestGreen!};
                \pgfpathcircle{\pgfpointorigin}{#1};
                \pgfusepath{fill};}
}}}}

\tikzset{grad4/.style={
    postaction={
        decorate,
        decoration={
            markings,
            mark=at position \pgfdecoratedpathlength-0.5pt with {\arrow[RoyalBlue!,line width=#1] {}; },
            mark=between positions 0 and \pgfdecoratedpathlength-0pt step 0.5pt with {
                \pgfmathsetmacro\myval{multiply(divide(
                    \pgfkeysvalueof{/pgf/decoration/mark info/distance from start}, \pgfdecoratedpathlength),100)};
                \pgfsetfillcolor{Violet!\myval!RoyalBlue!};
                \pgfpathcircle{\pgfpointorigin}{#1};
                \pgfusepath{fill};}
}}}}

\tikzset{grad5/.style={
    postaction={
        decorate,
        decoration={
            markings,
            mark=at position \pgfdecoratedpathlength-0.5pt with {\arrow[Violet!,line width=#1] {}; },
            mark=between positions 0 and \pgfdecoratedpathlength-0pt step 0.5pt with {
                \pgfmathsetmacro\myval{multiply(divide(
                    \pgfkeysvalueof{/pgf/decoration/mark info/distance from start}, \pgfdecoratedpathlength),100)};
                \pgfsetfillcolor{Black!\myval!Violet!};
                \pgfpathcircle{\pgfpointorigin}{#1};
                \pgfusepath{fill};}
}}}}

\tikzset{every picture/.style={line width=0.85pt}}
\begin{tikzpicture}[x=0.75pt,y=0.75pt,yscale=-1,xscale=1]
\tikzset{ma/.style={decoration={markings,mark=at position 0.5 with {\arrow[scale=0.7]{>}}},postaction={decorate}}}
\tikzset{ma2/.style={decoration={markings,mark=at position 0.3 with {\arrow[scale=0.7,Maroon!70!black]{>}}},postaction={decorate}}}
\tikzset{ma3/.style={decoration={markings,mark=at position 0.7 with {\arrow[scale=0.7,RoyalBlue!80!black]{>}}},postaction={decorate}}}
\tikzset{mar/.style={decoration={markings,mark=at position 0.5 with {\arrowreversed[scale=0.7]{>}}},postaction={decorate}}}

\draw[gradBlackToR=0.5]    (331.43,66.57) -- (331.43,81.57) ;
\draw[mar] [color=Maroon!  ,draw opacity=1 ]   (90.43,81.57) -- (331.43,81.57) 
node [pos=0.67, inner sep=0.75pt] [font=\small, color=Maroon!, opacity=1] {$\tikzxmark$}
node [pos=0.37, inner sep=0.75pt] [font=\small, color=Maroon!, opacity=1] {$\tikzxmark$}
node [pos=0, left][inner sep=0.75pt]  [font=\small,color=Maroon!  ,opacity=1 ]  {$\text{I}\;$}
node [midway, above, yshift=15pt, inner sep=0.75pt] [font=\small, color=black!, opacity=1] {$\bar{T}\{\textcolor{Violet}{KM}\}T\{\textcolor{RoyalBlue}{HIJ}\}\bar{T}\{\textcolor{ForestGreen}{FG}\}\bar{T}\{\textcolor{RedOrange}{CDE}\}T\{\textcolor{Maroon}{AB}\}$};

\draw[ma] [color=RedOrange!  ,draw opacity=1 ]   (90.43,95) -- (331.43,95) 
node [pos=0.67, inner sep=0.75pt] [font=\small, color=RedOrange!, opacity=1] {$\tikzxmark$}
node [pos=0.57, inner sep=0.75pt] [font=\small, color=RedOrange!, opacity=1] {$\tikzxmark$}
node [pos=0.27, inner sep=0.75pt] [font=\small, color=RedOrange!, opacity=1] {$\tikzxmark$}
node [pos=0, left][inner sep=0.75pt]  [font=\small,color=RedOrange!  ,opacity=1 ]  {$\text{II}\;$};
\draw[grad0=0.5]    (90.43,81.57) -- (90.43,95) ;

\draw [grad1=0.5]    (331.43,95) -- (331.43,108.43) ;
\draw[mar] [color=Dandelion!  ,draw opacity=1 ]   (90.43,108.43) -- (331.43,108.43) 
node [pos=0, left][inner sep=0.75pt]  [font=\small,color=Dandelion!  ,opacity=1 ]  {$\text{III}\;$};

\draw[grad2=0.5]    (90.43,108.43) -- (90.43,123.43) ;
\draw[ma] [color=ForestGreen!  ,draw opacity=1 ]   (90.43,123.43) -- (331.43,123.43)
node [pos=0.86, inner sep=0.75pt] [font=\small, color=ForestGreen!, opacity=1] {$\tikzxmark$}
node [pos=0.16, inner sep=0.75pt] [font=\small, color=ForestGreen!, opacity=1] {$\tikzxmark$}
node [pos=0, left][inner sep=0.75pt]  [font=\small,color=ForestGreen!  ,opacity=1 ]  {$\text{IV}\;$};

\draw[grad3=0.5]    (331.43,123.43) -- (331.43,138.43) ;
\draw[mar] [color=RoyalBlue!  ,draw opacity=1 ]   (90.43,138.43) -- (331.43,138.43) 
node [pos=0.61, inner sep=0.75pt] [font=\small, color=RoyalBlue!, opacity=1] {$\tikzxmark$}
node [pos=0.31, inner sep=0.75pt] [font=\small, color=RoyalBlue!, opacity=1] {$\tikzxmark$}
node [pos=0.27, inner sep=0.75pt] [font=\small, color=RoyalBlue!, opacity=1] {$\tikzxmark$}
node [pos=0, left] [inner sep=0.75pt]  [font=\small,color=RoyalBlue!  ,opacity=1 ]  {$\text{V}\;$};

\draw[grad4=0.5]    (90.43,138.43) -- (90.43,153.43) ;
\draw[ma] [color=Violet!, draw opacity=1 ] (90.43,153.43) -- (331.43,153.43) 
node [pos=0.67, inner sep=0.75pt] [font=\small, color=Violet!, opacity=1] {$\tikzxmark$}
node [pos=0.17, inner sep=0.75pt] [font=\small, color=Violet!, opacity=1] {$\tikzxmark$}
node [pos=0, left, inner sep=0.75pt]  [font=\small, color=Violet!, opacity=1] {$\text{VI}\;$};

\draw[grad5=0.5]    (331.43,153.43) -- (331.43,168.43) ;

\end{tikzpicture}}
\end{equation*}
\end{tcolorbox}
\end{minipage}\hfill
\begin{minipage}{0.50\textwidth}
\begin{tcolorbox}[colback=white, colframe=white, rounded corners]
\begin{equation*}
\hspace{-0.7cm}
\adjustbox{valign=c, width=\textwidth}{\tikzset{grad0/.style={
    postaction={
        decorate,
        decoration={
            markings,
            mark=at position \pgfdecoratedpathlength-0.5pt with {\arrow[Maroon!,line width=#1] {}; },
            mark=between positions 0 and \pgfdecoratedpathlength-0pt step 0.5pt with {
                \pgfmathsetmacro\myval{multiply(divide(
                    \pgfkeysvalueof{/pgf/decoration/mark info/distance from start}, \pgfdecoratedpathlength),100)};
                \pgfsetfillcolor{RedOrange!\myval!Maroon!};
                \pgfpathcircle{\pgfpointorigin}{#1};
                \pgfusepath{fill};}
}}}}

\newcommand\shadetext[2][]{%
  \setbox0=\hbox{{#2}}%
  \tikz[baseline=0]\path [#1] \pgfextra{\rlap{\copy0}} (0,-\dp0) rectangle (\wd0,\ht0);%
}

\tikzset{grad1/.style={
    postaction={
        decorate,
        decoration={
            markings,
            mark=at position \pgfdecoratedpathlength-0.5pt with {\arrow[RedOrange!,line width=#1] {}; },
            mark=between positions 0 and \pgfdecoratedpathlength-0pt step 0.5pt with {
                \pgfmathsetmacro\myval{multiply(divide(
                    \pgfkeysvalueof{/pgf/decoration/mark info/distance from start}, \pgfdecoratedpathlength),100)};
                \pgfsetfillcolor{Dandelion!\myval!RedOrange!};
                \pgfpathcircle{\pgfpointorigin}{#1};
                \pgfusepath{fill};}
}}}}

\tikzset{grad2/.style={
    postaction={
        decorate,
        decoration={
            markings,
            mark=at position \pgfdecoratedpathlength-0.5pt with {\arrow[Dandelion!,line width=#1] {}; },
            mark=between positions 0 and \pgfdecoratedpathlength-0pt step 0.5pt with {
                \pgfmathsetmacro\myval{multiply(divide(
                    \pgfkeysvalueof{/pgf/decoration/mark info/distance from start}, \pgfdecoratedpathlength),100)};
                \pgfsetfillcolor{ForestGreen!\myval!Dandelion!};
                \pgfpathcircle{\pgfpointorigin}{#1};
                \pgfusepath{fill};}
}}}}

\tikzset{grad3/.style={
    postaction={
        decorate,
        decoration={
            markings,
            mark=at position \pgfdecoratedpathlength-0.5pt with {\arrow[ForestGreen!,line width=#1] {}; },
            mark=between positions 0 and \pgfdecoratedpathlength-0pt step 0.5pt with {
                \pgfmathsetmacro\myval{multiply(divide(
                    \pgfkeysvalueof{/pgf/decoration/mark info/distance from start}, \pgfdecoratedpathlength),100)};
                \pgfsetfillcolor{RoyalBlue!\myval!ForestGreen!};
                \pgfpathcircle{\pgfpointorigin}{#1};
                \pgfusepath{fill};}
}}}}

\tikzset{grad4/.style={
    postaction={
        decorate,
        decoration={
            markings,
            mark=at position \pgfdecoratedpathlength-0.5pt with {\arrow[RoyalBlue!,line width=#1] {}; },
            mark=between positions 0 and \pgfdecoratedpathlength-0pt step 0.5pt with {
                \pgfmathsetmacro\myval{multiply(divide(
                    \pgfkeysvalueof{/pgf/decoration/mark info/distance from start}, \pgfdecoratedpathlength),100)};
                \pgfsetfillcolor{Violet!\myval!RoyalBlue!};
                \pgfpathcircle{\pgfpointorigin}{#1};
                \pgfusepath{fill};}
}}}}

\tikzset{grad5/.style={
    postaction={
        decorate,
        decoration={
            markings,
            mark=at position \pgfdecoratedpathlength-0.5pt with {\arrow[Violet!,line width=#1] {}; },
            mark=between positions 0 and \pgfdecoratedpathlength-0pt step 0.5pt with {
                \pgfmathsetmacro\myval{multiply(divide(
                    \pgfkeysvalueof{/pgf/decoration/mark info/distance from start}, \pgfdecoratedpathlength),100)};
                \pgfsetfillcolor{Black!\myval!Violet!};
                \pgfpathcircle{\pgfpointorigin}{#1};
                \pgfusepath{fill};}
}}}}

\tikzset{every picture/.style={line width=0.85pt}}
\begin{tikzpicture}[x=0.75pt,y=0.75pt,yscale=-1,xscale=1]
\tikzset{ma/.style={decoration={markings,mark=at position 0.5 with {\arrow[scale=0.7]{>}}},postaction={decorate}}}
\tikzset{ma2/.style={decoration={markings,mark=at position 0.3 with {\arrow[scale=0.7,Maroon!70!black]{>}}},postaction={decorate}}}
\tikzset{ma3/.style={decoration={markings,mark=at position 0.7 with {\arrow[scale=0.7,RoyalBlue!80!black]{>}}},postaction={decorate}}}
\tikzset{mar/.style={decoration={markings,mark=at position 0.5 with {\arrowreversed[scale=0.7]{>}}},postaction={decorate}}}

\draw[dash pattern=on 2pt off 1pt]    (251.93,81.57) -- (251.93,95) ;
\draw[dash pattern=on 2pt off 1pt]    (324.43,108.43) -- (324.43,123.43) ;
\draw[dash pattern=on 2pt off 1pt]    (203.43,108.43) -- (203.43,123.43) ;
\draw[dash pattern=on 2pt off 1pt]    (237.43,138.43) -- (237.43,153.43) ;

\draw[gradBlackToR=0.5]    (331.43,66.57) -- (331.43,81.57) ;
\draw[mar] [color=Maroon!  ,draw opacity=1 ]   (90.43,81.57) -- (331.43,81.57) 
node [pos=0.67, inner sep=0.75pt] [font=\small, color=Maroon!, opacity=1] {$\tikzxmark$}
node [pos=0.37, inner sep=0.75pt] [font=\small, color=Maroon!, opacity=1] {$\tikzxmark$}
node [pos=0, left][inner sep=0.75pt]  [font=\small,color=Maroon!  ,opacity=1 ]  {$\text{I}\;$}
node [midway, above, yshift=15pt, inner sep=0.75pt] [font=\small, color=black!, opacity=1] {$\textstyle R_F\{\textcolor{RoyalBlue}{F}{\shadetext[left color=RoyalBlue, right color=Violet, middle color=gray, shading angle=45]{$G$}}\}R_C\{\textcolor{Dandelion}{C}{\shadetext[left color=Dandelion, right color=ForestGreen, middle color=gray, shading angle=45]{$D$}}{\shadetext[left color=Dandelion, right color=ForestGreen, middle color=gray, shading angle=45]{$E$}}\}R_A\{\textcolor{Maroon}{A}{\shadetext[left color=Maroon, right color=RedOrange, middle color=gray, shading angle=45]{$B$}}\}$};
\draw[ma] [color=RedOrange!  ,draw opacity=1 ]   (90.43,95) -- (331.43,95) 
node [pos=0.67, inner sep=0.75pt] [font=\small, color=RedOrange!, opacity=1] {$\tikzxmark$}
node [pos=0, left][inner sep=0.75pt]  [font=\small,color=RedOrange!  ,opacity=1 ]  {$\text{II}\;$};
\draw[grad0=0.5]    (90.43,81.57) -- (90.43,95) ;

\draw [grad1=0.5]    (331.43,95) -- (331.43,108.43) ;
\draw[mar] [color=Dandelion!  ,draw opacity=1 ]   (90.43,108.43) -- (331.43,108.43) 
node [pos=0.47, inner sep=0.75pt] [font=\small, color=Dandelion!, opacity=1] {$\tikzxmark$}
node [pos=0.97, inner sep=0.75pt] [font=\small, color=Dandelion!, opacity=1] {$\tikzxmark$}
node [pos=0.27, inner sep=0.75pt] [font=\small, color=Dandelion!, opacity=1] {$\tikzxmark$}
node [pos=0, left][inner sep=0.75pt]  [font=\small,color=Dandelion!  ,opacity=1 ]  {$\text{III}\;$};

\draw[grad2=0.5]    (90.43,108.43) -- (90.43,123.43) ;
\draw[ma] [color=ForestGreen!  ,draw opacity=1 ]   (90.43,123.43) -- (331.43,123.43)
node [pos=0.97, inner sep=0.75pt] [font=\small, color=ForestGreen!, opacity=1] {$\tikzxmark$}
node [pos=0.47, inner sep=0.75pt] [font=\small, color=ForestGreen!, opacity=1] {$\tikzxmark$}
node [pos=0, left][inner sep=0.75pt]  [font=\small,color=ForestGreen!  ,opacity=1 ]  {$\text{IV}\;$};

\draw[grad3=0.5]    (331.43,123.43) -- (331.43,138.43) ;
\draw[mar] [color=RoyalBlue!  ,draw opacity=1 ]   (90.43,138.43) -- (331.43,138.43) 
node [pos=0.61, inner sep=0.75pt] [font=\small, color=RoyalBlue!, opacity=1] {$\tikzxmark$}
node [pos=0.31, inner sep=0.75pt] [font=\small, color=RoyalBlue!, opacity=1] {$\tikzxmark$}
node [pos=0, left] [inner sep=0.75pt]  [font=\small,color=RoyalBlue!  ,opacity=1 ]  {$\text{V}\;$};

\draw[grad4=0.5]    (90.43,138.43) -- (90.43,153.43) ;
\draw[ma] [color=Violet!, draw opacity=1 ] (90.43,153.43) -- (331.43,153.43) 
node [pos=0.61, inner sep=0.75pt] [font=\small, color=Violet!, opacity=1] {$\tikzxmark$}
node [pos=0, left, inner sep=0.75pt]  [font=\small, color=Violet!, opacity=1] {$\text{VI}\;$};

\draw[grad5=0.5]    (331.43,153.43) -- (331.43,168.43) ;

\end{tikzpicture}}
\end{equation*}
\end{tcolorbox}
\end{minipage}
\end{equation*}
    On the top left, we have a product of time-ordered products $T$, while on the top right, we have a product of anti-time-ordered products $\bar{T}$. The cartoon on the bottom left represents a mixed product of time- and anti-time-ordered products, while the cartoon on the bottom right represents a product of retarded products $R$. As before, time goes from right to left in each panel.
\end{mdexample}

The takeaway point here is that the set of operators spanned by multi-timefold Schwinger--Keldysh correlators such as \eqref{example 3fold} is, after $\theta$-functions bookkeeping, in 1-1 correspondence with the set identified in \eqref{eq:TOP}. Next, we will extend the familiar LSZ reduction formula revisited earlier to the OTOC we just introduced.

\subsubsection{Making manifest analyticity properties}

In what follows, we consider a few examples of interesting reduction formulae. As a warm-up, we consider once more conventional time-ordered amplitudes with the aim of establishing the so-called \defQ{optical theorem}/discontinuity/unitarity formula for $2\ot 2$ amplitudes. Later, applications at higher multiplicity will be discussed.

\begin{mdexample}
    \textbf{(Four-point amplitudes and discontinuities)}
In order to achieve our first goal, let us recall a useful implication of the stability condition
\begin{equation}
Sa^\dagger\ket{0}=a^\dagger\ket{0}=b^\dagger\ket{0}=S b^\dagger\ket{0}\,,
\end{equation}
following from the axioms introduced earlier, namely that 
\begin{equation}\label{eq:useful1}
    \textcolor{Maroon}{b_2^\dagger a_1^\dagger}\ket{0}=\textcolor{RoyalBlue}{b_1^\dagger a_2^\dagger}\ket{0}=b_1^\dagger b_2^\dagger\ket{0}\,.
\end{equation}
Using these, we can prove a useful intermediate result: in $34\ot 12$ kinematics, we have
\begin{equation}\label{eq:useful2}
    \begin{split}
        T(j_1j_2)\vacR&=T((a_1^\dagger-b_1^\dagger)(a_2^\dagger-b_2^\dagger))\vacR
        \\&=T(a_1^\dagger a_2^\dagger-a_1^\dagger b_2^\dagger-b_1^\dagger a_2^\dagger+ b_1^\dagger b_2^\dagger)\vacR\\&=(a_1^\dagger a_2^\dagger-\textcolor{Maroon}{b_2^\dagger a_1^\dagger}-\textcolor{RoyalBlue}{b_1^\dagger a_2^\dagger}+ b_1^\dagger b_2^\dagger)\vacR
        \\&=(a_1^\dagger a_2^\dagger- b_1^\dagger b_2^\dagger)\vacR\\&=\ket{12}-S^\dagger a_1^\dagger a_2^\dagger S\vacR\\&=(\mathbbm{1}-S^\dagger)\ket{12}\,.
    \end{split}
\end{equation}
We see from this simple calculation that $T(j_1j_2)\vacR$ simply picks up the interacting part of $S$ (it vanishes when the particles do not interact)! This identity can, of course, be applied to any state. For example, one can obtain the fully connected part of the amplitude as follows
\begin{equation}
    \begin{split}
       {}_\text{out\hspace{-0.1cm}}\bra{34} T(j_1j_2)\vacR&={}_\text{in\hspace{-0.1cm}}\bra{34}S(\mathbbm{1}-S^\dagger)\ket{12}\\&={}_\text{in\hspace{-0.1cm}}\bra{34}(S-\mathbbm{1})\ket{12}\\&=i\cM_{34\ot 12}\,
    \end{split}
\end{equation}
More interestingly, we can consider the correlator $\vac{\Tb{j_3j_4}\T{j_1j_2}}$. On the one hand, we can insert a complete basis of asymptotic states via the resolution of identity
\begin{equation}
    \mathbbm{1}=\sumint_X\ket{X}_{\mathrlap{\text{in}}}{\hspace{-0.4cm}\phantom{\bra{X}}}_{\mathrlap{\text{in}}\hspace{0.15cm}}\bra{X}=\sumint_X\ket{X}_{\mathrlap{\text{out}}}{\hspace{-0.2cm}\phantom{\bra{X}}}_{\mathrlap{\text{out}}\hspace{0.35cm}}\bra{X}\,,
\end{equation}
according to 
\begin{subequations}\label{eq:tc1}
    \begin{align}
\vac{\Tb{j_3j_4}\T{j_1j_2}}&=\sumint_X\vac{\Tb{j_3j_4}\ket{X}_{\mathrlap{\text{out}}}{\hspace{-0.2cm}\phantom{\bra{X}}}_{\mathrlap{\text{out}}\hspace{0.35cm}}\bra{X}\T{j_1j_2}}
        \\&=\sumint_X\bra{34}(\mathbbm{1}-S)\ket{X}_{\mathrlap{\text{out}}}{\hspace{-0.2cm}\phantom{\bra{X}}}_{\mathrlap{\text{out}}\hspace{0.35cm}}\bra{X}(\mathbbm{1}-S^\dagger)\ket{12}
        \\&=\sumint_X\bra{34}(\mathbbm{1}-S)S^\dagger\ket{X}_{\mathrlap{\text{in}}}{\hspace{-0.4cm}\phantom{\bra{X}}}_{\mathrlap{\text{in}}\hspace{0.15cm}}\bra{X}S(\mathbbm{1}-S^\dagger)\ket{12}\\&=\sumint_X\cM_{34\ot X}^\dagger\cM_{X\ot 12}\,.
    \end{align}
\end{subequations}
On the other hand, we can directly compute $\vac{\Tb{j_3j_4}\T{j_1j_2}}$ using \eqref{eq:useful2} twice:
\begin{equation}
    \begin{split}
        \vac{\Tb{j_3j_4}\T{j_1j_2}}&=(\vac{\Tb{j_3j_4})(\T{j_1j_2}})\\&=(\vac{(a_3 a_4- b_3 b_4))((a_1^\dagger a_2^\dagger- b_1^\dagger b_2^\dagger)})\,.
    \end{split}
\end{equation}
Dropping the forward terms and using
\begin{equation}\label{eq:MstarToMdag}
(\cM_{A\leftarrow B})^\ast=\<A|S|B\>^\ast=\<B|S^\dag|A\>=\cM^\dag_{B\leftarrow A}\,,
\end{equation}
we are left with 
\begin{equation}\label{eq:tc2}
    \begin{split}
        \vac{\Tb{j_3j_4}\T{j_1j_2}}&=-\vac{a_3 a_4b_1^\dagger b_2^\dagger}-\vac{b_3 b_4a_1^\dagger a_2^\dagger}\\&=-(\vac{b_1b_2 a_3^\dagger a_4^\dagger})^\dagger-\vac{b_3 b_4a_1^\dagger a_2^\dagger}\\&=i\cM_{34\ot 12}^\dagger-i\cM_{34\ot 12}\,.
    \end{split}
\end{equation}
Comparing both sides of \eqref{eq:tc1} and \eqref{eq:tc2} gives the optical theorem.
\end{mdexample}
\begin{mdexample}
\textbf{(The role of $R$-products in axiomatic field theory)}
Let us first illustrate why $R$-products are useful, again, with conventional time-ordered $2\ot 2$ scattering amplitudes. A first useful identity can be derived from the decomposition of the 2-point $R$-product in terms of $T$-products where both $p_1$ and $p_2$ have negative energies:
\begin{equation}\label{eq:rp1}
    \begin{split}
       \R{2}{j_1j_2} &=\T{j_1j_2}-\T{j_1}\T{j_2}\\&=T((a_1^\dagger-b_1^\dagger)(a_2^\dagger-b_2^\dagger))-(a_1^\dagger-b_1^\dagger)(a_2^\dagger-b_2^\dagger)\\&=(\cancel{a_1^\dagger a_2^\dagger}-b_2^\dagger a_1^\dagger-\cancel{b_1^\dagger a_2^\dagger}+\cancel{b_1^\dagger b_2^\dagger})-(\cancel{a_1^\dagger a_2^\dagger}- a_1^\dagger b_2^\dagger-\cancel{b_1^\dagger a_2^\dagger}+\cancel{b_1^\dagger b_2^\dagger})\\&=-[b_2^\dagger,a_1^\dagger]\,.
    \end{split}
\end{equation}
Note that physically, the terms involving $\bdag_1$  were expected to cancel because $j_1$ must be in the past lightcone of $j_2$ and therefore it cannot be an ``out'' measurement. Consequently, particle 1 cannot reach future infinity (it returns to past infinity). From there, an elementary computation using \eqref{eq:useful1} shows that
\begin{equation}\label{eq:ampRet1}
    \begin{split}
       {}_\text{out\hspace{-0.1cm}}\bra{34} \R{2}{j_1j_2}\vacR&=-{}_\text{out\hspace{-0.1cm}}\bra{34} [b_2^\dagger,a_1^\dagger]\vacR\\&=-{}_\text{out\hspace{-0.1cm}}\bra{34} b_2^\dagger a_1^\dagger\vacR+{}_\text{out\hspace{-0.1cm}}\bra{34} a_1^\dagger b_2^\dagger\vacR\\&=-\underbracket[0.4pt]{{}_\text{out\hspace{-0.1cm}}\bra{34} b_2^\dagger b_1^\dagger\vacR}_{\text{Forward!}}+{}_\text{out\hspace{-0.1cm}}\bra{34} a_1^\dagger a_2^\dagger\vacR\\&={}_\text{out\hspace{-0.1cm}}\bra{34} a_1^\dagger a_2^\dagger\vacR=\bra{34}S\ket{12}\,.
    \end{split}
\end{equation}
Thus, the conventional time-ordered scattering amplitude is also equal to the on-shell limit of some $\cR$ products.
What is interesting is that the $\cR$-product is supported over $x_2^0>x_1^0$. This guarantees that its Fourier transform 
\begin{equation}
    \begin{split}
        \mathcal{R}(p_1,p_2)=\int \d^\D x_1\d^\D x_2 \e^{-i(p_1\cdot x_1+p_2\cdot x_2)}{}_\text{out\hspace{-0.1cm}}\bra{34} \R{2}{j_1j_2}\vacR\,,
    \end{split}
\end{equation}
is analytic/converges in a certain domain, in which equal positive timelike imaginary parts are added to $p_2^\mu$ and $-p_1^\mu$. Indeed, under the relabeling $x_1=x_2+y$, we have
\begin{equation}
    \begin{split}
        \mathcal{R}(p_1,p_2)&=\int \d^\D y\e^{-ip_1\cdot y}\int \d^\D x_2 \e^{-i(p_1+p_2)\cdot x_2}{}_\text{out\hspace{-0.1cm}}\bra{34} \R{2}{j_1j_2}\vacR\,,
    \end{split}
\end{equation}

\be
x_2^0>x_1^0 \implies y^0=x_1^0-x_2^0<0 \implies y\in V^-\,.
\ee

Now, because $x_2^0>x_1^0$, we must have $y^0=x_1^0-x_2^0<0$ and we conclude that $y$ is in the past lightcone $V^-$ (past pointing). This means that the Fourier transform only converges provided that the first exponential decays, while the second oscillates. This happens precisely when 
\begin{equation}\label{eq:cond1}
     \text{Re}(-i y\cdot p_1)=y\cdot \text{Im}(p_1)<0 \quad \text{while} \quad p_1+p_2\in\mathbbm{R}^{1,3}\,.
\end{equation}

Since we work in mostly-plus signature, asking $y$ and $\text{Im}(p_1)$ to be timelike separated (implying that $\text{Im}(p_1)\in V^-$) is equivalent to the first condition above. Thus, \eqref{eq:ampRet1} indicates that the amplitude is \emph{at least} analytic provided  
that equal and opposite timelike imaginary parts are added to $p_1^\mu$ and $p_2^\mu$.

Of course, one could also represent the amplitude using the advanced product
\begin{equation}
    \begin{split}
        A_2(j_1j_2)=[j_1,j_2]\theta_{12}=-[j_2,j_1]\theta_{12}=-[j_2,j_1]+\R{2}{j_1j_2}\,.
    \end{split}
\end{equation}
Indeed, since it differs from the retarded product only by a commutator (which actually vanishes in this kinematic) involving no $\theta$-function,
we have 
\begin{equation}
   {}_\text{out\hspace{-0.1cm}}\bra{34} A_2(j_1j_2)\vacR={}_\text{out\hspace{-0.1cm}}\bra{34} \R{2}{j_1j_2}\vacR-{}_\text{out\hspace{-0.1cm}}\bra{34} \underbracket[0.4pt]{[j_2,j_1]\vacR}_{=0}=\eqref{eq:ampRet1}\,.
\end{equation}
Therefore, taking the Fourier transform of ${}_\text{out\hspace{-0.1cm}}\bra{34} A_2(j_1j_2)\vacR$, one would obtain a similar set of conditions to that in \eqref{eq:cond1}, but with the inequality sign reversed.
By ``edge-of-the-wedge'' type arguments, the equality of retarded and advanced representations suggests that the amplitude is analytic in a larger domain where equal and opposite imaginary parts are indeed added to $p_1^\mu$ and $p_2^\mu$, but with the timelike requirement relaxed---as required to be on-shell.

The existence of multiple reduction formulae for the same quantity has significant implications. In particular, they are often central in axiomatic proofs that scattering amplitudes are analytic, for example, in a neighborhood of the mass shell \cite{Bros:1964iho, Bros:1965kbd, Sommer:1970mr}. The argument also generalizes to $m\ot n$ scattering and has been used to demonstrate that amplitudes near the mass shell are essentially a finite sum of analytic functions \cite{Bros:1972jh}, with a single function sufficing for $2 \ot 2$ scattering. Moreover, while there exist only two distinct asymptotic observables at four points, namely $\cM$ and $\cM^\dag$, these can be written in terms of a significantly greater number of correlation functions which enjoy distinct analyticity properties. We expect that a solid understanding of these correlation functions will add fruitful insight into the analytic properties of amplitudes.

Another reduction formula can be obtained similarly to \eqref{eq:ampRet1} by inserting a retarded product between
one in and out \emph{single}-particle states. For example, focusing again on $34\ot12$ kinematics, inserting 
\begin{equation}\label{eq:rp2}
\begin{split}
    \cR_3\{j_3 j_2\}&= \T{j_3j_2}{-}\T{j_2}\T{j_3}\\&=\T{(b_3{-}a_3)(a_2^\dagger{-}b_2^\dagger)}{-}(a_2^\dagger{-}b_2^\dagger)(b_3{-}a_3)\\&= [b_3, a_2^\dag]\,,
\end{split}
\end{equation}
we get (after using stability once more)
\be
 \<4| \cR_3 \{ j_3j_2 \} |1\> =
 \<4| b_3\adag_2|1\> = 
 \<4| a_3 S\adag_2|1\> =
\<34|S|12\>\,.
\ee
This formula is relevant to analyticity at fixed-$t$ and crossing symmetry through the upper-half $s$-plane \cite{Bros:1965kbd}.  When the energies of particles $2$ and $3$ are flipped
\begin{equation}
\begin{split}
    \cR_3\{j_3 j_2\}&= \T{j_3j_2}-\T{j_2}\T{j_3}\\&=\T{(a^\dagger_3-b_3^\dagger)(b_2-a_2)}-(b_2-a_2)(a_3^\dagger-b_3^\dagger)\\&= [b_3^\dagger, a_2]\,,
\end{split}
\end{equation}
for the same correlator, we get
\be
\<4| \cR_3 \{ j_3j_2 \} |1\>=
 -\<4| a_2\bdag_3|1\> =
 -\<4| a_2 S^\dag \adag_3|1\> =
-\<4 2|S^\dag|13\>\,.
\ee
Thus, it reduces to $\cM^\dag$ instead of $\cM$. This is essentially the $2\to 2$ case of \defQ{crossing symmetry} in \eqref{eq:bdryCOnd} (see \eqref{eq:crossingFig2to2})!
The general case of \eqref{eq:bdryCOnd} is obtained in a similar way by not using the stability condition.

\end{mdexample}
We now describe for the first time a higher point application of retarded products.
\begin{mdexample}
    \textbf{(Higher point discontinuities and Steinmann relations)} In $234\ot 12$ kinematics, consider the on-shell limit:
\begin{equation}\label{5pt stein 1}
    \begin{split}
       -\vac{\R{5}{j_5j_4}\R{3}{j_3j_2}\adag_1} =~?\,
    \end{split}
\end{equation}
Its reduction in terms of familiar time-ordered objects combines \eqref{eq:rp1} (but without daggers) and \eqref{eq:rp2}. Indeed, using stability $\bra{0}a=\bra{0}b$ once again, we have
    \begin{align}
       -\vac{\R{5}{j_5j_4}\R{3}{j_3j_2}\adag_1} &=-\vac{(-[b_5,a_4])([b_3,\adag_2])\adag_1}\notag
       \\&=\vac{a_5a_4b_3\adag_2\adag_1}-\vac{b_4b_5b_3\adag_2\adag_1}
       \\&=\vac{a_5a_4S^\dag a_3 S\adag_2\adag_1}-\vac{a_4a_5a_3S\adag_2\adag_1}\notag\,.
    \end{align}
Now, we can make the following replacements: $S^\dagger=\mathbbm{1}-iT^\dagger$ and $S\mapsto iT$. The asymmetry between $S$ and $S^\dag$ has a simple interpretation: $\R{3}{j_3j_2}$ requires particle $3$ to be in the future lightcone of particle $2$; thus particles $1$ and $2$ must interact together \emph{before} particle $3$ is emitted, since all particles are stable and cannot freely radiate. Putting everything together, dropping the forward terms, and inserting a complete basis of asymptotic states $X$, we have 
\begin{equation}\label{eq:expandSteinmann}
    \begin{split}
        -\vac{\R{5}{j_5j_4}\R{3}{j_3j_2}\adag_1} &=i(-i)\vac{a_5a_4T^\dag a_3 T\adag_2\adag_1}\\&=\sumint_X\vac{a_5a_4T^\dag\ket{X}\bra{X}a_3 T\adag_2\adag_1}\\&=       \raisebox{-3em}{
\begin{tikzpicture}[line width=1]
\draw[line width = 1] (0,0.3) -- (-1.2,0.3) node[left] {\small$4$};
\draw[line width = 1] (0,-0.3) -- (-1.2,-0.3) node[left] {\small$5$};
\draw[line width = 1] (2,0.3) -- (3.2,0.3) node[right] {\small$2$};
\draw[line width = 1] (2,-0.3) -- (3.2,-0.3) node[right] {\small$1$};
\draw[line width = 1] (2,0) -- (1,-1) node[left] {\small$3$};
\filldraw[fill=gray!30, very thick](0,-0.3) rectangle (2,0.3);
\draw[] (1,0) node {$X$};
\filldraw[fill=gray!5, line width=1.2](0,0) circle (0.6) node {$-i\mathcal{M}^\dag$};
\filldraw[fill=gray!5, line width=1.2](2,0) circle (0.6) node {$i\mathcal{M}$};
\draw[dashed,Orange] (1,0.7) -- (1,-1.2);
\end{tikzpicture}}
    \\&\equiv {\text{Cut}}_{45}\boldsymbol{\delta}_5\,.
    \end{split}
\end{equation}
Thus, the correlator \eqref{5pt stein 1} is just a particular way of writing a unitarity cut.  It turns out that this representation has a very interesting feature. To see it, let us examine a bit its Fourier transform 
\begin{equation}\label{eq:momrep}
    \int \d^\D x_5\d^\D x_4\d^\D x_3\d^\D x_2 \e^{-i(p_5\cdot x_5+p_4\cdot x_4+p_3\cdot x_3+p_2\cdot x_2)}\vac{\R{5}{j_5j_4}\R{3}{j_3j_2}\adag_1}\,,
\end{equation}
and perform, say, the constant timelike shifts $x_3\mapsto x_3+\xi$ and $x_4\mapsto x_4+\xi$. Now, physically, if particles $3$ and $4$ are measured by a detector asymptotically far in the future ($\xi\to\infty$), we expect some $s_{34}$-channel non-analyticity in the momentum space representation \eqref{eq:momrep}.

However, because of the explicit dependence of the integrand on $\R{5}{j_5j_4}$ and $\R{3}{j_3j_2}$, the $\xi$-range is constrained as follows: 
\begin{equation}
    \R{3}{j_3j_2}\leadsto \adjustbox{valign=c}{\tikzset{every picture/.style={line width=0.75pt}}  
\begin{tikzpicture}[x=0.75pt,y=0.75pt,yscale=-1,xscale=1]
\draw    (276,104.14) -- (335.43,185.57) ; 
\draw    (276,185.57) -- (335.43,104.14) ;
\draw  [draw opacity=0] (306.4,96.05) .. controls (322.5,96.15) and (335.43,99.19) .. (335.43,102.93) .. controls (335.43,106.73) and (322.09,109.81) .. (305.64,109.81) .. controls (289.19,109.81) and (275.85,106.73) .. (275.85,102.93) .. controls (275.85,99.13) and (289.19,96.05) .. (305.64,96.05) .. controls (305.88,96.05) and (306.12,96.05) .. (306.36,96.05) -- (305.64,102.93) -- cycle ; \draw   (306.4,96.05) .. controls (322.5,96.15) and (335.43,99.19) .. (335.43,102.93) .. controls (335.43,106.73) and (322.09,109.81) .. (305.64,109.81) .. controls (289.19,109.81) and (275.85,106.73) .. (275.85,102.93) .. controls (275.85,99.13) and (289.19,96.05) .. (305.64,96.05) .. controls (305.88,96.05) and (306.12,96.05) .. (306.36,96.05) ;  
\draw  [draw opacity=0] (335.57,185.29) .. controls (335.57,185.36) and (335.58,185.43) .. (335.58,185.5) .. controls (335.58,189.3) and (322.24,192.38) .. (305.79,192.38) .. controls (289.43,192.38) and (276.16,189.34) .. (276,185.57) -- (305.79,185.5) -- cycle ; \draw   (335.57,185.29) .. controls (335.57,185.36) and (335.58,185.43) .. (335.58,185.5) .. controls (335.58,189.3) and (322.24,192.38) .. (305.79,192.38) .. controls (289.43,192.38) and (276.16,189.34) .. (276,185.57) ;  
\draw  [draw opacity=0][dash pattern={on 0.84pt off 2.51pt}] (276.19,186.29) .. controls (276.06,186.03) and (276,185.77) .. (276,185.5) .. controls (276,181.7) and (289.34,178.62) .. (305.79,178.62) .. controls (320.26,178.62) and (332.32,181.01) .. (335.02,184.17) -- (305.79,185.5) -- cycle ; \draw  [dash pattern={on 0.84pt off 2.51pt}] (276.19,186.29) .. controls (276.06,186.03) and (276,185.77) .. (276,185.5) .. controls (276,181.7) and (289.34,178.62) .. (305.79,178.62) .. controls (320.26,178.62) and (332.32,181.01) .. (335.02,184.17) ;  
\draw  [fill={rgb, 255:red, 0; green, 0; blue, 0 }  ,fill opacity=1 ] (303.5,144.86) .. controls (303.5,143.63) and (304.49,142.64) .. (305.71,142.64) .. controls (306.94,142.64) and (307.93,143.63) .. (307.93,144.86) .. controls (307.93,146.08) and (306.94,147.07) .. (305.71,147.07) .. controls (304.49,147.07) and (303.5,146.08) .. (303.5,144.86) node [left] [ color=RoyalBlue!, opacity=1] {$x_2$} -- cycle ;
\draw  [fill={rgb, 255:red, 0; green, 0; blue, 0 }  ,fill opacity=1 ] (309.5,122.86) .. controls (309.5,121.63) and (310.49,120.64) .. (311.71,120.64) .. controls (312.94,120.64) and (313.93,121.63) .. (313.93,122.86) .. controls (313.93,124.08) and (312.94,125.07) .. (311.71,125.07) .. controls (310.49,125.07) and (309.5,124.08) .. (309.5,122.86) node [left,xshift=0.1cm] [color=Maroon!, opacity=1] {$x_3$} -- cycle;

\end{tikzpicture}} \qquad \qquad \text{and} \qquad \qquad
    \R{5}{j_5j_4}\leadsto 
 \adjustbox{valign=c}{\tikzset{every picture/.style={line width=0.75pt}}        

\begin{tikzpicture}[x=0.75pt,y=0.75pt,yscale=-1,xscale=1]

\draw    (276,104.14) -- (335.43,185.57) ;

\draw    (276,185.57) -- (335.43,104.14) ;

\draw  [draw opacity=0] (306.4,96.05) .. controls (322.5,96.15) and (335.43,99.19) .. (335.43,102.93) .. controls (335.43,106.73) and (322.09,109.81) .. (305.64,109.81) .. controls (289.19,109.81) and (275.85,106.73) .. (275.85,102.93) .. controls (275.85,99.13) and (289.19,96.05) .. (305.64,96.05) .. controls (305.88,96.05) and (306.12,96.05) .. (306.36,96.05) -- (305.64,102.93) -- cycle ; \draw   (306.4,96.05) .. controls (322.5,96.15) and (335.43,99.19) .. (335.43,102.93) .. controls (335.43,106.73) and (322.09,109.81) .. (305.64,109.81) .. controls (289.19,109.81) and (275.85,106.73) .. (275.85,102.93) .. controls (275.85,99.13) and (289.19,96.05) .. (305.64,96.05) .. controls (305.88,96.05) and (306.12,96.05) .. (306.36,96.05) ;  
\draw  [draw opacity=0] (335.57,185.29) .. controls (335.57,185.36) and (335.58,185.43) .. (335.58,185.5) .. controls (335.58,189.3) and (322.24,192.38) .. (305.79,192.38) .. controls (289.43,192.38) and (276.16,189.34) .. (276,185.57) -- (305.79,185.5) -- cycle ; \draw   (335.57,185.29) .. controls (335.57,185.36) and (335.58,185.43) .. (335.58,185.5) .. controls (335.58,189.3) and (322.24,192.38) .. (305.79,192.38) .. controls (289.43,192.38) and (276.16,189.34) .. (276,185.57) ;  
\draw  [draw opacity=0][dash pattern={on 0.84pt off 2.51pt}] (276.19,186.29) .. controls (276.06,186.03) and (276,185.77) .. (276,185.5) .. controls (276,181.7) and (289.34,178.62) .. (305.79,178.62) .. controls (320.26,178.62) and (332.32,181.01) .. (335.02,184.17) -- (305.79,185.5) -- cycle ; \draw  [dash pattern={on 0.84pt off 2.51pt}] (276.19,186.29) .. controls (276.06,186.03) and (276,185.77) .. (276,185.5) .. controls (276,181.7) and (289.34,178.62) .. (305.79,178.62) .. controls (320.26,178.62) and (332.32,181.01) .. (335.02,184.17) ;  
\draw  [fill={rgb, 255:red, 0; green, 0; blue, 0 }  ,fill opacity=1 ] (303.5,144.86) .. controls (303.5,143.63) and (304.49,142.64) .. (305.71,142.64) .. controls (306.94,142.64) and (307.93,143.63) .. (307.93,144.86) .. controls (307.93,146.08) and (306.94,147.07) .. (305.71,147.07) .. controls (304.49,147.07) and (303.5,146.08) .. (303.5,144.86)  node [left] [ color=RoyalBlue!, opacity=1] {$x_5$} -- cycle ;
\draw  [fill={rgb, 255:red, 0; green, 0; blue, 0 }  ,fill opacity=1 ] (294.5,170.86) .. controls (294.5,169.63) and (295.49,168.64) .. (296.71,168.64) .. controls (297.94,168.64) and (298.93,169.63) .. (298.93,170.86) .. controls (298.93,172.08) and (297.94,173.07) .. (296.71,173.07) .. controls (295.49,173.07) and (294.5,172.08) .. (294.5,170.86)
node [right,xshift=0.1cm] [color=Maroon!, opacity=1] {$x_4$}
-- cycle ;

\end{tikzpicture}} 
\end{equation}
In other words, the integrand has support only for
\begin{subequations}
    \begin{align}
  \infty\ge x_3^\pm-x_2^\pm&>0\mapsto \xi^\pm>x_2^\pm-x_3^\pm\ge -\infty\,,\\
\infty\ge x_5^\pm-x_4^\pm&>0\mapsto \infty\ge  x_5^\pm-x_4^\pm>\xi^\pm\,,      
    \end{align}
\end{subequations}
such that 
\begin{equation}\label{eq:xiCond1}
    \infty\ge x_5^\pm-x_4^\pm>\xi^\pm>x_2^\pm-x_3^\pm\ge -\infty\,.
\end{equation}
Furthermore, since $\xi$ is timelike, we have a constraint on $\vec{\xi}_\perp$
\begin{equation}\label{eq:xiCond2}
  \xi^2=-\xi^+\xi^-+\vec{\xi}_\perp^2<0\implies -\sqrt{\xi^+\xi^-}<|\vec{\xi}_\perp|<\sqrt{\xi^+\xi^-}\,. 
\end{equation}
Conditions \eqref{eq:xiCond1} and \eqref{eq:xiCond2} put together prohibit past and future infinite timelike translations (i.e., limits $|\xi|\to \infty$). This indicates that we should not expect any $s_{34}$-channel singularity. In other words, we just argued that
\begin{equation}
    \text{Disc}_{s_{34}} \text{Cut}_{45}=0\,.
\end{equation}
This is an example of the so-called \defQ{Steinmann relations}. Of course, going through a similar reduction argument on the correlator $\vac{\R{4}{j_5j_4}\R{3}{j_3j_2}\adag_1}$, one could similarly argue that 
\begin{equation}
    \text{Disc}_{s_{35}} \text{Cut}_{45}=0\,.
\end{equation}
As before, this discussion suggests that the existence of multiple reduction formulae is central in revealing the full analytic properties of amplitudes.
\end{mdexample}
In the last few examples we examined, the Schwinger--Keldysh framework was not used. To demonstrate how it can be used to extract quantitative details about OTOCs, we next apply it to analyze the infrared (IR) divergences of the five-point in-in observable in \eqref{eq:5obs2}.

\subsection{Infrared divergences of in-in QED observables from the timefold}\label{sec:eikonal}

The purpose of this section is to analyze how the Schwinger--Keldysh picture introduced above can be used to understand the infrared (IR) properties of out-of-time-ordered observables. To do so, we will focus on a particular in-in QED observable, namely the inclusive measurement (``waveform'') of a photon in the background of two electrically charged particles (which we will take to be electrons); the analogous calculation in gravity is discussed in \cite{Caron-Huot:2023vxl} (see also \cite[Sec.~4]{chapterIsabellaEtAl}).
Using the blob pictures introduced earlier, this observable is represented as
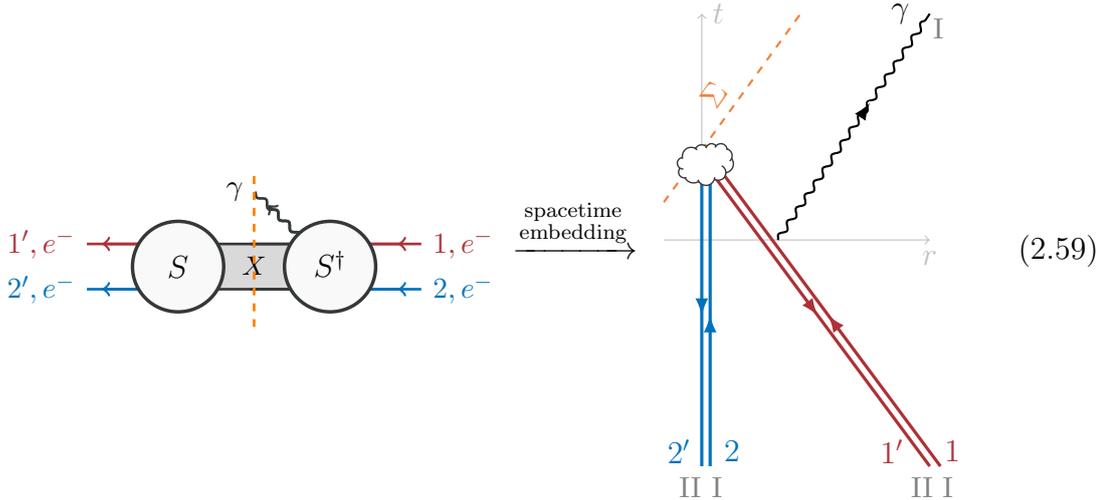
\begin{equation}
\adjustbox{valign=c}{\begin{tikzpicture}[line width=1,draw=charcoal]
    \draw[Maroon] (0,0.3) -- (-1.2,0.3) node[left] {\small$1', e^-$};
\draw[Maroon,->] (0,0.3) -- (-1,0.3);
\draw[RoyalBlue] (0,-0.3) -- (-1.2,-0.3) node[left] {\small$2', e^-$};
\draw[RoyalBlue,->] (0,-0.3) -- (-1,-0.3);
\draw[Maroon] (2,0.3) -- (3.2,0.3) node[right] {\small$1, e^-$};
\draw[Maroon,-<] (2,0.3) -- (3,0.3);
\draw[RoyalBlue] (2,-0.3) -- (3.2,-0.3) node[right] {\small$2, e^-$};
\draw[RoyalBlue,-<] (2,-0.3) -- (3,-0.3);
\draw[photon] (2,0) -- ++(135:1.40) node[left] {\small$\gamma$};
\draw[->,photon] (2,0) -- ++(135:1.15);
\filldraw[fill=gray!30](0,-0.3) rectangle (2,0.3);
\draw (2,0.3) -- (0.95,0.3);
\draw (2,-0.3) -- (0.95,-0.3);
\draw[] (1,0) node {$\medmath{X}$};
\filldraw[fill=gray!5, line width=1.3pt](0,0) circle (0.6) node[] {$S$};
\filldraw[fill=gray!5, line width=1.3pt](2,0) circle (0.6) node[yshift=1] {$S^\dag$};
\draw[dashed,orange] (1,1.2) -- (1,-0.8);
\end{tikzpicture}}
\xrightarrow{\substack{\text{spacetime}\\\text{embedding}}}
\adjustbox{valign=c}{
	\begin{tikzpicture}
		\tikzset{->-/.style={decoration={
					markings,
					mark=at position #1 with {\arrow{>}}},postaction={decorate}}}
	\draw[lightgray, ->] (-0.5,0) -- (3,0) node[below] {$r$};
	\draw[lightgray, ->] (0,-3) -- (0,3) node[right] {$t$};
	\draw[RoyalBlue, very thick, decoration={markings,
		mark=at position 0.5 with {\arrow{latex}},
	},postaction={decorate}] (0,1) -- (0,-3);
	\draw[RoyalBlue, very thick, decoration={markings,
		mark=at position 0.5 with {\arrow{latex}},
	},postaction={decorate}] (0.11,-3) -- (0.11,1);
	\draw[Maroon, very thick, decoration={markings,
		mark=at position 0.5 with {\arrow{latex}},
	},postaction={decorate}] (0.05,1) -- (3,-3);
	\draw[Maroon, very thick, decoration={markings,
		mark=at position 0.5 with {\arrow{latex}},
	},postaction={decorate}] (3.14,-3) -- (0.19,1);
	\draw[Orange, thick, dashed] (-0.5,0.5) -- (1.3,3) node[midway, rotate=55,above] {$\Sigma$};
	\node[scale=0.7] at (0,1) {\cloud};
	\draw[photon, thick] (1,0) -- (3,3) node[left] {$\gamma\;$};
	\draw[white, very thick, decoration={markings,
		mark=at position 0.5 with {\arrow[black]{latex}},
	},postaction={decorate}] (2.2,1.8) -- (2.22,1.83);
	\draw[RoyalBlue] (-0.3,-2.8) node {$2'$};
	\draw[RoyalBlue] (0.4,-2.8) node {$2$};
	\draw[Maroon] (2.5,-2.8) node {$1'$};
	\draw[Maroon] (3.3,-2.8) node {$1$};
  	\draw[gray] (-0.15,-3) node [below] {\small$\rtwo$};
	\draw[gray] (0.2,-3) node [below] {\small$\rone$};
    \draw[gray] (2.9,-3) node [below] {\small$\rtwo$};
    \draw[gray] (3.24,-3) node [below] {\small$\rone$};
    \draw[gray] (2.9,2.8) node [right] {\small$\rone$};
  \end{tikzpicture}}
\end{equation}
Here, the spacelike surface $\Sigma$ is arbitrary and need only to be in the causal past of photon measurement; in the Schwinger--Keldysh picture (right panel above), it plays the role of the ``turning point'' from fold I to II.

Our goal now is to extract the IR divergence of this observable from the Schwinger--Keldysh formalism. However, before we go into the quantitative details, let us spell out the physical expectation. Physically, we expect the IR divergences to arise from \emph{asymptotic} regions where the electrons follow eikonal lines and interact weakly via soft-photon exchanges. More precisely, these regions are associated to the $\textcolor{ForestGreen}{2}+\textcolor{Purple}{2}=4$ Feynman diagrams that describe the long-range photon exchanged between the electrons on the same timefold \emph{and} those on different timefolds 
\begin{equation}\label{eq:regions}
    \adjustbox{valign=c}{\begin{tikzpicture}[line width=1,draw=charcoal]
    \draw[Maroon] (0,0.3) -- (-1.5,0.3) node[left] {\small$\textcolor{gray}{(\text{II})}~1'$};
\draw[Maroon,->] (0,0.3) -- (-1,0.3);
\draw[RoyalBlue] (0,-0.3) -- (-1.5,-0.3) node[left] {\small$\textcolor{gray}{(\text{II})}~2'$};
\draw[RoyalBlue,->] (0,-0.3) -- (-1,-0.3);
\draw[ForestGreen,photon] (-1.2,0.3) -- (-1.2,-0.3);
\draw[Maroon] (2,0.3) -- (3.5,0.3) node[right] {\small$1~\textcolor{gray}{(\text{I})}$};
\draw[Maroon,-<] (2,0.3) -- (3,0.3) node[fill=white,pos=0.7]{};
\draw[RoyalBlue] (2,-0.3) -- (3.5,-0.3) node[right] {\small$2~\textcolor{gray}{(\text{I})}$};
\draw[RoyalBlue,-<] (2,-0.3) -- (3,-0.3) node[fill=white,pos=0.7]{};
\draw[ForestGreen,photon] (3.3,0.3) -- (3.3,-0.3);
\draw[photon] (2,0) -- ++(135:1.40) node[left] {\small$\textcolor{gray}{(\text{I})}~\gamma$};
\draw[->,photon] (2,0) -- ++(135:1.15);
\filldraw[fill=gray!30](0,-0.3) rectangle (2,0.3);
\draw (2,0.3) -- (0.95,0.3);
\draw (2,-0.3) -- (0.95,-0.3);
\draw[] (1,0) node {$\medmath{X}'$};
\filldraw[fill=gray!5, line width=1.3pt](0,0) circle (0.6) node[] {$S$};
\filldraw[fill=gray!5, line width=1.3pt](2,0) circle (0.6) node[yshift=1] {$S^\dag$};
\draw[dashed,Orange] (1,1.6) -- (1,-1.6) node[pos=0.7,right]{$\Sigma$};
\begin{scope}[xshift=-5,yshift=0]
    \draw[thick,photon,Purple] (-0.5,-0.3) .. controls (0.5,-1.9) and (1.5,-1.9) .. (3.3,0.3);
    \draw[thick,photon,Purple] (-0.5,0.3) .. controls (0.5,1.9) and (1.5,1.9) .. (3.3,-0.3);
\end{scope}
\end{tikzpicture}}
\end{equation}
where $X'$ is the complete set of states, apart from the photons that cross the cut at $\Sigma$.

Note that self-energy diagrams, which account for particles moving in their own fields, are not included. This will be explained in more detail later on. Moreover, soft interactions involving the $X'$ states do not need to be considered since these cancel out upon inclusively summing over $X'$---note that $X'$ never even appears in the Schwinger--Keldysh calculation if we place the timefold $\Sigma$ early enough---this is the essence of the celebrated Bloch--Nordsieck theorem \cite{Bloch:1937pw}. 

Now, a canonical way to extract the IR divergence is therefore to compute these one-loop Feynman diagrams, sum them up, and use QED exponentiation identities to get the all-order result. We can check that this procedure leads to the same result as the one we are about to derive by considering Wilson lines in the Schwinger--Keldysh formalism (see \eqref{eq:ExpCdef5} below). 

As hinted above, what we need to compute is the vacuum expectation value of the electrically charged eikonal lines on the Schwinger--Keldysh contour $\mathcal{C}$:
\begin{equation}\label{eq:ExpCdef}
\begin{split}
\text{Exp}^\mathcal{C}&=\bra{0}\mathcal{C}\exp\Big[\frac{ie}{\hbar}\sum_{j\in \{1,2,1',2'\}}\int\d \tau p_j^\mu A_\mu(\tau p_j)\Big]\ket{0}\\&=\bra{0}\mathcal{C}\exp\Big[\frac{ie}{\hbar}\sum_{j\in\{1,2\}}\int_{-\infty}^\Sigma\d \tau (p_j^\mu A_\mu(\tau p_j)- p_{j'}^\mu A_\mu(\tau p_{j'}))\Big]\ket{0}\,.
\end{split}
\end{equation}
Above, $e$ and $p_j^\mu$ are the electrons' charge and momenta (respectively), $A_\mu$ denotes the gauge field of the soft radiation, while $\tau$ parameterizes the electrons' straight line trajectories ($x_j^\mu=\tau p_j^\mu$). We will see later how the four Feynman diagrams mentioned above (see \eqref{eq:regions}) arise in the leading-order term in the eikonal regime (which we will refer to below as the ``\defQ{classical part}'') in the Shwinger--Keldysh picture.

We will focus here on the ``classical regime'' where $p_i-p_i'=\mathcal{O}(\hbar)$ (physically related to the width of wavepackets in \eqref{eq:5obs1a}); this approximation could be relaxed if desired.

Thus, we first extract the classical part of the integrand in \eqref{eq:ExpCdef}. To do so, it is convenient to introduce the following kinematics
\begin{equation}
    \begin{tikzpicture}[scale=0.6,thick,
baseline={([yshift=-0.4ex]current bounding box.center)}]
    \draw[line width=2] (-1,1) -- (0,0) -- (1,1)  (-1,-1) -- (0,0) -- (1,-1);
    \draw[photon] (-0.5,0) node[]{} -- +(-1.2,0);
    \filldraw[fill=gray!5, line width=1.2](0,0) circle (0.8);
    \node[] at (2,1.5) {$p_1=\bar{p}_1 + \frac{q_1}{2}$};
    \node[] at (-2,1.5) {$p_1'=\bar{p}_1 - \frac{q_1}{2}$};
    \node[] at (2,-1.5) {$p_2=\bar{p}_2 + \frac{q_2}{2}$};
    \node[] at (-2,-1.5) {$p_2'=\bar{p}_2 - \frac{q_2}{2}$};
    \node[] at (-3.5,0) {$k=q_1+q_2$};
\end{tikzpicture}
\end{equation}
In the classical regime, the momentum transfer is small, meaning that $q_i=\mathcal{O}(\hbar)$.  Similarly, the difference in gauge fields is correspondingly small
\begin{equation}
    A_\mu^{\text{diff}}(x)=A_\mu^{\text{I}}(x)-A_\mu^{\text{II}}(x)=\mathcal{O}(\hbar)\,.
\end{equation}
 Note that by momentum conservation, the barred momenta are orthogonal to the respective momentum transfer: $q_j\cdot \barp_j=0$. 

Using the above results, an elementary Taylor expansion around $q_j^\mu=0$ gives
\begin{equation}
    p_j^\mu A_\mu(\tau p_j)- p_{j'}^\mu A_\mu(\tau p_{j'})=\underbracket[0.4pt]{\barp_j^\mu A_\mu^{\text{diff}}(\tau \barp_j)+q_j\cdot \frac{\partial}{\partial \barp_{j}}\barp_j^\mu A_\mu^{\text{avg}}(\tau \barp_j)}_{\mathcal{O}(\hbar)}+\mathcal{O}(\hbar^2)\,,
\end{equation}
where we introduce the average of fields $A_\mu^{\text{avg}}(x)=\frac{1}{2}(A_\mu^{\text{I}}(x)+A_\mu^{\text{II}}(x))$. Plugging this back into \eqref{eq:ExpCdef}, we have (up to effects suppressed in $\hbar$)
\begin{equation}\label{eq:ExpCdef2}
\text{Exp}^\mathcal{C}=\bra{0}\mathcal{C}\exp\Big[\frac{ie}{\hbar}\sum_{j\in\{1,2\}}\int_{-\infty}^\Sigma\d \tau \Big(\barp_j^\mu A_\mu^{\text{diff}}(\tau \barp_j)+q_j\cdot \frac{\partial}{\partial \barp_{j}}\barp_j^\mu A_\mu^{\text{avg}}(\tau \barp_j)\Big)\Big]\ket{0}\,.
\end{equation}
Now, since both terms in the integrand are of $\mathcal{O}(\hbar)$, it is clear that the exponential argument is purely classical.
Consequently, from now on we set $\hbar=1$.

From there we proceed by making the following observation: in the IR, the gauge fields are Gau\ss ian and thus \emph{fully} characterized by their two-point functions. This means that the average of the exponential is equal to the exponential of half the ``variance'' 
\begin{equation}\label{eq:ExpCdef22}
\begin{split}
\text{Exp}^\mathcal{C}&=\exp\Big[\frac{-e^2}{2}\sum_{j\in\{1,2\}}\int_{-\infty}^\Sigma\d \tau\d \tau' \bra{0}\mathcal{C} \Lambda_j(\tau)\Lambda_j(\tau')\ket{0}\Big]\,.
\end{split}
\end{equation}
where we introduce the shorthand $\Lambda_j(\tau)=\barp_j^\mu A_\mu^{\text{diff}}(\tau \barp_j)+q_j\cdot \frac{\partial}{\partial \barp_{j}}\barp_j^\mu A_\mu^{\text{avg}}(\tau \barp_j)$.  This is the usual mechanism underlying the exponentiation of infrared divergences in QED and here we are simply applying it to the waveform.

Next, expanding $\bra{0}\mathcal{C} \Lambda_j(\tau)\Lambda_j(\tau')\ket{0}$ leads to 16 terms, many of which (14) vanish. For some terms, the vanishing is mathematically obvious. For example, from the largest time equation in \eqref{largest time equation}, we have 
\begin{equation}
   \bra{0} \cC A_\mu^{\text{diff}}(x) A_\mu^{\text{diff}}(y)\ket{0}=0\,,
\end{equation}
while the $A^{\text{avg}} A^{\text{avg}}$ terms come with two powers of $q$ which makes them sub-classical compared to the terms to be considered shortly.
For the remaining $A^{\text{diff}} A^{\text{avg}}$ terms, a calculation is needed.
Two of them involve ``fields moving in their own fields'' which vanish in dimensional regularization as they give scaleless integrals, as will be made explicit shortly.
Thus the calculation reduces to
\begin{equation}\label{eq:ExpCdef3}
\begin{split}
\text{Exp}^\mathcal{C}&{=}\exp\Big[{-}e^2\sum_{j\in\{1,2\}}q_j{\cdot} \frac{\partial}{\partial \barp_j}\barp_j^\mu\barp_{\bar{j}}^\nu\int_{-\infty}^\Sigma\d \tau\d \tau' \bra{0}\mathcal{C} A_\mu^{\text{avg}}(\tau \barp_j)A_\nu^{\text{diff}}(\tau' \barp_{\bar{j}})\ket{0}\Big]\,,
\end{split}
\end{equation}
where $\bar{j}\in \{1,2\}\setminus j$ and the factor of $\frac{1}{2}$ is cancelled by the factors of $2$ in the surviving cross-terms. Let us point out that expanding the bracket in the I-II basis would return a total of $2+2=4$ terms, which are in one-to-one correspondence with the four physically relevant Feynman diagrams anticipated earlier in \eqref{eq:regions}.  Using retarded propagators reduces the number of diagrams
that need to be considered in the classical limit.

Next, in order to perform the $\tau$ integrals in \eqref{eq:ExpCdef3}, we need to evaluate the leftover retarded propagators (defined earlier in \eqref{eq:retProp}) in, e.g., the Feynman gauge
\begin{equation}
    \bra{0}\mathcal{C} A_\mu^{\text{avg}}(x)A_\nu^{\text{diff}}(y)\ket{0}=\eta_{\mu\nu}\int \frac{\d^\D p}{(2\pi)^\D}\frac{-i\e^{-i p\cdot r}}{p^2-i\varepsilon p^0}\qquad \text{with}~r=x-y \,.
\end{equation}
A useful observation to perform this integral is to note that the result is Lorentz invariant. Assuming that $r$ is timelike, we can therefore boost in a frame where $\Vec{r}=0$, perform the integral in $p^0$ using the residue theorem, use spherical coordinates to perform the remaining spatial integrations and then boost the answer back to $\Vec{r}\neq0$. Setting $\D=4$ at the very end of this procedure gives
\begin{equation}
    \bra{0}\mathcal{C} A_\mu^{\text{avg}}(x)A_\nu^{\text{diff}}(y)\ket{0}=\frac{\eta_{\mu\nu}\theta_{xy}}{2\pi i}\delta[(x-y)^2]\,.
\end{equation}
Now using this expression, the (divergent) $\tau$ integrals can be performed, leading to
\begin{equation}\label{eq:ExpCdef4}
\begin{split}
\text{Exp}^\mathcal{C}&{=}\exp\Big[\frac{-i e^2}{4\pi}\sum_{j\in\{1,2\}}q_j\cdot \frac{\partial}{\partial \barp_j}\frac{\bary}{\sqrt{\bary^2-1}}\int_{-\infty}^{\Sigma}\d\log \tau\Big]\,,
\end{split}
\end{equation}
where we have used $\barp_i=\barm_i \barv_i$ as well as $\bary=-\barv_1\cdot\barv_2$. To regulate the logarithmically divergent integral, we introduce a cutoff by replacing the lower bound with the spatial separation between the incoming states $(\Delta x^\mu_{\text{in}}=\tau_{\text{min}} \Delta p^\mu)$ at the moment when they are shot into the bulk, while the upper bound is replaced by the impact parameter $(b^\mu=\tau_{\text{max}} \Delta p^\mu)$. This leads to
\begin{equation}\label{eq:ExpCdef42}
\begin{split}
\text{Exp}^\mathcal{C}&{=}\exp\Big[\frac{-ie^2}{4\pi}\sum_{j\in\{1,2\}}q_j\cdot \frac{\partial}{\partial \barp_j}\frac{\bary}{\sqrt{\bary^2-1}}\log\frac{\tau_{\text{max}}}{\tau_{\text{min}}}\Big]\,.
\end{split}
\end{equation}
Finally, computing the derivatives yields the logarithmically divergent time shift
\begin{equation}\label{eq:ExpCdef5}
\begin{split}
\text{Exp}^\mathcal{C}&{=}\exp\Big[-i 
\frac{e^2}{4\pi}\frac{k\cdot(\barp_1+\barp_2)}{\barm_1\barm_2}\frac{1}{(\bary^2-1)^{3/2}}\log\frac{\tau_{\text{max}}}{\tau_{\text{min}}}\Bigg]
\equiv \exp\Big[-i\Delta t \Delta E \Big]\,,
\end{split}
\end{equation}
where in the rest frame $\Delta E=k^0$.
We see that if we shot the initial states from infinitely far in the past at fixed energy, the in-in observable experiences a large logarithmic time delay (for a positron-electron pair, it would be a time advance).
This is due to the repulsive $\propto e^2/r^2$ between them, which delays the moment of collision relative to the straight-line trajectories of non-interacting particles which would be sent with identical initial conditions.
We point out that the conventional (time-ordered) amplitude $\mathcal{M}_{e^-e^-\gamma\ot e^-e^-}$ does not experience this time delay; this is because the phase in \eqref{eq:ExpCdef5} is exactly canceled by that from interactions between the out states (which comes with an opposite sign by momentum conservation).  Instead, the time-ordered amplitude would exhibit a large \emph{real} exponent $\e^{-\frac12 N_\gamma}$ that represents the probability of \emph{not} emitting any photon.  We see that infrared divergences for the inclusive waveform and exclusive 5-point amplitude describe vastly different physics. A similar setup in gravity is discussed in the contribution \cite[Sec.~4]{chapterIsabellaEtAl}.

\section[The FBI local transform and analyticity of multi-point scattering amplitudes]{The FBI local transform and analyticity of multi-point scattering amplitudes \\ \normalfont{\textit{Simon Caron-Huot}}}\label{sec:FBI}
The purpose of this final lecture is to provide a rough overview of \cite{Bros:1972jh,Bros:1971ghu} and to discuss the implications of causality on the analyticity of $m\ot n$ scattering amplitudes. 

One of the main historical hurdles in this program is the fact that conventional Fourier transforms of position-space correlators do not show clear convergence on the mass shell (this was pointed out earlier in Sec.~\ref{sec:genRed}). The challenge, therefore, involves devising a variant of the traditional Fourier transform that not only converges on the mass shell but also possesses convergence properties robust enough to study analyticity deep in the complex kinematic space.

To date, a clear and practical understanding of how to realize such a construction remains blurry. However, we believe that a key tool in achieving this goal is the so-called \defQ{Fourier--Bros--Iagolnitzer} (FBI) transform,
or \defQ{local Fourier transform} \cite{Bros:1971ghu}.

Below, we sketch a (very!) brief yet comprehensive overview of this intricate topic, which we believe could be useful for anyone targeting this goal. See also \cite[App.~A]{Caron-Huot:2023ikn}.

\subsection{Why (and when) are amplitudes analytic?} \label{sec:why?}

The story begins with the conventional Fourier transform of the (initially off-shell) position-space correlator:
\begin{equation}\label{eq:corr0}
    G(p_1,...,p_n)=\int \vac{T(j_1 \dots j_n)}\prod_{i=1}^n \d^\D x_i \e^{-ip_i\cdot x_i}\,.n
\end{equation}
We assume that the first singularity of the two-point functions of currents $j_i$ is at some mass $\tilde{m}>m$ which is strictly separated from the mass shell, i.e., that the two-point function of unamputated fields looks like:
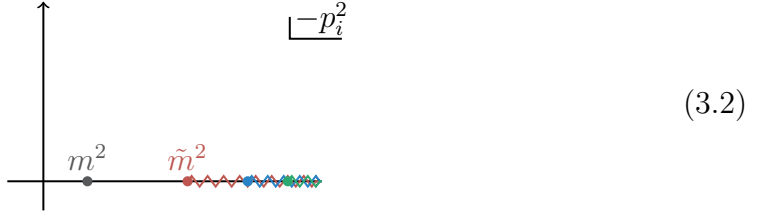
\begin{equation}\label{eq:path1}
         \adjustbox{valign=c}{
\tikzset{every picture/.style={line width=0.75pt}}  
\begin{tikzpicture}[x=0.75pt,y=0.75pt,yscale=-1,xscale=1]
\draw[<-]    (178,73.5) -- (178,178.5) ;
\draw[->]    (160,163.86) -- (317,163.86) ;
\draw[yshift=-13]   (327,109.5) -- (301,109.5) -- (301,98.5); 
\begin{scope}[yshift=-7]
    \draw[yshift=-13] (302,98.9) node [anchor=north west][inner sep=0.75pt]  [font=\normalsize]  {$-p_i^2$};
\end{scope}
  \draw[decorate, decoration={zigzag, segment length=6, amplitude=2}, Maroon!80]    (250,163.86) -- (317,163.86);
   \draw[Maroon!80,fill=Maroon!80,thick] (250,163.86) circle (2) node[above]{$\Tilde{m}^2$};

     \draw[Black!80,fill=Black!80,thick] (200,163.86) circle (2) node[above]{$m^2$};

  \draw[decorate, decoration={zigzag, segment length=6, amplitude=2}, RoyalBlue!80]    (280,163.86) -- (317,163.86);
   \draw[RoyalBlue!80,fill=RoyalBlue!80,thick] (280,163.86) circle (2);

     \draw[decorate, decoration={zigzag, segment length=6, amplitude=2}, ForestGreen!80]    (300,163.86) -- (317,163.86);
   \draw[ForestGreen!80,fill=ForestGreen!80,thick] (300,163.86) circle (2);
\end{tikzpicture}
    }
\end{equation}
To streamline the discussion, we first set $n=4$ and focus on $34\ot 12$ kinematics.
\begin{mdexample}
\textbf{(Four-point)} Under this assumption, stability ensures, for example
\begin{equation}\label{eq:vanish1}
\mathcal{G}_1=\vac{T(j_2j_3j_4)T(j_1)}=\sumint_X\bra{0}T(j_2j_3j_4)\ket{X}\bra{X}T(j_1)\ket{0}=0\,.
\end{equation}
Therefore, near the mass shell
\eqref{eq:corr0} can also be written as
    \begin{equation}\label{eq:corr0a}
    G(p_1p_2p_3p_4)=\int \left(\vac{T(j_1j_2j_3j_4)}-\mathcal{G}_1\right)\prod_{i=1}^4 \d^\D x_i\, \e^{-ip_i\cdot x_i}\equiv G_1\,.
\end{equation}
The interesting thing about \eqref{eq:corr0a} is that its support is smaller because the right-hand side manifestly vanishes unless 1 is in the future of either $2,3,4$:
\begin{equation}
    \text{supp}(G_1)=\{x:x_1\succcurlyeq x_2 \vee x_3 \vee x_4\}\,.
\end{equation}
Now, we could write \eqref{eq:corr0} in various other ways that have different threshold properties, by subtracting different products that vanish below the threshold.
There are in total eight terms similar to \eqref{eq:vanish1}, namely
\begin{equation}
    \mathcal{G}_{1\le i\le 4}=\vac{
T(\prod_{\substack{k=1\\ k\neq i}}^4j_k)T(j_i)} \quad \text{and} \quad \mathcal{G}_{5\le i\le 8}=\vac{T(j_i)T(\prod_{\substack{k=1\\ k\neq i-4}}^4j_k)}\,.
\end{equation}
In $34\ot 12$ kinematics, we further have
\begin{subequations}
    \begin{align}
        \mathcal{G}_9=\vac{T(j_1j_3)T(j_2j_4)}&=0\,,\\
        \mathcal{G}_{10}=\vac{T(j_2j_4)T(j_1j_3)}&=0\,,\\
        \mathcal{G}_{11}=\vac{T(j_1j_4)T(j_2j_3)}&=0\,,\\
        \mathcal{G}_{12}=\vac{T(j_2j_3)T(j_1j_4)}&=0\,,\\
        \mathcal{G}_{13}=\vac{T(j_1j_2)T(j_3j_4)}&=0\,.
    \end{align}
\end{subequations}
Note that $\mathcal{G}_{14}=\vac{T(j_3j_4)T(j_1j_2)}\neq0$ in these kinematics. Let us emphasize once more that 
\begin{equation}\label{eq:corr1}
   G_{i}=\int \left(\vac{T(j_1j_2j_3j_4)}-\mathcal{G}_i\right)\prod_{i=1}^4 \d^\D x_i \e^{-ip_i\cdot x_i}\,
\end{equation}
all define different functions and they only agree for real momenta which happen to be below certain thresholds.

What comes next relies on picking a point $p$ in
$34\ot 12$ kinematics.  
Define the \defQ{essential support} near $p$ as the intersection of the support of all representations which agree near $p$:
\begin{equation}\label{eq:ES2}
    \text{ES}_p=\bigcap_{i=1}^{13} \text{supp}(G_i(x))\,.
\end{equation}
The amazing claim is that the analyticity properties of
$G$ in a complex neighborhood of $p$ are the same as those of a Fourier transform of a function supported in $\text{ES}_p$.

The calculation of such intersections is an
interesting exercise in Boolean mathematics (try it!).
Some $G_i$ force certain points to be in the future of others, and some other $G_i$'s impose the opposite constraint. The result is that we need: $x_1=x_2$; $x_3=x_4$; and a certain causality relation between these two points:
\begin{equation} \label{eq:ES3}
    \text{ES}_{34\ot 12}=\{x:x_4\asymp x_3\succcurlyeq x_1\asymp x_2\}
\qquad
    \adjustbox{valign=c}{
\begin{tikzpicture}[baseline= {($(current bounding box.base)+(10pt,10pt)$)},line width=1, scale=0.7]
\coordinate (a) at (0,0) ;
\coordinate (b) at ($(b)+(-0:1)$);
\draw[postaction={decorate,decoration={markings,
		mark=at position 0.6 with {\arrow[black]{latex}}}}] (b) -- (a);
\draw[black] (b) -- ++ (30:1) node[right]{\footnotesize$2$};
\draw[black] (b) -- ++ (-30:1) node[right] {\footnotesize$1$};
\draw[black] (a) -- (-150:1) node[left] {\footnotesize$4$};
\draw[black] (a) -- (150:1) node[left] {\footnotesize$3$};
\fill[black,thick] (a) circle (0.07);
\fill[black,thick] (b) circle (0.07);
\end{tikzpicture}
}
\end{equation}
The physical interpretation of \eqref{eq:ES3} is that near any point $p$ in $34 \ot 12$ kinematics, $G(p_1p_2p_3p_4)$ has the same analyticity properties as the Fourier transform of the shown tree-level $s$-channel Feynman diagram, in which the internal state is allowed to propagate in any direction inside the light cone.

To see the implied analyticity properties, write $y\asymp x_3\asymp x_4$ and $x\asymp x_1\asymp x_2$ such that
\begin{equation}\label{eq:corr2}
    \begin{split}
        G(p_1p_2p_3p_4)&\simeq \int_{y\succcurlyeq x}\d^\D x\,\d^\D y\,\e^{-i(p_3+p_4)\cdot (y-x)} (\ldots)\,.
    \end{split}
\end{equation}
The integral converges exponentially provided
$\text{Re}(-i(p_3+p_3)\cdot (y-x))<0$ throughout, which is guaranteed if $\text{Im}(p_3+p_3)^\mu\in V^+$.
In $34\ot 12$ kinematics, this only amounts to a constraint on $s$:
\begin{equation}
    \begin{split}
        \text{Im}(s)&=\text{Im}(-(p_3+p_4)^2)\\&=-2\underbracket[0.4pt]{\text{Im}(p_3+p_4)}_{\in V^+}\cdot \underbracket[0.4pt]{\text{Re}(p_3+p_4)}_{\in V^+}>0\,.
    \end{split}
\end{equation}
Since the amplitude is a function of Mandelstam invariants, this means that \eqref{eq:corr2} converges, provided that we have $s+i\varepsilon$, but we can have either $t\pm i\varepsilon$. 
This is a statement about a small neighborhood of an arbitrary real point $p$.
\end{mdexample}

Note that the integrand $(\cdots)$ in \eqref{eq:corr2} is not quite the same as the original correlator, however it will be constructed explicitly below as part of the proof.

Of course, a similar story persists at higher multiplicities.  By automating the intersection calculus, we find the following essential supports:
\begin{mdexample}
\textbf{(Five-point)} Repeating this exercise for $n=5$, we find that the essential support of $\braket{345|S|12}$ is contained within
\begin{equation}
    \adjustbox{valign=c}{
\begin{tikzpicture}[baseline= {($(current bounding box.base)+(10pt,10pt)$)},line width=1, scale=0.7]
\coordinate (a) at (0,0) ;
\coordinate (b) at (1,0) ;
\coordinate (c) at ($(b)+(-0:1)$);
\draw[postaction={decorate,decoration={markings,
		mark=at position 0.6 with {\arrow[black]{latex}}}}] (b) -- (a);
\draw[postaction={decorate,decoration={markings,
		mark=at position 0.6 with {\arrow[black]{latex}}}}] (c) -- (b);
\draw[black] (b) -- ++ (110:1) node[right] {\footnotesize$3$};
\draw[black] (c) -- ++ (30:1) node[right]{\footnotesize$2$};
\draw[black] (c) -- ++ (-30:1) node[right] {\footnotesize$1$};
\draw[black] (a) -- (-150:1) node[left] {\footnotesize$5$};
\draw[black] (a) -- (150:1) node[left] {\footnotesize$4$};
\fill[black,thick] (a) circle (0.07);
\fill[black,thick] (b) circle (0.07) node[below,yshift=-15]{$x_4{\asymp} x_5 {\succcurlyeq} x_3{\succcurlyeq} x_1{\asymp} x_2$};
\fill[black,thick] (c) circle (0.07);
\end{tikzpicture}
}
\cup 
    \adjustbox{valign=c}{
\begin{tikzpicture}[baseline= {($(current bounding box.base)+(10pt,10pt)$)},line width=1, scale=0.7]
\coordinate (a) at (0,0) ;
\coordinate (b) at (1,0) ;
\coordinate (c) at ($(b)+(-0:1)$);
\draw[postaction={decorate,decoration={markings,
		mark=at position 0.6 with {\arrow[black]{latex}}}}] (b) -- (a);
\draw[postaction={decorate,decoration={markings,
		mark=at position 0.6 with {\arrow[black]{latex}}}}] (c) -- (b);
\draw[black] (b) -- ++ (110:1) node[right] {\footnotesize$4$};
\draw[black] (c) -- ++ (30:1) node[right]{\footnotesize$2$};
\draw[black] (c) -- ++ (-30:1) node[right] {\footnotesize$1$};
\draw[black] (a) -- (-150:1) node[left] {\footnotesize$5$};
\draw[black] (a) -- (150:1) node[left] {\footnotesize$3$};
\fill[black,thick] (a) circle (0.07);
\fill[black,thick] (b) circle (0.07) node[below,yshift=-15]{$ x_3{\asymp} x_5{\succcurlyeq} x_4{\succcurlyeq} x_1{\asymp} x_2$};
\fill[black,thick] (c) circle (0.07);
\end{tikzpicture}
}
\cup 
    \adjustbox{valign=c}{
\begin{tikzpicture}[baseline= {($(current bounding box.base)+(10pt,10pt)$)},line width=1, scale=0.7]
\coordinate (a) at (0,0) ;
\coordinate (b) at (1,0) ;
\coordinate (c) at ($(b)+(-0:1)$);
\draw[postaction={decorate,decoration={markings,
		mark=at position 0.6 with {\arrow[black]{latex}}}}] (b) -- (a);
\draw[postaction={decorate,decoration={markings,
		mark=at position 0.6 with {\arrow[black]{latex}}}}] (c) -- (b);
\draw[black] (b) -- ++ (110:1) node[right] {\footnotesize$5$};
\draw[black] (c) -- ++ (30:1) node[right]{\footnotesize$2$};
\draw[black] (c) -- ++ (-30:1) node[right] {\footnotesize$1$};
\draw[black] (a) -- (-150:1) node[left] {\footnotesize$4$};
\draw[black] (a) -- (150:1) node[left] {\footnotesize$3$};
\fill[black,thick] (a) circle (0.07);
\fill[black,thick] (b) circle (0.07) node[below,yshift=-15]{$ {x_3{\asymp} x_4{\succcurlyeq} x_5{\succcurlyeq} x_1{\asymp} x_2}$};
\fill[black,thick] (c) circle (0.07);
\end{tikzpicture}
}
\end{equation}
Each contribution is analytic in some complex cone around real momenta.  When the intersection of these three cones and the mass shell is nonempty, the amplitude is analytic; otherwise, it is \emph{a priori} just the sum of three functions which are each analytic in different cones; this example is further discussed in \cite{Bros:1972jh}.

We notice that once again the essential support diagrams are tree-level diagrams.
The situation changes at six-point.
\end{mdexample}
\begin{mdexample}
\textbf{(Six-point)} Repeating this exercise for $n=6$ is a more challenging combinatorial exercise. We find that the essential support of $\braket{3456|S|12}$ contains loop-like diagrams, e.g., 

\begin{equation}
        \adjustbox{valign=c}{
\begin{tikzpicture}[baseline= {($(current bounding box.base)+(10pt,10pt)$)},line width=1, scale=0.7]
\coordinate (a) at (0,0) ;
\coordinate (b) at (1,1) ;
\coordinate (c) at (2,0) ;
\coordinate (d) at (1,-1) ;
\draw[postaction={decorate,decoration={markings,mark=at position 0.6 with {\arrow[black]{latex}}}}] (b) -- (a);
\draw[postaction={decorate,decoration={markings,mark=at position 0.6 with {\arrow[black]{latex}}}}] (c) -- (b);
\draw[postaction={decorate,decoration={markings,mark=at position 0.6 with {\arrow[black]{latex}}}}] (c) -- (d);
\draw[postaction={decorate,decoration={markings,mark=at position 0.6 with {\arrow[black]{latex}}}}] (d) -- (a);
\draw[black] (b) -- ++ (180:1) node[left] {\footnotesize$3$};
\draw[black] (d) -- ++ (-180:1) node[left] {\footnotesize$6$};
\draw[black] (c) -- ++ (30:1) node[right]{\footnotesize$2$};
\draw[black] (c) -- ++ (-30:1) node[right] {\footnotesize$1$};
\draw[black] (a) -- (-150:1) node[left] {\footnotesize$5$};
\draw[black] (a) -- (150:1) node[left] {\footnotesize$4$};
\fill[black,thick] (a) circle (0.07);
\fill[black,thick] (d) circle (0.07);
\fill[black,thick] (b) circle (0.07) node[below,yshift=-45]{$x_4{\asymp} x_5 {\succcurlyeq} x_{3,6}{\succcurlyeq} x_1{\asymp} x_2$};
\fill[black,thick] (c) circle (0.07);
\end{tikzpicture}
}
\end{equation}
The physical significance of such diagrams remains unclear.
\end{mdexample}

\subsection{The FBI transform} 

How can we obtain Fourier representations like \eqref{eq:corr2} that converge on the mass shell?
The FBI (Fourier--Bros--Iagolnitzer) transform gives a nice constructive procedure \cite{Bros:1971ghu} that is essentially a judicious application of Stoke's theorem.
The procedure has been used by \cite{Bros:1972jh} to prove analyticity
in the vicinity of a real point $p$ but
we suspect that it may also be a good tool to study physics deep into the complex plane.

\def\FBI{\mathcal{F}_{\Phi,\tau}}

The construction uses the following ingredients. Let
\begin{equation}
    \begin{split}
        R&=\text{an open set in real momentum space}\,,\\
        \Phi(p)&=\text{a localizing function such that}~\text{Re}(\Phi(p))\begin{sqcases}
            <1 & \text{inside}~R=\stackrel{\circ}{R}\,,\\
            =1 &\text{on}~\partial R\,,
        \end{sqcases}\\
        \{G_i(p)\}&= \text{a set of functions that agree on $R$ (see, e.g., \eqref{eq:corr1})}.
    \end{split}
\end{equation}
Denote $\text{ES}_R$ the intersection of the supports of all $G_i(x)$. Then, for any $\tau>0$, the FBI transform is defined by:
\begin{equation}\label{eq:FBIdef}
    \FBI(x)=\int \frac{\d^\D p}{(2\pi)^\D} \e^{ip\cdot x-\tau \Phi(p)}G_i(p)\theta(p\in R)\,.
\end{equation}
There are a few important comments to make about this definition: 
\begin{enumerate}
    \item The transform $\FBI(x)$ does not depend on the index $i$; this is because of the explicit step function, which is only non-zero on the region $R$ where all the $G_i$'s agree. For instance, for the four-point example discussed in Sec.~\ref{sec:why?}, we would take $\theta(p\in R)=\theta(p_1^2+\Tilde{m}_1^2)\theta(p_2^2+\Tilde{m}_2^2)...$
    \item  Naively, we might think that $\mathcal{F}_{\Phi,\tau=0}(x)$ enjoys the support properties of each $G_{i}(x)$; that is, that $\mathcal{F}_{\Phi,\tau=0}(x)=0$ for $x\not\in \text{ES}_R$.
    The only obstruction to this is that $\theta(p\in R)$ is not an analytic function near the boundary of $R$: this prevents $\FBI(x)$ from decaying exponentially in any direction.
    \item Technically, it is straightforward to replace $\theta(p\in R)$ by a smooth approximation to a step function, so as to avoid ill-defined products of distributions.  However, no amount of smoothness will grant us exponential decay in directions outside $\text{ES}_R$.  The key will be to use $\tau$.
    Only by increasing $\tau$ proportionally to the distance from $\text{ES}_R$ can we suppress the contributions from the sharp transition of the step function at the boundary $\partial R$:
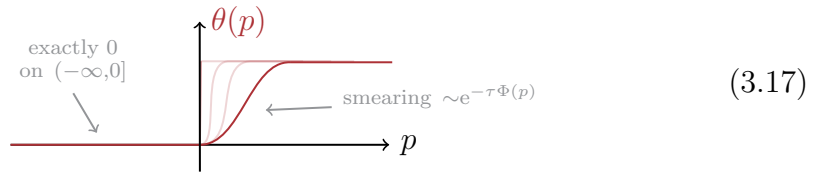
    \begin{equation}
 \adjustbox{valign=c}{\tikzset{every picture/.style={line width=0.75pt}}
\begin{tikzpicture}[x=0.75pt,y=0.75pt,yscale=-1,xscale=1]

\draw[->]    (71,172.5) -- (260,172.5) node[right]{$p$};
 
\draw [color=Maroon  ,draw opacity=1 ]   (71,172.5) -- (165.5,172.5) ;
 
\draw [color=Maroon  ,draw opacity=1 ]   (210.5,131) -- (261.5,131) ;

\draw [color=Maroon ,draw opacity=1 ]   (165.5,172.5) .. controls (186,172.5) and (190.5,130.5) .. (210.5,131) ;

\draw [color=Maroon  ,draw opacity=0.2 ]   (165.5,172.5) .. controls (186,172.5) and (171.5,130) .. (191.5,130.5) ;

\draw [color=Maroon  ,draw opacity=0.2 ]   (191.5,130.5) -- (242.5,130.5) ;

\draw [color=Maroon  ,draw opacity=0.2 ]   (165.5,172.5) .. controls (176,172.5) and (165,130.5) .. (180,130.5) ;

\draw [color=Maroon  ,draw opacity=0.2 ]   (180,130.5) -- (231,130.5) ;

\draw [color=Maroon  ,draw opacity=0.2 ]   (166,130.5) -- (217,130.5) ;

\draw [color=Maroon  ,draw opacity=0.5 ]   (166,130.5) -- (165.5,172.5) ;
 
\draw[->]    (165.5,186) -- (165.4,110.6) node[right]{$\textcolor{Maroon}{\theta(p)}$};

\draw[<-] [color=Gray  ,draw opacity=1 ]   (198.4,155.03) -- (231.29,153.57) 
node[pos=1, right]{$\substack{\text{smearing}~\sim\e^{-\tau \Phi(p)}\\ \textcolor{white}{dummy}}$};
 
\draw[->] [color=Gray  ,draw opacity=1 ]   (102.29,146.57) -- (114,166.43) node[pos=0, above]{$\substack{\text{exactly 0}\\\text{on}~(-\infty,0]}$};
\end{tikzpicture}}
\end{equation}
The boundary contributions are then manifestly suppressed by
\begin{equation}\label{eq:howFar}
    \e^{-\tau \Phi(p)}|_{p\in\partial R}=\e^{-\tau}\,.
\end{equation}
\item This will turn into get exponential decay outside of $\text{ES}_R$ by choosing suitable $\tau(x)\propto |x|$.
\end{enumerate}
Physically, for Gau\ss ian $\Phi(p)$, $\tau(x)\propto |x|$ means roughly that we convolve the correlator against wavepackets whose width in $x$-space grows like $\sqrt{|x|}$. This exponentially suppresses  unwanted contributions from $\partial R$ because these oscillate at the ``wrong frequency'' (compared with frequencies $p\in R$ which we are interested in).

The second step in the construction is to find an
inverse transform for a non-constant $\tau(x)$.

Thanks to standard Fourier inversion theorems, $G(p)$ for $p\in R$ can be recovered by integrating \eqref{eq:FBIdef} along any constant-$\tau$ slice, and then multiplying the result by $\e^{\tau \Phi(p)}$:
\begin{equation} \label{inverse FBI}
    G(p)=\int \d^\D x\, \e^{-ip\cdot x+\tau \Phi(p)}\mathcal{F}_{\Phi,\tau}(x)\, \qquad\mbox{($\tau$=constant)}.
\end{equation}
This does not yet grant us any analyticity in $p$, because $\mathcal{F}_{\Phi,\tau}(x)$ decays with $\tau$ but not with $x$.

Remarkably, the inverse transform \eqref{inverse FBI} can now be deformed to a general surface $\tau(x)$.
The construction is surprisingy simple and amounts to turning the integrand into the pullback of a closed $n$-form on $(\tau,x)\in\mathbbm{R}^{\D+1}$ space:
\begin{equation} \label{FBI deformed}
    G(p)=\underbracket[0.4pt]{\int_{\tau=\tau(x)}
    \e^{-ip\cdot x+\tau(x) \Phi(p)}
    \left(\mathcal{F}_{\Phi,\tau(x)}(x)\d^\D x +
    \sum_{i=1}^\D \mathcal{F}_{\Phi,\tau(x)}^i(x) \d\tau \d^{\D-1}x_i\right)}_{\text{(any admissible $\tau(x)$)}}
\,.
\end{equation}
We invite the reader to look at \cite{Bros:1971ghu} for details on the construction of $\mathcal{F}_{\Phi,\tau(x)}^i(x)$ ensuring that the parenthesis is a \emph{closed} form. By Stokes' theorem, the integral is then independent of the surface $\tau(x)$
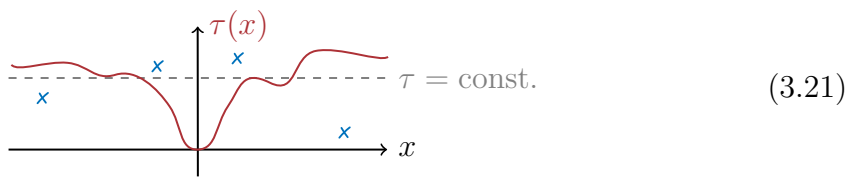
\begin{equation}
 \adjustbox{valign=c}{\tikzset{every picture/.style={line width=0.75pt}}
\begin{tikzpicture}[x=0.75pt,y=0.75pt,yscale=-1,xscale=1]
\draw [color=white!0]    (165.5,186) -- (165.4,110.6) 
node [xshift=-15,yshift=-15] [font=\small, color=RoyalBlue!, opacity=1] {$\tikzxmark$} node [xshift=15,yshift=-12] [font=\small, color=RoyalBlue!, opacity=1] {$\tikzxmark$}  node [xshift=55,yshift=-40] [font=\small, color=RoyalBlue!, opacity=1] {$\tikzxmark$} node [xshift=-58,yshift=-27] [font=\small, color=RoyalBlue!, opacity=1] {$\tikzxmark$};

\draw[->]    (71,172.5) -- (260,172.5) node[right]{$x$};
\begin{scope}[xshift=0,yshift=-27]
    \draw[dashed,color=gray]    (71,172.5) -- (260,172.5) node[right]{$\tau=\text{const.}$};
\end{scope}
 
\draw[->]    (165.5,186) -- (165.4,110.6) node[right]{$\textcolor{Maroon}{\tau(x)}$};

\begin{scope}[xshift=-15,yshift=-15]
\draw [color=Maroon  ,draw opacity=1 ]   (92.43,149.86) .. controls (95.68,153.65) and (109.91,147.83) .. (120.43,148.86) .. controls (130.95,149.89) and (133.88,158.07) .. (144.43,154.86) .. controls (154.98,151.65) and (165.25,161.66) .. (171.43,170.86) .. controls (177.61,180.05) and (175.97,192.84) .. (185.5,192.5) .. controls (195.03,192.16) and (194.17,180.22) .. (200.43,170.86) .. controls (206.69,161.49) and (207.43,151.86) .. (221.43,158.86) .. controls (235.43,165.86) and (232.42,145.14) .. (244.43,142.86) .. controls (256.43,140.58) and (276.52,148.79) .. (280.43,145.86) ;
\end{scope}
\end{tikzpicture}}
    \end{equation}
and therefore equal to $G(p)$ by the above comments, provided that the surface is homologous to a constant-$\tau$ slice. 

This is a practical result because it implies that we have the freedom to select the \emph{optimal} $\tau(x)$---i.e., that which produces the strongest exponential decay with $x$---to make the analyticity properties of $G(p)$ manifest in possibly large (complex) domains in $p$.
Below, we will illustrate how the choice of $\Phi(p)$ can greatly influence the size of the analyticity domains. 
\begin{mdexample}
    \textbf{(Gau\ss ian $\Phi$ and edge-of-the-wedge theorem)}
    As a simple application, to be continued below, consider two functions $G_1(p_1,p_2)$ and $G_2(p_1,p_2)$ of two complex variables $(G_i:\mathbb{C}^2\to \mathbb{C})$, which coincide in a real region $R$ which we take to be the unit ball:
 \begin{equation}
\tikzset{
pattern size/.store in=\mcSize, 
pattern size = 10pt,
pattern thickness/.store in=\mcThickness, 
pattern thickness = 0pt,
pattern radius/.store in=\mcRadius, 
pattern radius = 100pt}
\makeatletter
\pgfutil@ifundefined{pgf@pattern@name@_quadrillage}{
\pgfdeclarepatternformonly[\mcThickness,\mcSize]{_quadrillage}
{\pgfqpoint{0pt}{-\mcThickness}}
{\pgfpoint{\mcSize}{\mcSize}}
{\pgfpoint{\mcSize}{\mcSize}}
{
\pgfsetcolor{\tikz@pattern@color}
\pgfsetlinewidth{\mcThickness}
\pgfpathmoveto{\pgfqpoint{0pt}{\mcSize}}
\pgfpathlineto{\pgfpoint{\mcSize+\mcThickness}{-\mcThickness}}
\pgfusepath{stroke}
}}
\makeatother
\begin{tikzpicture}[x=0.75pt,y=0.75pt,yscale=-1,xscale=1]

\draw[->]    (71,172.5) -- (260,172.5) node[right]{$\text{Re}(p_1)$};
 
\draw[->]    (165.5,216) -- (165.4,110.6) node[right]{$\text{Re}(p_2)$};

\begin{scope}[xshift=-15,yshift=-15]

\draw  [color=Gray  ,draw opacity=1 ][fill=White  ,fill opacity=1 ] (160.5,192.5) .. controls (160.5,178.69) and (171.69,167.5) .. (185.5,167.5) .. controls (199.31,167.5) and (210.5,178.69) .. (210.5,192.5) .. controls (210.5,206.31) and (199.31,217.5) .. (185.5,217.5) .. controls (171.69,217.5) and (160.5,206.31) .. (160.5,192.5) -- cycle ;

\draw  [color=Maroon  ,draw opacity=1 ]
[pattern=_quadrillage,pattern size=6pt,pattern thickness=0.25pt,pattern radius=0pt, pattern color=Maroon]
(160.5,192.5) node[fill=white,xshift=18.7pt]{$R$} .. controls (160.5,178.69) and (171.69,167.5) .. (185.5,167.5) .. controls (199.31,167.5) and (210.5,178.69) .. (210.5,192.5) .. controls (210.5,206.31) and (199.31,217.5) .. (185.5,217.5) .. controls (171.69,217.5) and (160.5,206.31) .. (160.5,192.5) -- cycle ;
 \end{scope}
\end{tikzpicture}
    \end{equation}
For $R$ the unit ball, we can choose $\Phi$ to be a Gau\ss ian $\Phi(p_1,p_2)= p_1^2+p_2^2$, such that $\Phi(\partial R)=1$ and $0\le \Phi(\stackrel{\circ}{R})<1$. 
The FBI transform and its inverse read explicitly
\begin{subequations}\label{eq:FBIs}
\begin{align}
\label{FBI Gaussian direct}
 \mathcal{F}_{\tau}(x_1,x_2) &= \int_{p_1^2+p_2^2\leq 1} \frac{\d p_1 \d p_2}{(2\pi)^2} \e^{ip_1x_1+ip_2x_2-(p_1^2+p_2^2)\tau} G_i(p_1,p_2), \\
G(p_1,p_2) &= \int \d x_1\d x_2\, \e^{-ip_1x_1-ip_2x_2+(p_1^2+p_2^2)\tau}
\mathcal{F}_{}^*(x_1,x_2)
\,,\label{FBI Gaussian inverse}
\end{align}
\end{subequations}
where
\be
\mathcal{F}_{}^*(x_1,x_2)=\left( 1 + \sum_{j=1}^2 \frac{\d\tau}{\d x_j}\left(ip_j-\frac{\partial}{\partial x_j}\right)\right) \mathcal{F}_{}(x_1,x_2)
\Big|_{\tau=\tau(x_1,x_2)}.
\ee
The $\d\tau/\d x_j$ term originates from the correction term in
\eqref{FBI deformed} and ensures invariance under small deformation of the surface $\tau(x_1,x_2)$. (One should set $\tau\mapsto \tau(x)$ in ${\cal F}$ only after taking the partial derivative.)
\end{mdexample}

We will now use the FBI transform and its inverse to give a proof of the celebrated \emph{edge-of-the-wedge} theorem. For this we will suppose, moreover, that the conventional Fourier transforms of $G_1(p_1,p_2)$ and $G_2(p_1,p_2)$ are supported, respectively, on the first and third quadrants in $x$-space (and thus analytic in corresponding quadrants of ${\rm Im}\, p$):
\begin{equation}
\tikzset{every picture/.style={line width=0.75pt}} 
\begin{tikzpicture}[x=0.75pt,y=0.75pt,yscale=-1,xscale=1]
 
\draw[->]    (91,192.5) -- (280,192.5) node[right]{$x_1$};
 
\draw[->]    (185.7,255.4) -- (185.7,133.8) node[left]{$x_2$};

\draw  [draw opacity=0][fill=RoyalBlue  ,fill opacity=0.25 ] (185.7,133.8) node[fill opacity=1,xshift=35,yshift=-22]{$\tilde{G}_1(x_1,x_2)$} -- (279.91,133.8) -- (279.91,192.4) -- (185.7,192.4) -- cycle;

\draw  [draw opacity=0][fill=Maroon  ,fill opacity=0.25 ] (91,192.5) node[fill opacity=1,xshift=35,yshift=-22]{$\tilde{G}_2(x_1,x_2)$} -- (185.21,192.5) -- (185.21,255) -- (91,255) -- cycle ;
\end{tikzpicture}
\qquad
\begin{tikzpicture}[x=0.75pt,y=0.75pt,yscale=-1,xscale=1]
 
\draw[->]    (91,192.5) -- (280,192.5) node[right]{${\rm Im}\,p_1$};
 
\draw[->]    (185.7,255.4) -- (185.7,133.8) node[left]{${\rm Im}\,p_2$};

\draw  [draw opacity=0][fill=RoyalBlue  ,fill opacity=0.25 ] (185.7,133.8) node[fill opacity=1,xshift=35,yshift=-22]{$G_2\, \rm analytic$} -- (279.91,133.8) -- (279.91,192.4) -- (185.7,192.4) -- cycle;

\draw  [draw opacity=0][fill=Maroon  ,fill opacity=0.25 ] (91,192.5) node[fill opacity=1,xshift=35,yshift=-22]{$G_1\, \rm analytic$} -- (185.21,192.5) -- (185.21,255) -- (91,255) -- cycle ;
\end{tikzpicture} \label{eow rectangles}
\end{equation}
The basic \emph{edge-of-the-wedge} question is: what analyticity near $R$ does the agreement between $G_1(p_1,p_2)$ and $G_2(p_1,p_2)$ grants us?

Since the intersection of the $x$-supports is a single point ${\rm ES}_R=\{0\}$, the logic of the essential support suggests that the function should enjoy near $R$ the analytic properties of a Fourier integral supported on a point, i.e., be analytic in a full complex neighborhood of $R$. This is a non-trivial extension of the analyticity domain in the right panel of  
\eqref{eow rectangles}, where we add some small disk around the origin.  This is exactly the content of the \emph{edge-of-the-wedge theorem}, which we will now explain from the FBI perspective.

The key question to ask is: how small can we make
$\e^{-\tau(x_1,x_2)}$ outside the first quadrant, if we take the transform \eqref{FBI Gaussian direct} of $G_1$?
And how small can we make $\e^{-\tau(x_1,x_2)}$ outside the third quadrant if we use the transform with $G_2$?
Since the two transforms agree, our estimate of 
$\mathcal{F}_{\tau}$ will be the best of both.

Consider the transform starting with $G_1(p_1,p_2)$ in \eqref{FBI Gaussian direct} and suppose $x_1<0$.
Because $G_1$ is analytic in the third quadrant of ${\rm Im}\,p$, the factor $\e^{ip_1x_1}$ can be made smaller by deforming the contour into the lower-half $p_1$-plane,
where the integrand is analytic.
As usual with Fourier integrals, we estimate that it is ``as small as we can make it by deforming the contour.''

Since the contour is pinned to the real boundary of $\partial R$ where $\Phi(\partial R)=1$,
the \emph{optimal} decay we can hope to achieve is $\e^{-\tau}$.  Thus, the useful complex integration contours are contours of uniform suppression:
\begin{equation}\label{eq:master}
        \text{Re}(-ip{\cdot}x+\tau \Phi(p))=\tau\,.
\end{equation}
Following \cite{Bros:1971ghu}, we restrict attention to contours where the imaginary part of $p$ is along a single direction $q$,
such that $p_j = \bar{p}_j + i  \tau \lambda(\bar{p}_j)q_j$,
where $\bar{p}_j$, $q_j$ and $\lambda$ are all real
(and we always keep $x$'s and $\tau$ real).
The relation \eqref{eq:master} becomes
\begin{equation}\label{eq:master1}
        \lambda(\bar{p})\frac{q{\cdot}x}{\tau} + {\rm Re}\,\Phi(\bar{p}+iq\lambda(\bar{p}))=1\,.
\end{equation}
We can eliminate $\frac{q{\cdot}x}{\tau}$ by rescaling $\lambda$ and setting $\tilde{q}=q\frac{\tau}{q{\cdot}x}$, to characterize the constant-suppression contours as:
\begin{equation}\label{eq:master2}
        \Gamma_{\Phi,\tilde{q}} = \{p=\bar{p}+i \lambda(\bar{p})\tilde{q} :\,\, \bar{p}\in R\,
        \mbox{ and }\, \lambda(\bar{p})=1-{\rm Re}\,\Phi(\bar{p}+i\tilde{q}\lambda(\bar{p}))\}\,.
\end{equation}
We say that $\Gamma_{\Phi,\tilde{q}}$ is \emph{admissible} if it is a sensible contour (i.e., $\lambda$ real and positive everywhere) and is a deformation of the original one with $\lambda=0$ in \eqref{FBI Gaussian direct}. Contours with small enough $\tilde{q}$ are, of course, always admissible. With increasing magnitude of $\tilde{q}$ we may need to stop at a singularity of $G_i(p)$ or if $\lambda$ becomes complex.
This is summarized in the following cartoon:
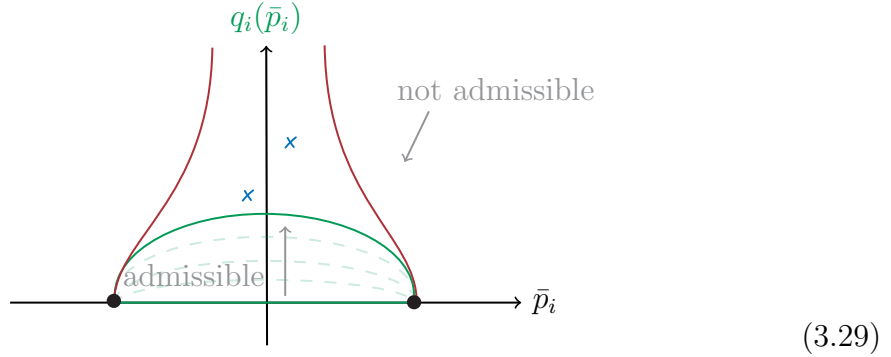
\begin{equation}
\tikzset{every picture/.style={line width=0.75pt}} 
\begin{tikzpicture}[x=0.75pt,y=0.75pt,yscale=-1.35,xscale=1.35]
\draw  [->]  (95,239.5) -- (284,239.5) node[right]{$\barp_i$};
\draw  [->]  (190,255.4) -- (189.7,143.8) node[above,color=ForestGreen]{$q_i(\barp_i)$};
\draw [draw opacity=0.2] (190,255.4) -- (189.7,143.8) node[pos=0.5,xshift=-7] [font=\small, color=RoyalBlue!, opacity=1] {$\tikzxmark$} node[pos=0.5,xshift=9,yshift=20] [font=\small, color=RoyalBlue!, opacity=1] {$\tikzxmark$};
\draw  [draw opacity=0][dash pattern={on 4.5pt off 4.5pt}] (133.71,236.71) .. controls (133.71,236.71) and (133.71,236.71) .. (133.71,236.71) .. controls (133.71,224.72) and (158.48,215) .. (189.04,215) .. controls (219.59,215) and (244.36,224.72) .. (244.36,236.71) -- (189.04,236.71) -- cycle ; \draw  [color=ForestGreen  ,draw opacity=0.2 ][dash pattern={on 4.5pt off 4.5pt}] (133.71,236.71) .. controls (133.71,236.71) and (133.71,236.71) .. (133.71,236.71) .. controls (133.71,224.72) and (158.48,215) .. (189.04,215) .. controls (219.59,215) and (244.36,224.72) .. (244.36,236.71) ;  
\draw  [draw opacity=0][dash pattern={on 4.5pt off 4.5pt}] (133.68,239.5) .. controls (133.68,239.5) and (133.68,239.5) .. (133.68,239.5) .. controls (133.68,230.91) and (158.67,223.95) .. (189.5,223.95) .. controls (220.33,223.95) and (245.32,230.91) .. (245.32,239.5) -- (189.5,239.5) -- cycle ; \draw  [color=ForestGreen  ,draw opacity=0.2 ][dash pattern={on 4.5pt off 4.5pt}] (133.68,239.5) .. controls (133.68,239.5) and (133.68,239.5) .. (133.68,239.5) .. controls (133.68,230.91) and (158.67,223.95) .. (189.5,223.95) .. controls (220.33,223.95) and (245.32,230.91) .. (245.32,239.5) ;  
\draw  [draw opacity=0][dash pattern={on 4.5pt off 4.5pt}] (133.36,240.35) .. controls (133.36,240.35) and (133.36,240.35) .. (133.36,240.35) .. controls (133.36,235.19) and (158.35,231) .. (189.18,231) .. controls (220.01,231) and (245,235.19) .. (245,240.35) -- (189.18,240.35) -- cycle ; \draw  [color=ForestGreen  ,draw opacity=0.2 ][dash pattern={on 4.5pt off 4.5pt}] (133.36,240.35) .. controls (133.36,240.35) and (133.36,240.35) .. (133.36,240.35) .. controls (133.36,235.19) and (158.35,231) .. (189.18,231) .. controls (220.01,231) and (245,235.19) .. (245,240.35) ;  
\draw  [draw opacity=0] (133.71,236.71) .. controls (133.71,236.71) and (133.71,236.71) .. (133.71,236.71) .. controls (133.71,219.99) and (158.48,206.43) .. (189.04,206.43) .. controls (219.59,206.43) and (244.36,219.99) .. (244.36,236.71) -- (189.04,236.71) -- cycle ; \draw [color=ForestGreen]  (133.71,236.71) .. controls (133.71,236.71) and (133.71,236.71) .. (133.71,236.71) .. controls (133.71,219.99) and (158.48,206.43) .. (189.04,206.43) .. controls (219.59,206.43) and (244.36,219.99) .. (244.36,236.71) ;  
\draw  [color=Maroon]  (133.36,240.35) .. controls (133.03,215.5) and (168.87,202.5) .. (169.79,144.5) ;
\draw  [color=Maroon]   (245.32,239.5) .. controls (245.63,214.64) and (212.14,201.64) .. (211.29,143.64) ;
\draw [color=ForestGreen]   (133.36,239.5) -- (245.32,239.5) ;
\draw [<-] [color=Gray  ,draw opacity=1 ]   (196.71,211) -- (196.71,237) node[pos=0.7,left,xshift=-3.5]{admissible};
\begin{scope}[yshift=-10,xshift=30]
    \draw [->] [color=Gray  ,draw opacity=1 ]   (209.86,181.57) -- (200.86,200.43) node[pos=0,above,xshift=25]{not admissible};
\end{scope}
\draw  [draw opacity=0][fill=Black  ,fill opacity=1 ] (130.71,238.86) .. controls (130.71,237.4) and (131.9,236.21) .. (133.36,236.21) .. controls (134.82,236.21) and (136,237.4) .. (136,238.86) .. controls (136,240.32) and (134.82,241.5) .. (133.36,241.5) .. controls (131.9,241.5) and (130.71,240.32) .. (130.71,238.86) -- cycle ;
\draw  [draw opacity=0][fill=Black  ,fill opacity=1 ] (241.71,239.36) .. controls (241.71,237.9) and (242.9,236.71) .. (244.36,236.71) .. controls (245.82,236.71) and (247,237.9) .. (247,239.36) .. controls (247,240.82) and (245.82,242) .. (244.36,242) .. controls (242.9,242) and (241.71,240.82) .. (241.71,239.36) -- cycle ;
\end{tikzpicture}
\end{equation}
Finally, for a given $x$, we find the optimal decay $e^{-\tau}$ by scanning over admissible $\tilde{q}$:
\begin{equation} \label{max tau}
    \tau_{\rm max}(x) = \max_{\tilde{q}\,\rm admissible} \tilde{q}{\cdot}x\,.
\end{equation}

In the case of Gau\ss ian $\Phi$,
the contours in \eqref{eq:master2} can be found analytically and are sensible provided that
$\sqrt{\tilde{q}_1^2+\tilde{q}_2^2}\leq \frac12$.
In order to be admissible for the $G_1$ version of \eqref{FBI Gaussian direct}, the vector $q$ must also lie within the third quadrant; for the $G_2$ version, it must lie in the first quadrant.
Since the two versions agree, the set of admissible contours is the union:
\begin{equation} \label{gaussian admissible}
 \Gamma_{\Phi,\tilde{q}} \mbox{ is admissible } \Leftrightarrow \sqrt{\tilde{q}_1^2+\tilde{q}_2^2}<\frac12 \mbox{ and } \tilde{q}_1\tilde{q}_2\geq0\,.
\end{equation}
The optimal decay in \eqref{max tau} is then
$\e^{-\tau_\text{max}(x)}=\e^{-C|x|}$
where $C$ is an $\mathcal{O}(1)$ constant depending on the direction of $x$.  Crucially, $C>0$ in all directions.

Let us reiterate how that happened.
If we only knew about the function $G_1(x_1,x_2)$, which is supported within the first $x$-quadrant, we could use the contours \eqref{eq:master2} to conclude that $\tau$ can be turned on to make ${\cal F}_{\Phi,\tau(x)}(x_1,x_2)$ decay in all directions \emph{outside} the first quadrant. We would not deduce any decay within the first quadrant.

Similarly, if we only knew about $G_2(x_1,x_2)$, we could prove exponential decay of ${\cal F}_{\Phi,\tau(x)}(x_1,x_2)$ for suitable $\tau$ in any direction
outside the third quadrant:
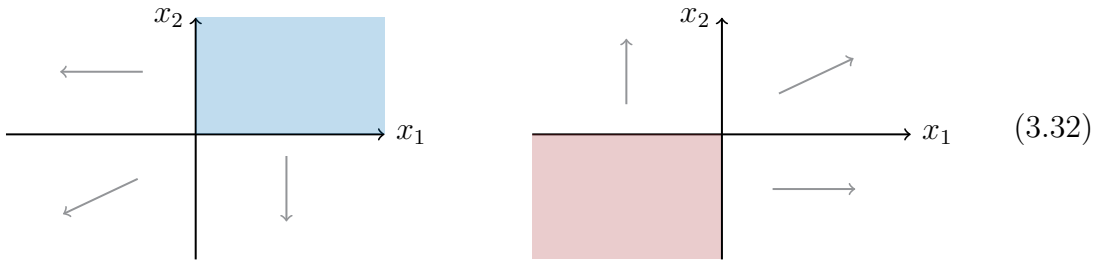
\begin{equation}
\adjustbox{valign=c}{\tikzset{every picture/.style={line width=0.75pt}} 
\begin{tikzpicture}[x=0.75pt,y=0.75pt,yscale=-1,xscale=1]
\draw[->]    (91,192.5) -- (280,192.5) node[right]{$x_1$};
\draw[->]    (185.7,255.4) -- (185.7,133.8) node[left]{$x_2$};
\draw  [draw opacity=0][fill=RoyalBlue  ,fill opacity=0.25 ] (185.7,133.8) 
-- (279.91,133.8) -- (279.91,192.4) -- (185.7,192.4) -- cycle;
\draw  [<-][color=Gray]  (118,161) -- (159.29,161) ;
\draw[<-][color=Gray]    (231,236.34) -- (231,203.38) ;
\draw[<-][color=Gray]    (119.46,232.61) -- (156.75,214.89) ;
\end{tikzpicture}}
~
\qquad
~
\adjustbox{valign=c}{\tikzset{every picture/.style={line width=0.75pt}} 
\begin{tikzpicture}[x=0.75pt,y=0.75pt,yscale=-1,xscale=1]
\draw[->]    (91,192.5) -- (280,192.5) node[right]{$x_1$};
\draw[->]    (185.7,255.4) -- (185.7,133.8) node[left]{$x_2$};
\draw  [draw opacity=0][fill=Maroon  ,fill opacity=0.25 ] (91,192.5) 
-- (185.21,192.5) -- (185.21,255) -- (91,255) -- cycle ;
\draw[->][color=Gray]    (138,177.34) -- (138,144.38) ;
\draw  [->][color=Gray]  (211,220) -- (252.29,220) ;
\draw [->][color=Gray]   (214.16,171.96) -- (251.45,154.24) ;
\end{tikzpicture}}
    \end{equation}
The magic of the argument is that since the two functions $G_1(p_1,p_2)$ and $G_2(p_1,p_2)$ agree on $R$, the two functions ${\cal F}_{\Phi,\tau}(x_1,x_2)$ defined by \eqref{FBI Gaussian direct} are in fact the \emph{same}, which hence must decay in every direction!

The final step is to evaluate the inverse transform \eqref{FBI Gaussian inverse} and determine for which complex $(p_1,p_2)$ it converges.
The main claim is that it does so for any sensible $p$ which lies on a contour $\Gamma_{\Phi,\tilde{q}}$
where $\tilde{q}$ is in the convex hull of the admissible set (i.e., \eqref{gaussian admissible}) \cite{Bros:1971ghu}.  This region can be plotted and looks like a slightly distorted four-ball with $|p|\lesssim 1$.

In comparison with other approaches to the edge-of-the-wedge theorem, an advantage of this method (stressed in
\cite{Bros:1971ghu}) is that the original functions $G_i(p_1,p_2)$ need not be analytic in any open cone, and isolated directions suffice. (For example, if we have a representation $G_3(x_1,x_2)$ that vanishes for $x_1>0$, then
$G_3(p_1,p_2)$ is not necessarily an analytic function of two variables, we only know that it is analytic in $p_1$.)  This feature is essential for combining the representations which appear in \eqref{eq:corr1}.
Also, the fact that we have to replace the set of admissible $\tilde{q}$ by its convex hull means that the FBI also generalizes the so-called ``tube theorems''.
This is quite impressive for a method which amounts to Stokes' theorem combined with single-variable contour deformations!

An intriguing feature is that the deduced domain of analyticity depends on the initial choice of $\Phi(p)$.  
This seems to be relatively unexplored.
Although Gau\ss ians suffice to prove local statements, the regions which arise in physics
tend to be large regions defined by inequalities of the type $-p_I^2\leq m_I^2$.
One wonders if judicious choices of $\Phi(p)$ could grant us access to deeper regions of the complex plane, for example simplifying proofs of crossing and its generalizations.

\newpage
\bibliographystyle{jhep}
\bibliography{references}

\providecommand{\href}[2]{#2}\begingroup\raggedright\begin{thebibliography}{10}

\bibitem{RecordsBook}
N.~Arkani-Hamed, P.~Benincasa, S.~Caron-Huot, M.~Correia, S.~Curry, M.~Giroux, F.~M. Haehl, H.~S. Hannesdottir, M.~T. Hansen, A.~Hebbar, G.~Isabella, J.~Lebl, M.~H.~G. Lee, S.~Mizera, E.~Pajer, C.~Pasiecznik, E.~Passemar, M.~Rangamani, B.~C. van Rees, F.~Vazão, A.~M. Wolz and Z.~Zhou, \emph{{Records from the S-Matrix Marathon: Selected Topics on Scattering Amplitudes}}. Springer Lecture Notes in Physics, 2025.

\bibitem{Caron-Huot:2023vxl}
S.~Caron-Huot, M.~Giroux, H.~S. Hannesdottir and S.~Mizera, \emph{{What can be measured asymptotically?}}, \href{https://doi.org/10.1007/JHEP01(2024)139}{\emph{JHEP} {\bfseries 01} (2024) 139} [\href{https://arxiv.org/abs/2308.02125}{{\ttfamily 2308.02125}}].

\bibitem{Caron-Huot:2023ikn}
S.~Caron-Huot, M.~Giroux, H.~S. Hannesdottir and S.~Mizera, \emph{{Crossing beyond scattering amplitudes}},  \href{https://arxiv.org/abs/2310.12199}{{\ttfamily 2310.12199}}.

\bibitem{Weinberg:1995mt}
S.~Weinberg, \emph{{The Quantum theory of fields. Vol. 1: Foundations}}. Cambridge University Press, 6, 2005, \href{https://doi.org/10.1017/CBO9781139644167}{10.1017/CBO9781139644167}.

\bibitem{Omnes:1960oya}
R.~Omn\`es, \emph{{D\'emonstration des relations de dispersion}}, {\emph{Les Houches Lect. Notes} {\bfseries 10} (1960) 317}.

\bibitem{Kosower:2018adc}
D.~A. Kosower, B.~Maybee and D.~O'Connell, \emph{{Amplitudes, Observables, and Classical Scattering}}, \href{https://doi.org/10.1007/JHEP02(2019)137}{\emph{JHEP} {\bfseries 02} (2019) 137} [\href{https://arxiv.org/abs/1811.10950}{{\ttfamily 1811.10950}}].

\bibitem{Hawking:1975vcx}
S.~W. Hawking, \emph{{Particle Creation by Black Holes}}, \href{https://doi.org/10.1007/BF02345020}{\emph{Commun. Math. Phys.} {\bfseries 43} (1975) 199}.

\bibitem{Maldacena:2015waa}
J.~Maldacena, S.~H. Shenker and D.~Stanford, \emph{{A bound on chaos}}, \href{https://doi.org/10.1007/JHEP08(2016)106}{\emph{JHEP} {\bfseries 08} (2016) 106} [\href{https://arxiv.org/abs/1503.01409}{{\ttfamily 1503.01409}}].

\bibitem{Bros:1964iho}
J.~Bros, H.~Epstein and V.~J. Glaser, \emph{{Some rigorous analyticity properties of the four-point function in momentum space}}, \href{https://doi.org/10.1007/BF02733596}{\emph{Nuovo Cim.} {\bfseries 31} (1964) 1265}.

\bibitem{Bros:1965kbd}
J.~Bros, H.~Epstein and V.~Glaser, \emph{{A proof of the crossing property for two-particle amplitudes in general quantum field theory}}, \href{https://doi.org/10.1007/BF01646307}{\emph{Commun. Math. Phys.} {\bfseries 1} (1965) 240}.

\bibitem{Bros:1972jh}
J.~Bros, V.~Glaser and H.~Epstein, \emph{{Local analyticity properties of the n particle scattering amplitude}}, \href{https://doi.org/10.5169/seals-114374}{\emph{Helv. Phys. Acta} {\bfseries 45} (1972) 149}.

\bibitem{Bros:1985gy}
J.~Bros, \emph{{Derivation of asymptotic crossing domains for multiparticle processes in axiomatic quantum field theory: A general approach and a complete proof for $2 \to 3$ particle processes}}, \href{https://doi.org/10.1016/0370-1573(86)90056-6}{\emph{Phys. Rept.} {\bfseries 134} (1986) 325}.

\bibitem{Mizera:2021fap}
S.~Mizera, \emph{{Crossing symmetry in the planar limit}}, \href{https://doi.org/10.1103/PhysRevD.104.045003}{\emph{Phys. Rev. D} {\bfseries 104} (2021) 045003} [\href{https://arxiv.org/abs/2104.12776}{{\ttfamily 2104.12776}}].

\bibitem{LSZ}
H.~Lehmann, K.~Symanzik and W.~Zimmermann, \emph{Zur {F}ormulierung quantisierter {F}eldtheorien}, {\emph{Nuovo Cimento} {\bfseries 1} (1955) 205}.

\bibitem{ruelle1961}
D.~Ruelle, \emph{Connection between wightman functions and green functions inp-space}, {\emph{Il Nuovo Cimento (1955-1965)} {\bfseries 19} (1961) 356}.

\bibitem{araki1960properties}
H.~Araki and N.~Burgoyne, \emph{{Properties of the Momentum Space Analytic Function}}, \href{https://doi.org/10.1007/BF02725943}{\emph{Nuovo Cim.} {\bfseries 18} (1960) 342}.

\bibitem{chapterHaehlRangamani}
F.~M. Haehl and M.~Rangamani, \emph{{Records from the S-Matrix Marathon: Schwinger-Keldysh Formalism}},  10, 2024, \href{https://arxiv.org/abs/2410.10602}{{\ttfamily 2410.10602}}.

\bibitem{britto2024cuttingedge}
R.~Britto, C.~Duhr, H.~S. Hannesdottir and S.~Mizera, \emph{{Cutting-Edge Tools for Cutting Edges}},  \href{https://arxiv.org/abs/2402.19415}{{\ttfamily 2402.19415}}.

\bibitem{Sommer:1970mr}
G.~Sommer, \emph{{Present state of rigorous analytic properties of scattering amplitudes}}, \href{https://doi.org/10.1002/prop.19700181102}{\emph{Fortsch. Phys.} {\bfseries 18} (1970) 577}.

\bibitem{chapterIsabellaEtAl}
M.~Correia, H.~S. Hannesdottir, G.~Isabella, A.~M. Wolz, Z.~Zhou, M.~Giroux, S.~Mizera and C.~Pasiecznik, \emph{{Records from the S-Matrix Marathon: Gravitational Physics from Scattering Amplitudes}},  12, 2024, \href{https://arxiv.org/abs/2412.11649}{{\ttfamily 2412.11649}}.

\bibitem{Bloch:1937pw}
F.~Bloch and A.~Nordsieck, \emph{{Note on the Radiation Field of the electron}}, \href{https://doi.org/10.1103/PhysRev.52.54}{\emph{Phys. Rev.} {\bfseries 52} (1937) 54}.

\bibitem{Bros:1971ghu}
J.~Bros and D.~Iagolnitzer, \emph{{Causality and local analyticity - mathematical study}}, {\emph{Ann. Inst. H. Poincare Phys. Theor.} {\bfseries 18} (1973) 147}.

\end{thebibliography}\endgroup

\end{document}